\newif\iffigs\figstrue
\documentclass[a4paper,11pt]{article}
\usepackage{latexsym,amssymb,lscape,graphics}
\usepackage{graphicx}        
\usepackage{longtable}
\usepackage{youngtab}
\usepackage{multirow}
\usepackage{color}
\usepackage{slashed,epsfig}
\usepackage{amsfonts}

\newtheorem{definizione}{Definition}[section]

\newcommand{\bd}{\begin{definizione}}
\newcommand{\ed}{\end{definizione}}
\def\Gslat{\relax{\slash\kern-.58em G}}
\def\Dslat{\relax{\slash\kern-.68em D}}
\def\Fslat{\relax{\slash\kern-.68em F}}
\def\Phislat{\relax{\slash\kern-.65em \Phi}}

\def\IC{\relax\,\hbox{$\inbar\kern-.3em{\rm C}$}}
\def\IG{\relax\,\hbox{$\inbar\kern-.3em{\rm G}$}}
\def\IB{\relax{\rm I\kern-.18em B}}
\def\ID{\relax{\rm I\kern-.18em D}}
\def\IL{\relax{\rm I\kern-.18em L}}
\def\IF{\relax{\rm I\kern-.18em F}}
\def\IH{\relax{\rm I\kern-.18em H}}
\def\II{\relax{\rm I\kern-.17em I}}
\def\IN{\relax{\rm I\kern-.18em N}}
\def\IP{\relax{\rm I\kern-.18em P}}
\def\IQ{\relax\,\hbox{$\inbar\kern-.3em{\rm Q}$}}
\def\bfzero{\relax\,\hbox{$\inbar\kern-.3em{\rm 0}$}}
\def\IK{\relax{\rm I\kern-.18em K}}
\def\IG{\relax\,\hbox{$\inbar\kern-.3em{\rm G}$}}
 \font\cmss=cmss10 \font\cmsss=cmss10 at 7pt
\def\IR{\relax{\rm I\kern-.18em R}}
\def\ZZ{\relax\ifmmode\mathchoice
{\hbox{\cmss Z\kern-.4em Z}}{\hbox{\cmss Z\kern-.4em Z}} {\lower.9pt\hbox{\cmsss Z\kern-.4em Z}}
{\lower1.2pt\hbox{\cmsss Z\kern-.4em Z}}\else{\cmss Z\kern-.4em Z}\fi}
\def\bfone{\relax{\rm 1\kern-.35em 1}}

\def\inbar{\vrule height1.5ex width.4pt depth0pt}
\def\bfzero{\relax{\rm I\kern-.18em 0}}
\def\bfone{\relax{\rm 1\kern-.35em 1}}

\DeclareFontFamily{U}{rsf}{} \DeclareFontShape{U}{rsf}{m}{n}{
  <5> <6> rsfs5 <7> <8> <9> rsfs7 <10-> rsfs10}{}
\DeclareMathAlphabet\Scr{U}{rsf}{m}{n}

\newcommand{\ft}[2]{{\textstyle\frac{#1}{#2}}}
\def\tilde{\widetilde}

\def\1bar{1\hskip -.275cm -}
\def\2bar{2\hskip -.275cm -}
\def\3bar{3\hskip -.275cm -}

\newsavebox{\uuunit}
\sbox{\uuunit}
                 {\setlength{\unitlength}{0.825em}
                      \begin{picture}(0.6,0.7)
                                      \thinlines
                                      \put(0,0){\line(1,0){0.5}}
                                      \put(0.15,0){\line(0,1){0.7}}
                                      \put(0.35,0){\line(0,1){0.8}}
                                     \multiput(0.3,0.8)(-0.04,-0.02){10}{\rule{0.5pt}{0.5pt}}
                      \end {picture}}

\makeatletter \@addtoreset{equation}{section} \makeatother

\def\bfone{\relax{\rm 1\kern-.35em 1}}

\def\bfone{\relax{\rm 1\kern-.35em 1}}
\font\cmss=cmss10 \font\cmsss=cmss10 at 7pt

\newcommand{\so}{\mathfrak{so}}

\newcommand{\usp}{\mathfrak{usp}}

\newcommand{\osp}{\mathfrak{osp}}

\newcommand{\rmi}{\mathrm{i}}

\begin{document}
\begin{titlepage}
\begin{center}
\vskip 0.2cm
\vskip 0.2cm
{\Large\sc Minimal $D=7$ Supergravity \\
\vskip 0.2cm
and \\
\vskip 0.3cm
the supersymmetry of Arnold-Beltrami Flux branes  }\\[1cm]
{\sc
P.~Fr\'e${}^{\; a,b,e}$,  { {\sc P.A.~Grassi}${}^{\; c,b}$}, L. Ravera${}^{\; d,b}$, and M. Trigiante$^{\;d,b}$}\\[10pt]
{${}^{a}$\sl\small Dipartimento di Fisica\footnote{Prof. Fr\'e is
presently fulfilling the duties of Scientific Counselor of the Italian Embassy in the Russian Federation,
Denezhnij pereulok, 5, 121002 Moscow, Russia. \emph{e-mail:} \quad {\small {\tt pietro.fre@esteri.it}}},
Universit\`a di Torino\\${}^{b}$INFN -- Sezione di Torino \\
via P. Giuria 1, \ 10125 Torino \ Italy\\}
\emph{e-mail:} \quad {\small {\tt fre@to.infn.it}}\\
\vspace{5pt}
{${}^c$\sl\small Dipartimento di Scienze e Innovazione Tecnologica,\\
Viale T. Michel 11, 15121 Alessandria, Italy
Universit\`a del Piemonte Orientale,}\\
\emph{e-mail:} \quad {\small {\tt pietro.grassi@uniupo.it}}\\
 \vspace{5pt}
{{\em $^d$\sl\small  DISAT, Politecnico di Torino,}}\\
{\em C.so Duca degli Abruzzi, 24, I-10129 Torino, Italy}~\quad\\
\emph{e-mail:}\quad   {\small {\tt lucrezia.ravera@polito.it, mario.trigiante@polito.it}}
\\
 \vspace{5pt}
{{\em $^{e}$\sl\small  National Research Nuclear University MEPhI\\ (Moscow Engineering Physics Institute),}}\\
{\em Kashirskoye shosse 31, 115409 Moscow, Russia}~\quad\\
\vspace{5pt}
\vspace{15pt}
\begin{abstract}
In this paper we study some properties of the newly found Arnold-Beltrami flux-brane solutions to the minimal $D=7$ supergravity. To this end we first single out the appropriate Free Differential Algebra
containing both a gauge $3$-form $\mathbf{B}^{[3]}$ and a gauge $2$-form $\mathbf{B}^{[2]}$: then we present
the complete rheonomic parametrization of all the generalized curvatures. This allows us to identify two-brane
configurations with Arnold-Beltrami fluxes in the transverse space with exact solutions of supergravity and
to analyze the Killing spinor equation in their background. We find that there is no preserved supersymmetry
if there are no additional translational Killing vectors. Guided by this principle we explicitly construct
Arnold-Beltrami flux two-branes that preserve $0$, $\ft 18$ and $\ft 14$ of the original supersymmetry.
Two-branes without fluxes are instead $\mathrm{BPS}$ states and preserve $\ft 12$ supersymmetry. For each
two-brane solution we carefully study its discrete symmetry  that is always given by some appropriate
crystallographic group $\Gamma$. Such symmetry groups $\Gamma$  are transmitted to the $D=3$ gauge theories on
the brane world--volume that occur in the gauge/gravity correspondence. Furthermore we illustrate the
intriguing relation between gauge fluxes in  two-brane solutions and hyperinstantons in $D=4$ topological
sigma-models.
\end{abstract}
\end{center}
\end{titlepage}
\tableofcontents \noindent {}
\newpage
\section{Introduction}
Minimal Supergravity in $D=7$ contains 16 supercharges and it is usually named $\mathcal{N}=2$ since the 16
supercharges are arranged into a pair of pseudo-Majorana spinors.
\par
The Poincar\'e (ungauged) version of the theory has been constructed independently by  Townsend and van
Nieuwenhuizen in \cite{PvNT} and by Salam and Sezgin in \cite{SalamSezgin} in two different formulations that
use respectively a three-form gauge field $\mathbf{B}^{[3]}_{\mu\nu\rho}$ and a two-form gauge field
$\mathbf{B}^{[2]}_{\mu\nu}$, in addition to the graviton $g_{\mu\nu}$, the gravitino $\Psi_{A|\mu}^{\alpha}$
($\alpha=1,\dots,8$, $\mu\,=\,0,1,\dots,6$, $A\, = \, 1,2$), the dilatino $\chi_{A}^\alpha$, three gauge
fields $\mathcal{A}^\Lambda_\mu$ ($\Lambda \, = \, 1,2,3$) and the dilaton $\phi$, that are common to both
formulations. From the on-shell point of view the number of degrees of freedom described by either
$\mathbf{B}^{[3]}_{\mu\nu\rho}$ or $\mathbf{B}^{[2]}_{\mu\nu}$ is the same and the two types of gauge fields
are electric-magnetic dual to each other.
\par
The gauging of the theory was also independently considered both in \cite{PvNT} and in \cite{SalamSezgin}. The
coupling of minimal $D=7$ supergravity to $n$ vector multiplets was constructed by Bergshoeff et al in
\cite{bershoffo1} on the basis of the two-form formulation and shown to be based on the use of the coset
manifold:
\begin{equation}\label{targuccio}
    \mathcal{M}_{3n+1} \, = \, \mathrm{SO(1,1)} \,\times \,\frac{\mathrm{SO(3,n)}}{\mathrm{SO(3)\times SO(n)}}
\end{equation}
as scalar manifold that encodes the spin zero degrees of freedom of the theory.
\par
In all the quoted references the construction was done using the Noether coupling procedure, up to
four-fermion terms in the Lagrangian and up to two-fermion and three-fermion terms in the transformation
rules. Correspondingly the on-shell closure of the supersymmetry algebra was also checked only up to such
terms.
\par
There is a renewed interest in this supergravity theory in relation with the classification of Arnold-Beltrami
fields \cite{beltramus} recently obtained by one of us, in a different collaboration, in \cite{arnolderie}. These fields, originally
introduced by Beltrami as solutions of the first order equation that bears his name \cite{beltramus}, were
shown to have high relevance in mathematical hydrodynamics by Arnold who proved a famous theorem according to
which  the only flows capable of admitting chaotic streamlines are the Beltrami flows
\cite{arnoldus,ArnoldBook,contactgeometria}. This theorem originated a vast literature on the so named
ABC-flows that correspond to the simplest solutions of Beltrami equation \cite{Henon,ABCFLOW}. The Beltrami
vector fields live on three-dimensional tori and in mathematical hydrodynamics are interpreted as velocity
fields of some fluid. They can also be used as compactification fluxes in the transverse space to the world
volume of $2$-brane solutions of $D=7$ supergravity theory. This new interpretation of Beltrami fields,
jocosely described by the authors as a Sentimental Journey from Hydrodynamics to Supergravity, was proposed in
\cite{arnoldtwobranes}. In this way the rich discrete symmetries of Arnold-Beltrami fields that are now turned
\textit{from flows into fluxes} can be transmitted to the three dimensional gauge theories living on the world
volume of the two-brane. Another intriguing relation of this type of $\mathrm{3D}$-vector fields with the
\textit{tri-holomorphic hyperinstantons}, namely with the instanton configurations of four-dimensional
sigma-models that are singled out by the topological twist, was recently pointed in \cite{pietroantoniosorin}.
\par
\begin{figure}[!hbt]
\begin{center}
\iffigs
\includegraphics[height=105mm]{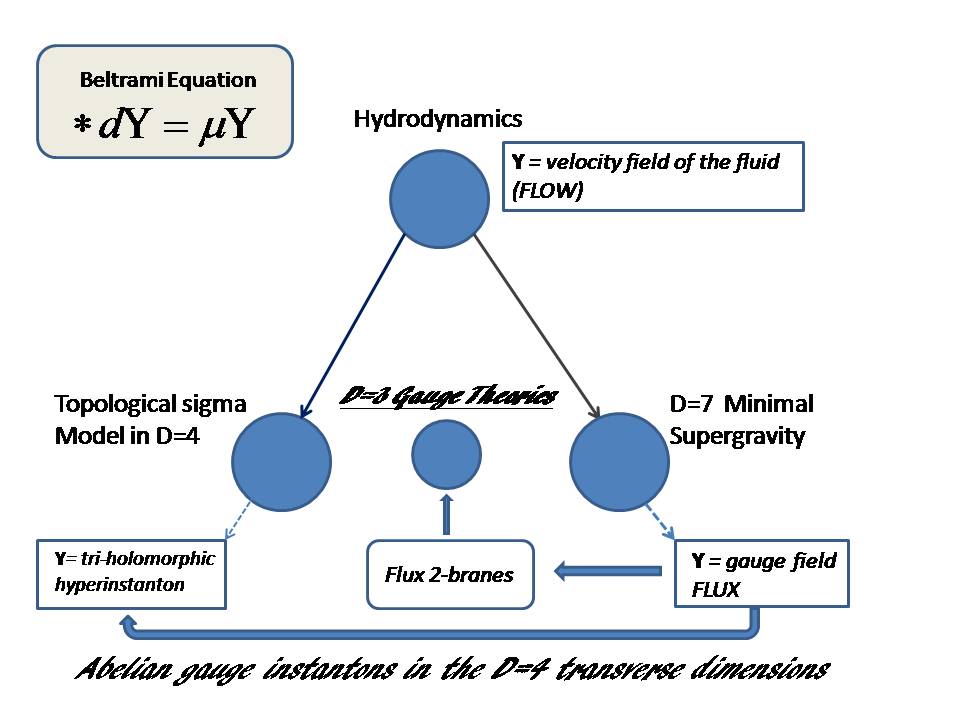}
\else
\end{center}
 \fi
\caption{\it \label{figuroide}}
 \iffigs
 \hskip 1cm \unitlength=1.1mm
 \end{center}
  \fi
\end{figure}
The intriguing set of multi-sided relations implied by  different interpretations of  Beltrami vector fields
is graphically summarized in fig.\ref{figuroide} which provides a sort of conceptual map for the present
paper.
\par
 In \cite{arnoldtwobranes} the explicit construction of $2$-brane solutions with Arnold-Beltrami fluxes was performed
but their embedding in $d=7$ supergravity was not discussed and what is the most relevant issue, namely the
residual supersymmetry that they might preserve, was not explored. This is the main goal of the present paper.
\par
With this motivation,  we have first reconsidered the construction of minimal $D=7$ supergravity  in the
approach based on Free Differential Algebras and rheonomy (for reviews see \cite{castdauriafre} and also the
second volume of \cite{pietrobook}). The goal is that of clarifying the algebraic structure underlying the
theory, thus providing a solid basis for the analysis
of the $2$-branes mentioned above. In this systematic revisitation of  $D=7$ supergravity we have found
several subtleties whose clarification was in our opinion extremely important, in particular in relation with
the double formulation in terms of  $\mathbf{B}^{[3]}_{\mu\nu\rho}$ and $\mathbf{B}^{[2]}_{\mu\nu}$ which
obviously plays a primary role for brane solutions.
\par
In this paper we present the complete rheonomic solution of Bianchi identities which, as it is well known,
implicitly implies the fermionic and bosonic field equations of all the fields. The request that the rheonomic
parameterizations of the $2$-form curvature $\mathfrak{G}^{[3]}$ and of the 3-form curvature
$\mathfrak{G}^{[4]}$ should be compatible completely fixes all the coefficients in the rheonomic
parameterizations and therefore determines all supersymmetry transformation rules including higher order  terms in the fermion fields. As we show, upon suitable rescalings, these transformation rules
fully coincide with those derived (up to linear order in the fermions) by the authors of
\cite{PvNT,bershoffo1}. This is a very significant consistency test that goes hand in hand with another
important test already obtained in \cite{arnoldtwobranes}. There it was shown that Beltrami flux $2$ brane
solutions of a bosonic theory with the same content as $D=7$ supergravity can exist \textit{if and only if}
the ratios between the coefficients in the action are exactly the same as those determined by the authors of
\cite{PvNT}. This leads to an exact prediction on the bosonic subset of the coefficients appearing in the
geometric lagrangian of $D=7$ supergravity, whose explicit form is still under construction. We plan to
present it in a forthcoming paper.
\par
The information mentioned above is sufficient to embed the Arnold-Beltrami flux-branes into $D=7$ supergravity
and to write down the precise form of the \textit{Killing spinor equation} in general terms and to polarize on
this type of backgrounds.
\par
The second main result of this paper is the analysis of the supersymmetry preserved by $2$-branes and flux
$2$-branes. Without fluxes the $2$-branes preserve $\ft 12$ of the original supersymmetry and they always
admit eight Killing spinors. With Arnold-Beltrami fluxes supersymmetry is usually completely broken, unless
the solution, besides discrete symmetries has also extra translational Killing vectors. With two translational
Killing vectors one can preserve $\ft 14$ of the original supersymmetry, corresponding to the presence of four
Killing spinors. With one translational Killing vector one can preserve $\ft 18$ of the original
supersymmetry, corresponding to the presence of two Killing spinors. The presence of the translational Killing
vectors is a necessary, yet not sufficient condition. Accurate choices of the fluxes have to be made which
lead to certain precise discrete symmetries illustrated in our worked out examples.
\par
Our paper is organized as follows
\begin{description}
  \item[a)] In section \ref{twobranastoria} we review the construction, introduced in
  \cite{arnoldtwobranes} of two-branes in seven dimensions with Arnold Beltrami Fluxes in the transverse
  space.
  \item[b)] In section \ref{algebrabasata} we discuss the algebraic basis of $D=7$ supergravity. In
  particular, utilizing crucial Fierz identities we derive the underlying Free Differential Algebra and
  we analyze its properties.
  \item[c)] In section \ref{construzia} we address the geometric construction of $D=7$ supergravity
  introducing the rheonomic parameterizations  of the curvatures and the general form of the action. The actual
  determination of the coefficients is provided in Appendix \ref{dettaglione}.
  \item[d)] In section \ref{boselagravaialetto} we discuss the explicit embedding of the flux brane solutions
  into supergravity. This is a necessary essential intermediate step in order to be able to discuss the
  residual supersymmetry.
  \item[e)] In section \ref{chilluspina} we write the Killing spinor equation and investigate its general
  properties. There we present the logic of a computerized algorithm devised to investigate the presence or absence of
  Killing spinors.
  \item[f)] In section \ref{unesempiopertutti} we present three explicit cases of flux $2$-brane solutions
  with zero, $\ft 14$ and $\ft 12$ preserved supersymmetry, respectively. We carefully discuss the discrete
  symmetries of these solutions.
  \item[g)]In section \ref{mitirosu} we briefly discuss the uplifting of Arnold Beltrami flux 2-branes to $D=11$ supergravity.
  \item[e)] Section \ref{zakliuchenie} contains our conclusions.
\end{description}
\section{D=7 two-branes with Arnold Beltrami Fluxes in the transverse directions}
\label{twobranastoria} In this section we review the construction of  \cite{arnoldtwobranes} based on the
general form of $p$-brane actions which is described in many places in the literature (in particular we refer
the reader to chapter 7, Volume Two of \cite{pietrobook} and to all the papers  there cited) and we focus on
the the case $p\, = \, 2$ in $D=7$. The  concern of \cite{arnoldtwobranes} was the elementary $2$-brane
solution in $D=7$. It was shown in \cite{arnoldtwobranes} that this latter exists for all values of the
exponential coupling parameter $a$ defined below. Each value of $a$ corresponds to a different value of the
dimensional reduction invariant parameter $\Delta$ also defined below. Obviously $D=7$ supergravity
corresponds to a unique value of $\Delta$ which, as we recall in section \ref{trovoDelta}, is the magic
$\Delta \, = \, 4$ for which the solution becomes particularly simple and elegant and typically preserves one
half of the supersymmetries.
\par
Subsequently, in \cite{arnoldtwobranes}, on the background of the $2$-brane solution it was considered  the
inclusion of fluxes of an additional triplet of vector fields, in this way mimicking the bosonic field content
of $D=7$ supergravity. In presence of a topological interaction term between the triplet of gauge fields and
the $3$-form which defines the $2$-brane, it was shown that the  fluxes can be introduced into the framework
of an exact solution if they are Arnold Beltrami vector fields satisfying Beltrami equation. The only
conditions for the existence of such a solution is $\Delta \, = \, 4$ plus  a precise relation between the
coefficients of the kinetic terms in the lagrangian and the coefficient of the topological interaction term.
Clearly this relation is precisely satisfied by the coefficients of minimal $D=7$ supergravity as we show in
the present paper.
\subsection{The general form of a $2$-brane action in $D=7$}
In the mostly minus metric that we utilize, the correct form of the action in $D=7$ admitting an electric
$2$-brane solution is the following one:
\begin{eqnarray}
  \mathcal{A}_{2brane} &=& \int \, d^7x \, \mathcal{L}_{2brane} \nonumber\\
  \mathcal{L}_{2brane} &=&  \mbox{det} V \, \left(- R[g] \, -
  \, \ft 14 \,\partial^\mu \varphi \, \partial_\mu\varphi
  \, + \, \ft {1}{96} \, e^{-a\,\varphi}
  \, \mathbf{F}_{\lambda\mu\nu\rho}\,\mathbf{F}^{\lambda\mu\nu\rho}\right)\label{braneaction}
\end{eqnarray}
where $a$ is a free parameter, $\varphi$ denotes the dilaton field with a canonically normalized kinetic
term\footnote{Note that in the notations adopted in this paper and in all the literature on rheonomic
supergravity the normalization of the curvature scalar and of the Ricci tensor is one half of the
normalization used in most textbooks of General Relativity. Hence the relative normalization of the Einstein
term $R[g]$ and of the dilaton term $\partial^\mu \varphi \, \partial_\mu\varphi$ is $\ft 14$ and not $\ft
12$.} and \footnote{Note also that in the notations of all the literature on rheonomic supergravity the
components of the form $Q^{[p]}\,= \,\mathrm{d} \Omega^{[p-1]}$ are defined with strength one, namely
$Q_{\lambda_1 \dots \lambda_{p}} \, = \, \ft{1}{p!} \left( \partial_{\lambda_1} \Omega_{\lambda_2 \dots
\lambda_{p}}\, + \, (p! -1)\mbox{-terms} \right)$.}:
\begin{equation}\label{turacciolo}
    \mathbf{F}_{\lambda\mu\nu\rho} \, \equiv \, \partial_{[\lambda } \, \mathbf{A}_{\mu\nu\rho]}
\end{equation}
is the field-strength of the three-form $\mathbf{A}^{[3]}$ which couples to the world volume of the two-brane.
\par
The field equations following from (\ref{braneaction}) can be put into the following convenient form:
\begin{eqnarray}
  \Box_{cov} \, \varphi  &=&
  \frac{a}{48} e^{-a \varphi}\, \mathbf{F}_{\lambda\mu\nu\rho}\,\mathbf{F}^{\lambda\mu\nu\rho} \label{stdilato}\\
  \mathrm{d} \star\left[ e^{-a\varphi} \, \star\mathbf{F}^{[4]}\right] &=& 0 \label{st4form}\\
  \mbox{Ric}_{\mu\nu}&=& \frac{1}{4}\partial_\mu\varphi\, \partial_\nu\varphi
  \, + \, S_{\mu\nu} \label{stEinstein}\\
  S_{\mu\nu}  &=& - \, \frac{1}{24} e^{-a\varphi} \left(\mathbf{F}_{\mu...} \,\mathbf{F}_{\nu}^{\phantom{\nu}...}
  \, - \, \ft {3}{20} \, g_{\mu\nu} \, \mathbf{F}_{....}\,\mathbf{F}^{....}\right)
\end{eqnarray}
and they admit the following exact electric $2$-brane solution:
\begin{eqnarray}
  ds^2 &=& H(y)^{-\frac{8}{5\Delta}}\, d\xi^\mu\otimes d\xi^\nu
  \, - \, H(y)^{\frac{12}{5\Delta}} \, dy^I\otimes dy^J \, \delta_{IJ} \nonumber\\
  \varphi  &=& - \frac{2a}{\Delta} \, \log \, H(y) \nonumber\\
  \mathbf{F}_{[4]} &=& 4\,\mathrm{d} \left[ H(y)^{-1} \,
  \mathrm{d}\xi^\mu\wedge \mathrm{d}\xi^\nu
  \wedge \mathrm{d}\xi^\rho \, \epsilon_{\mu\nu\rho}  \right] \label{branamet}
\end{eqnarray}
where the seven coordinates have been separated into two sets $\xi^\mu$ ($\mu=0,1,2$) spanning the $2$-brane
world volume and $y^{I}$ ($I=3,4,5,6$) spanning the transverse space to the brane. In the above solution
$H(y)$ is any harmonic function living on the $4$-dimensional transverse space to the brane whose metric is
assumed to be flat:
\begin{equation}\label{harmonica}
 \Box_{\mathbb{R}^4} \, H(y) \, \equiv \,   \sum_{I=1}^4\frac{\partial^2}{\partial (y^I)^2} H(y) \, = \, 0
\end{equation}
and the parameters $a$ and $\Delta$ are related by the celebrated formula:
\begin{equation}\label{simonbocca}
\Delta \, = \, a^2 \, + \, 2 \frac{d \, \tilde{d}}{D-2} \, = \, a^2 \, + \, \frac{12}{5}
\end{equation}
which follows from $d=3, \quad\tilde{d}\,=\,2$ and $D=7$. Physically $d$ is the dimension of the electric
$2$-brane world volume, while $\tilde{d}$ is the dimension of the world-sheet spanned by the magnetic string
which is dual to the $2$-brane.
\par
In section \ref{trovoDelta} we will discuss the relation of the brane action (\ref{braneaction}) with the
bosonic action of Minimal ungauged $D=7$ supergravity and show that the specific coefficients of the kinetic
terms appearing in this latter determine the value of $\Delta$. Indeed the supersymmetry of the action imposes
$\Delta \, = \, 4$. In a later section we discuss the Killing spinors admitted by the solution
(\ref{branamet}).
\subsection{The two-brane with Arnold Beltrami Fluxes}
\label{flussibrane} As a next step, in  \cite{arnoldtwobranes}  the two-brane action (\ref{braneaction}) was
generalized  introducing also a triplet of one-form fields $\mathbf{A}^{\Lambda}$, ($\Lambda \, = \, 1,2,3$)
whose field strengths are denoted $\mathbf{F}^{\Lambda} \, \equiv \, \mathrm{d}\mathbf{A}^{\Lambda}$. In this
way we mimic the field-content of Minimal $D=7$ supergravity. Explicitly one has the new bosonic action:
\begin{eqnarray}
  \mathcal{A}^{flux}_{2brane} &=& \int \, d^7x \, \mathcal{L}^{flux}_{2brane} \nonumber\\
  \mathcal{L}^{flux}_{2brane} &=&  \mbox{det} V \, \left(- R[g] \,
   - \, \ft 14 \,\partial^\mu \varphi \, \partial_\mu\varphi
  \, + \, \ft {1}{96} \, e^{-a\,\varphi} \, \mathbf{F}_{\lambda\mu\nu\rho}\,
  \mathbf{F}^{\lambda\mu\nu\rho}\right.\nonumber\\
  &&\left. + \, \ft{\omega}{8}\, e^{\ft{a}{2} \varphi} \,
  \mathbf{F}^\Lambda_{\lambda\mu\ }\,\mathbf{F}^{\Lambda|\lambda\mu}\right)
   \, - \, \kappa \,\, \mathbf{F}_{\lambda_1\dots\lambda_4}
  \, \mathbf{F}^\Lambda_{\lambda_5\lambda_6} \,
  \mathbf{A}^\Lambda_{\lambda_7} \, \epsilon^{\lambda_1\dots \lambda_7}\label{fluxbraneaction}
\end{eqnarray}
where  two new real parameters $\omega$ and $\kappa$ do appear. Crucial for the consistent insertion of fluxes
is the topological interaction term with coefficient $\kappa$.
\par
The modified field equations associated with the new action (\ref{fluxbraneaction}) can be written in the
following way:
\begin{eqnarray}
  \Box_{cov} \, \varphi  &=& \frac{a}{48} e^{-a \varphi}\,
  \mathbf{F}_{\lambda\mu\nu\rho}\,\mathbf{F}^{\lambda\mu\nu\rho}
  \, - \, \omega \frac{a}{8} e^{a \varphi}\,
  \mathbf{F}^\Lambda_{\lambda\mu}\,\mathbf{F}^{\Lambda|\lambda\mu}\label{stdilatoflux}\\
  \mathrm{d} \left[ e^{-a\varphi} \, \star\mathbf{F}^{[4]}\right]
  &=& 1152 \, \kappa \, \,\mathbf{F}^\Lambda \, \wedge \, \mathbf{F}^\Lambda \label{st4formflux}\\
  \mathrm{d}\left[ e^{\ft a 2 \varphi} \, \star\mathbf{F}^{\Lambda}\right]
  &=& 8 \, \frac{\kappa}{\omega} \, \,\mathbf{F}^{[4]} \, \wedge \, \mathbf{F}^\Lambda \label{st2formflux}\\
  \mbox{Ric}_{\mu\nu}&=& \frac{1}{4}\partial_\mu\varphi\, \partial_\nu\varphi
  \, + \, S^{[4]}_{\mu\nu} \, + \, S^{[2]}_{\mu\nu}\label{stEinsteinflux}\\
  S^{[4]}_{\mu\nu}  &=& - \, \frac{1}{24} e^{-a\varphi} \left(\mathbf{F}_{\mu...}
  \,\mathbf{F}_{\nu}^{\phantom{\nu}...}\, - \, \ft {3}{20} \, g_{\mu\nu} \,
  \mathbf{F}_{....}\,\mathbf{F}^{....}\right) \label{st4stressflux}\\
  S^{[2]}_{\mu\nu}  &=& - \, \omega \, \frac{1}{4} e^{\ft a 2 \varphi}
  \left(\mathbf{F}^\Lambda_{\mu.}
  \,\mathbf{F}_{\nu}^{\Lambda|\phantom{\nu}.}\, - \, \ft {1}{10}
  \, g_{\mu\nu} \, \mathbf{F}^\Lambda_{..} \,\mathbf{F}^{\Lambda|..}\right)\label{st2stressflux}
\end{eqnarray}
In \cite{arnoldtwobranes} the above equations were solved with the same ansatz as  in the previous case for
the metric, the dilaton and the $4$-form, introducing  also a non trivial  $\mathbf{F}^\Lambda$ in the
transverse space spanned by the coordinates $y$. Explicitly, the ansatz considered  in \cite{arnoldtwobranes}
is the following one.
\begin{eqnarray}
  ds^2 &=& H(y)^{-\frac{8}{5\Delta}}\, d\xi^\mu\otimes d\xi^\nu
  \, - \, H(y)^{\frac{12}{5\Delta}} \, dy^I\otimes dy^J \, \delta_{IJ} \nonumber\\
  \varphi  &=& - \frac{2a}{\Delta} \, \log \, H(y) \nonumber\\
  \mathbf{F}_{[4]} &=& 4\,\mathrm{d} \left[ H(y)^{-1} \, \mathrm{d}\xi^\mu\wedge \mathrm{d}\xi^\nu \wedge
   \mathrm{d}\xi^\rho \, \epsilon_{\mu\nu\rho}  \right] \nonumber\\
  \mathbf{F}^{\Lambda} &=&\mathrm{d} \left[ \mathbf{W}^\Lambda_I(y) \, dy^I \right]
  \label{branametflux}
\end{eqnarray}
\subsubsection{Arnold Beltrami vector fields on the torus $\mathrm{T}^3$ as fluxes}
In order to solve the above equations a change of topology was put forward in \cite{arnoldtwobranes}. In the
brane solutions without fluxes the transverse space to the brane volume was chosen flat and non compact,
namely $\mathbb{R}^4$. To introduce the fluxes one mantains  it flat but one compactifies three of its
dimensions by identifying them with those of a three-torus $\mathrm{T}^3$. In other words one performs the
replacement:
\begin{equation}\label{corinto1}
    \mathbb{R}^4 \, \rightarrow \, \mathbb{R}_+\otimes \mathrm{T}^3
\end{equation}
Secondly, on the abstract $\mathrm{T}^3$-torus one utilizes the flat metric consistent with octahedral
symmetry, namely according to the setup of \cite{arnolderie} one  identifies:
\begin{equation}\label{torellocubico}
    \mathrm{T}^3 \, \simeq \, \frac{\mathbb{R}^3}{\Lambda_{cubic}}
\end{equation}
where $\Lambda_{cubic}$ denotes the cubic lattice, \textit{i.d.} the abelian group of discrete translations of
the euclidian three-coordinates $\left\{ X,Y,Z\right\}$, defined below:
\begin{equation}\label{traslodiscreto}
\Lambda_{cubic} \, \ni \, \gamma_{n_1,n_2,n_3} \quad : \quad
\left\{ X,Y,Z\right\}\, \rightarrow \, \left\{
X+n_1,Y+n_2,Z+n_3\right\} \quad ; \quad n_{1,2,3} \, \in \, \mathbb{Z}
\end{equation}
Functions on $\mathrm{T}^3$ are periodic functions of $X,Y,Z$, with respect to the translations
(\ref{traslodiscreto}).
\par
According to (\ref{corinto1}) one splits the four coordinates $y^I$ as follows:
\begin{equation}\label{kastriula}
    y^I \, = \, \left\{\underbrace{U}_{\in \,\mathbb{R}}\, ,\,
    \underbrace{X,Y,Z}_{\equiv \,\mathbf{X} \,\in \,\mathrm{T}^3}\right\}
\end{equation}
In \cite{arnolderie}, one of us, in a different collaboration, has classified and constructed all the solutions of Beltrami equation:
\begin{equation}\label{Beltramino}
    \star \, \mathrm{d} \mathbf{Y}^{[1]} \, = \, \mu \, \mathbf{Y}^{[1]}
\end{equation}
for one-forms  $\mathbf{Y}^{[1]}$ defined over the three-torus (\ref{torellocubico}) outlining the strategy to
construct the same solutions also in the case of other crystallographic lattices like, for instance, the
hexagonal one. These solutions are organized in orbits with respect to the cubic lattice point group, namely
the 24-elements octahedral group $\mathcal{O}_{24}$ and their parameter space is decomposed into irreducible
representations of appropriate subgroups of a universal classifying group with $1536$ elements
\cite{arnolderie}. Using such one-forms $\mathbf{Y}^{[1]}$ as building blocks for the brane fluxes appeared
very appealing in \cite{arnoldtwobranes} since it introduces the corresponding discrete symmetries into the
brane solution.
\par
Explicitly  the last of the ans\"atze  (\ref{branametflux})  was specialized in the following way:
\begin{eqnarray}\label{corrido}
 \mathbf{F}^{\Lambda} & = & \lambda \,\mathrm{d}\, \left[ e^{2\pi\mu U} \,
 \mathbf{W}^{\Lambda}\left(\mathbf{X}\right)  \right] \\
     \mathbf{W}^{\Lambda}\left(\mathbf{X}\right)
     & = & \mathcal{E }^\Lambda_{\phantom{\Lambda}A} \,  \mathbf{Y}^{A}\left(\mathbf{X}\right)
\end{eqnarray}
where $\mathbf{Y}^{A}\left(\mathbf{X}\right)$ denotes a basis of solutions of Beltrami equation
(\ref{Beltramino}) pertaining to eigenvalue $\mu$ and the \textit{embedding matrix} $\mathcal{E
}^\Lambda_{\phantom{\Lambda}A}$ is a constant matrix which constructs three linear independent combinations of
such fields. Furthermore $\lambda$ is some numerical parameter.
\par
It was shown in \cite{arnoldtwobranes} that all field equations (\ref{stdilatoflux}-\ref{st2stressflux}) are
solved if the following conditions are verified
\begin{eqnarray}
  \Delta &=& 4 \, \Leftrightarrow \, a \, = \, 2 \, \sqrt{\frac{2}{5}}\nonumber \\
   \kappa  &=& \frac{\omega}{384} \nonumber \\
 \Box_{\mathbb{R}_{+} \times \mathrm{T}^3} \, H(U,\mathbf{X}) &=& - \, \frac{\lambda^2}{24}\,
 \exp\left[2 \, \pi \, \mu \, U\right]\,\mu^2 \, J(\mathbf{X}) \nonumber \\
  J(\mathbf{X})  &\equiv& \sum_{\Lambda=1}^3 \, \sum_{i=1}^3 \,
  \mathbf{W}^\Lambda(\mathbf{X})_i  \mathbf{W}^\Lambda(\mathbf{X})_i \label{sicurmorio}
\end{eqnarray}
The first two conditions of (\ref{sicurmorio}) are a specification of the parameters in the brane lagrangian. It
was already noted in \cite{arnoldtwobranes} that such a specification corresponds to selecting a bosonic
lagrangian that, up to field redefinitions, is equivalent to the bosonic lagrangian of minimal $D=7$
supergravity. The third equation is the only differential condition that solves the entire system of field
equations. The function $H(y)$ appearing in the metric, in the dilaton and in the three-form
$\mathbf{B}^{[3]}$ needs to satisfy a inhomogeneous Laplace equation whose source $J(\mathbf{X})$ is entirely
determined by the Beltrami vector fields according to the formula displayed in the last of eq.s
(\ref{sicurmorio}).
\subsection{Relation of the Arnold-Beltrami Fluxes with Hyperinstantons}
In the recent paper \cite{pietroantoniosorin} the relation between Beltrami equation (\ref{Beltramino}) and
the defining equation of tri-holomorphicity was explored. It was shown in the past in \cite{Anselmi:1992tz}
that a suitable definition of what we can name a tri-holomorphic map from a flat HyperK\"ahler
four--dimensional manifold $\mathcal{HK}_{4}$ to any HyperK\"ahler manifold $\mathcal{HK}_{4n}$:
\begin{equation}\label{copinus}
    q \, : \, \mathcal{HK}_{4} \, \rightarrow \, \mathcal{HK}_{4n}
\end{equation}
naturally emerges from the topological twist of an $\mathcal{N}=2$ supersymmetric sigma model in $D=4$. The
following first order differential equation:
\begin{equation}
q^\star-\sum_{x=1}^3 \,J_x\circ q^\star\circ j_x=0, \label{afeq4}
\end{equation}
where $J^x$ denote the three complex structures of the target manifold $\mathcal{HK}_{4n}$   and $j^x$ those
of the base manifold is obtained as the BRST-variation of the antighost produced by the twist. Henceforth
eq.(\ref{afeq4}) defines in a unique algebraic way the \textit{instantonic} maps on which the functional
integral should be localized in the topological version of the sigma-model. For this reason the maps
satisfying eq.(\ref{afeq4}) were dubbed \textit{hyperinstantons} in \cite{Anselmi:1992tz} and it was also
observed that they are tri-holomorphic since eq.(\ref{afeq4}) can be interpreted as the statement that they
are holomorphic with respect to the average of the three complex structures. In \cite{pietroantoniosorin} the
base manifold was chosen to be $\mathbb{R}_+ \times \mathrm{T^3}$ while the target manifold was simply chosen
to be $\mathbb{R}^4$. In this way the equation of tri-holomorphicity was applied to maps:
\begin{equation}
    q \, : \, \mathbb{R}_+ \times \mathrm{T^3} \, \rightarrow \, \mathbb{R}^4
\end{equation}
It was shown in \cite{pietroantoniosorin} that, under very mild assumptions, the general solution of equation
(\ref{afeq4}) is as follows. Let $G(\mathbf{X})$ be a generic function on the $\mathrm{T^3}$ torus , let
$\mathbf{Y}(\mathbf{X})$ be a solution of Beltrami equation (\ref{Beltramino}) corresponding to eigenvalue
$\mu$ and define:
\begin{eqnarray}
 \Phi\left(\mathrm{U},\mathbf{X}\right)&=& e^{ -2\,\mu \,\mathrm{U}}\,
  G\left(\mathbf{X}\right) \nonumber\\
  \mathbf{A}\left(\mathrm{U},\mathbf{X}\right)&=&e^{-\,2\,\mu \,\mathrm{U}}\,\mathbf{Y}\left(\mathbf{X}\right)
  \label{parlandoconte}
\end{eqnarray}
where $U$ is the positive real variable spanning $\mathbb{R}_+$. Then the image of the point
$\{U,\mathbf{X}\}\in \mathbb{R}_+\times \mathrm{T^3} $ with respect to a map $q$ that satisfies the
tri-holomorphic constraint (\ref{afeq4}) is given by $\{q_0,\mathbf{q}\}\in \mathbb{R}^4$, where:
\begin{eqnarray}
  q_0\left(\mathrm{U},\mathbf{X}\right)&=& \, - \,\partial_\mathrm{U}
  \,\Phi\left(\mathrm{U},\mathbf{X}\right)\nonumber\\
\mathbf{q}\left(\mathrm{U},\mathbf{X}\right)&=&
  \,\nabla \,\Phi\left(\mathrm{U},\mathbf{X}\right)\,
   +\, \mathbf{A}\left(\mathrm{U},\mathbf{X}\right)\label{generasolata}
\end{eqnarray}
the operator $\nabla$ representing the derivatives with respect to the torus coordinates.
\par
Next, if we interpret the four  components $\{q_0,\mathbf{q}\}$ as the components of a gauge $1$-form in
$\mathbb{R}_- \times \mathrm{T^3}$ (where $\mathrm{U}\to -U $), namely if we set:
\begin{equation}\label{consuelo}
    \mathcal{A} \, = \, q_0 \mathrm{d}U \, + \, \mathbf{q}\cdot \mathrm{d}\mathbf{X}
\end{equation}
we obtain:
\begin{equation}\label{canzovallo}
    \mathcal{A} \, = \, \mathrm{d}\Phi\left(U,\mathbf{X}\right)\, + \,
    e^{ -2\,\mu\,\mathrm{U}}\,\mathbf{Y}
\end{equation}
We recall also that this gauge connection satisfies a suitable gauge fixing (see \cite{pietroantoniosorin} for
a complete discussion). It appears clearly from eq. (\ref{canzovallo}) that the function $\Phi\left(U,\mathbf{X}\right)$ is just an
irrelevant gauge transformation which has no influence on the gauge field strengths appearing in supergravity.
Apart from it the gauge fields entering the brane solutions as fluxes are just hyperinstantons in the
transverse directions to the brane, namely on $\mathbb{R}_- \times \mathrm{T^3}$. The restriction to
$\mathbb{R}_- \Leftrightarrow \mathbb{R}_+$ on the sigma-model side of this correspondence is greatly
illuminated by it. Indeed on the supergravity side $U$ has to be negative in order to keep the metric real.
Choosing the parameter $\lambda$ appropriately we can arrange that $U=0$, which is a boundary in the sigma
model, corresponds to a metric singularity in supergravity. This singularity is the brane itself, since $U$ is
nothing else but the distance from the brane.
\section{The algebraic basis of $D=7$ supergravity}
\label{algebrabasata} Motivated by $2$-branes  with Arnold-Beltrami fluxes that we have summarized in the
previous section, we turn to the reconstruction of $D=7$ supergravity in a systematic algebro-geometric
approach. Our final aim is to embed the considered $2$-branes in supergavity and to investigate their
supersymmetries.
\par
As announced in the introduction, in the present section we clarify the algebraic basis of minimal $D=7$
supergravity in terms of Free Differential Algebras, preparing the stage for its \textit{ex novo}
reconstruction in the rheonomic approach.
\subsection{Pseudo Majorana spinors in $D=7$}
The main property of the Clifford algebra in $D=7$ with Minkowski signature (see eq.(\ref{clifford7})) is that
there is only one type of conjugation matrix, namely $\mathcal{C}_{-}$ (see
\cite{castdauriafre},\cite{pietrobook}) and that this latter is symmetric:
\begin{equation}\label{chargematrix}
 \mathcal{ C}_{-} \, \Gamma_a \,\mathcal{ C}_{-}^{-1} \, = \, - \, \Gamma_a^T \quad ;
 \quad \mathcal{ C}_{-} \, = \, \mathcal{ C}_{-}^T \quad ; \quad \mathcal{ C}_{-}^2 \, = \, \mathbf{1}_{8\times8}
\end{equation}
This being the case one can always choose a basis where $ \mathcal{ C}_{-}$ is just the identity matrix in
eight-dimensions and the gamma--matrices are all antisymmetric as described in appendix \ref{gammola} Hence
there are no Majorana spinors but, just as in $d=5$, we can introduce doublets of pseudo-Majorana gravitino
one-forms. Minimal $D=7$ supergravity corresponds to the case where we have just one such doublet that we name
$\Psi_A$ ($A \, = \, 1,2$):
\begin{equation}\label{gravitino1form}
    \Psi_A^c \, \equiv \, (\overline{\Psi}^A)^T \, = \, -\, \Gamma_0 \Psi_A^\star \, = \, \epsilon^{AB} \, \Psi_A
\end{equation}
An explicit solution of the pseudo-Majorana constraint in the gamma matrix basis described in appendix
\ref{gammola} is shown below:
\begin{equation}\label{psioni}
    \Psi_1 \, = \, \left(
\begin{array}{l}
 \alpha _1+i \beta _1 \\
 \alpha _2+i \beta _2 \\
 \alpha _3+i \beta _3 \\
 \alpha _4+i \beta _4 \\
 \alpha _5+i \beta _5 \\
 \alpha _6+i \beta _6 \\
 \alpha _7+i \beta _7 \\
 \alpha _8+i \beta _8
\end{array}
\right) \quad ; \quad \Psi_2 \, = \, \left(
\begin{array}{l}
 -i \alpha _7-\beta _7 \\
 -i \alpha _3-\beta _3 \\
 i \alpha _2+\beta _2 \\
 -i \alpha _8-\beta _8 \\
 i \alpha _6+\beta _6 \\
 -i \alpha _5-\beta _5 \\
 i \alpha _1+\beta _1 \\
 i \alpha _4+\beta _4
\end{array}
\right)
\end{equation}
where $\alpha_{1,\dots ,8}$ and $\beta_{1,\dots,8}$ are  real components. This explicitly shows that minimal
$D=7$ supergravity is based on a superalgebra with $16$ supercharges, just one half of the maximum $32$.
\par
When we discuss Killing spinors for the $2$-brane solutions we utilize another gamma matrix basis well adapted
to the split of $7$-dimensions in $3+4$. Such a basis is described in appendix \ref{splittorio}. The explicit
form of a pair of pseudo-Majorana spinors in this basis is provided here below:
\begin{equation}\label{splittiepsi}
  \begin{array}{ccccccc}
     \epsilon_1 & = & \left(
\begin{array}{c}
 \xi _1-i \xi _2 \\
 \xi _3+i \xi _4 \\
 \theta _1-i \theta _6 \\
 \theta _2+i \theta _5 \\
 \xi _5-i \xi _6 \\
 \xi _7+i \xi _8 \\
 \theta _3-i \theta _8 \\
 \theta _4+i \theta _7 \\
\end{array}
\right) & ; & \epsilon_2 & = & \left(
\begin{array}{c}
 \xi _4+i \xi _3 \\
 \xi _2-i \xi _1 \\
 \theta _5+i \theta _2 \\
 \theta _6-i \theta _1 \\
 \xi _8+i \xi _7 \\
 \xi _6-i \xi _5 \\
 \theta _7+i \theta _4 \\
 \theta _8-i \theta _3 \\
\end{array}
\right)
   \end{array}
\end{equation}
where $\xi_{1},\dots , \xi_8$ and $\theta_1,\dots,\theta_8$ are two octets of real anticommuting parameters.
The particular form of this parameterization is already adapted to the projection that will be enforced by the
spin one-half fermion transformation rules in the Killing spinor equation. This projection will simply delete
the eight parameters $\theta$.
\subsection{Fierz identities} As usual, the core of  any supergravity construction is
provided by the $4$-$\Psi$ and $3$-$\Psi$ Fierz identities. Indeed from the $4$-$\Psi$ Fierz identities one
obtains the available Chevalley cocycles  that give rise to the Free-Differential Algebra extension of the
super Poincar\'e algebra. This latter encodes  the $p$-form gauge fields that complete the gravitational
multiplet. On the other hand $3$-$\Psi$ Fierz are crucial in the construction of a rheonomic parameterization
of the curvature which solves Bianchi identities.
\par
The first step in this analysis is provided by counting the $2$-$\Psi$ independent components and arranging
them into a complete set of bosonic-currents. In this case, since we have $16$-supercharges, the number of
independent components of the symmetric wedge product is
\begin{equation}\label{2psicompo}
 \mbox{\# of components of}  \,\, \Psi^\alpha_A \, \wedge \, \Psi^\beta_B \, = \, \ft 12 \, 16 \times 17 \, = \, 136
\end{equation}
Introducing the three Pauli matrices $\sigma^{\Lambda|A}_{\phantom{\Lambda|A}B}$ ($\Lambda \, =\,1,2,3$,
$A,B\, = \, 1,2$) according to the conventions of appendix \ref{sigmotta} we can distribute the 136 components
in the following exhaustive set of fermionic currents:
\par
\begin{center}
\begin{tabular}{|rc|l|r|}
  \hline
   name & \null &current& \# of components \\
  \hline
 $\mathbf{J}^a$& = & $\overline{\Psi}^A \, \wedge \, \Gamma^a \,\Psi_A$ & 7 \\
  $\mathbf{J}^{ab}$& = &$ {\rm i} \,\overline{\Psi}^A \, \wedge \, \Gamma^{ab} \,\Psi_A$ & 21 \\
  $\mathbf{J}^\Lambda$ & = &  ${\rm i}\, \sigma^{\Lambda|B}_{\phantom{\Lambda|B}A} \,
  \overline{\Psi}^A \, \wedge \,\Psi_B$ & 3 \\
  $\mathbf{J}^\Lambda_{pqr}$& =&{\rm i}\, $\sigma^{\Lambda|B}_{\phantom{\Lambda|B}A} \,
  \overline{\Psi}^A \, \wedge \,  \Gamma_{pqr} \Psi_B$ & 105 \\
  \hline
  \null & \null & \null & 136 \\
  \hline
\end{tabular}
\end{center}
The factors ${\rm i}$ have been placed in the above formulae in such a way as to make the corresponding
fermion currents real. There are two fundamental  $4$-$\Psi$ Fierz identities that might be deduced by means
of group theory, counting the number of singlet representations that appear in the symmetric product of
$4$-$\Psi$ but which we have simply verified with  a computer programme by direct evaluation.  They are the
following ones:
\begin{eqnarray}\label{4pissi}
    \mathbf{J}^a \, \wedge \, \mathbf{J}_a & = & - \,  \mathbf{J}^\Lambda \, \wedge
    \, \mathbf{J}^\Lambda \label{bergsFierz}\\
    \mathbf{J}^{ab} \, \wedge \, \mathbf{J}_a & = &0 \label{riccarFierz}
\end{eqnarray}
The above two identities are the basis for the existence of two distinct  FDAs both able to describe the
degrees of freedom of the $D=7$ graviton multiplet in the Poincar\'e case. As we will illustrate below the FDA
associated with identity (\ref{bergsFierz}) is the one implicitly chosen by Bergshoeff et al in their
construction of the minimal theory in \cite{bershoffo1}. The FDA associated with the second identity is
associated with the formulation of \cite{PvNT} in terms of a gauge three-form $\mathbf{B}^{[3]}$.
\par
Besides the above $4$-$\Psi$ Fierz identities there are also some $3$-$\Psi$ ones that are quite relevant in
the supergravity construction.
\par
The basic $3$-$\Psi$ Fierz identity is the  one below and it is related with the closure of the anti de Sitter
superalgebra. Let us define the following three structures:
\begin{eqnarray}
  \Pi^{(1)}_A &=& \Gamma_a \, \Psi_A \, \wedge \, \overline{\Psi}^B \, \wedge \, \Gamma^a \, \Psi_B \nonumber \\
   \Pi^{(2)}_A &=&  \Gamma_{ab} \, \Psi_A \, \wedge \, \overline{\Psi}^B \, \wedge \, \Gamma^{ab} \, \Psi_B \nonumber\\
  \Pi^{(0)}_A  &=& {\rm i} \, \sigma^{\Lambda|B}_{\phantom{\Lambda|B}A} \,\Psi_B\,
  \wedge \, {\rm i} \, \sigma^{\Lambda|D}_C\, \, \overline{\Psi}^C \, \wedge \,  \Psi_D
  \label{tripsistrut}
\end{eqnarray}
By explicit evaluation or by more lengthy group theoretical methods one can prove that the following linear
combination vanishes identically if and only if the here mentioned condition on the coefficients is satisfied:
\begin{equation}\label{3firzo}
    \mu \,  \Pi^{(1)}_A  \, + \, \nu \,   \Pi^{(2)}_A \, + \, \rho \,
    \Pi^{(0)}_A  \, = \, 0 \quad \stackrel{\mbox{iff}}{\Leftrightarrow} \quad \mu \, + \, 6 \, \nu \, + \rho \, = \, 0
\end{equation}
Another important Fierz identity which we will use in the solution of the Bianchi identities is obtained as
follows. Define the following structures:
\begin{eqnarray}
  \Sigma^{(1)}_b &=& \overline{\Psi}^A \Gamma_b \, \Gamma_p \, \chi_A \, \wedge \, \overline{\Psi}^C \, \wedge \,
  \Gamma^p \, \Psi_C \nonumber \\
 \Sigma^{(2)}_b &=& \overline{\Psi}^A \Gamma_b \, \Gamma_{pq} \, \chi_A \, \wedge \, \overline{\Psi}^C \, \wedge \,
  \Gamma^{pq} \, \Psi_C \nonumber \\
\Sigma^{(3)}_b &=& {\rm i} \, \sigma^{\Lambda|A}_{\phantom{\Lambda|A}B}\, \overline{\Psi}^B \Gamma_b \, \chi_A
\, \wedge \,  {\rm i} \, \sigma^{\Lambda|D}_{\phantom{\Lambda|D}C}\,\wedge \, \overline{\Psi}^C \, \wedge \,
\Psi_D \nonumber \\
\Sigma^{(4)}_b &=& {\rm i} \, \sigma^{\Lambda|A}_{\phantom{\Lambda|A}B}\, \overline{\Psi}^B \Gamma_b \,
\Gamma_{pqr} \,\chi_A \, \, \wedge \,  {\rm i} \, \sigma^{\Lambda|D}_{\phantom{\Lambda|D}C}\,\wedge \,
\overline{\Psi}^C \, \wedge \, \Gamma^{pqr} \Psi_D
 \label{minestrina3psi}
\end{eqnarray}
where $\chi_A$ is a generic (anticommuting) pseudo-Majorana spin $\ft 12$ zero-form.
\par
By explicit evaluation we find that the linear combination:
\begin{equation}\label{elleb}
    \ell_b \, \equiv \, g_1 \,  \Sigma^{(1)}_b\, + \, g_2 \,  \Sigma^{(2)}_b \, + \, g_3 \,
    \Sigma^{(3)}_b\, + \, g_4 \,  \Sigma^{(4)}_b
\end{equation}
vanishes if and only if :
\begin{equation}\label{baraccone}
   \ell_b \, = \, 0 \quad \stackrel{\mbox{iff}}{\Leftrightarrow} \quad  g_3 \, = \,
   \frac{1}{6} \left(-5\,g_1-14\, g_2\right)\quad ; \quad g_4 \, = \,
   \frac{1}{36} \left(2\,
   g_2-g_1\right)
\end{equation}
\subsection{The orthosymplectic super Lie algebra $\osp\mathrm(2,6|2)$}
In $D=7$ we have not only Poincar\'e supergravity but also anti--de--Sitter supergravity and it turns out that
it is not only convenient but, for a deeper understanding of the underlying structure of the theory,  it is
even essential to start from the simple super Lie algebra case.
\par
The relevant superalgebra for the $\mathrm{AdS_7}$-case is the orthosymplectic superalgebra
$\osp\mathrm(2,6|2)$ which contains, as bosonic subalgebra, the anti de Sitter algebra of isometries in 7 dimensions
$\so(2,6)$ times $\usp(2)$ which is the automorphism algebra of the pseudo Majorana spinors.
\par
The curvatures of $\osp\mathrm(2,6|2)$ can be written as follows:
\begin{eqnarray}
  \hat{\mathfrak{T}}^a &\equiv& \underbrace{\mathrm{d}V^a
  \, - \, \omega^{ab} \, \wedge \,V^b \,}_{\mathcal{D}V^a} - \, \ft 12 \,
   \overline{\Psi}^A \, \wedge \, \Gamma^a \,\Psi_A \nonumber\\
  \hat{\mathfrak{R}}^{ab} &\equiv& \mathrm{d}\omega^{ab}
  \, - \, \omega^{ac} \, \wedge \, \omega^{cb} + \, 4\, g^2 \, V^a \, \wedge \, V^b
  \, - \, {\rm i}\, g \, \overline{\Psi}^A \, \wedge \, \Gamma^{ab} \,\Psi_A \nonumber\\
  \hat{\rho}_A &\equiv& \underbrace{\mathrm{d}\Psi_A \, - \, \ft 14 \, \omega^{ab}
  \, \Gamma_{ab} \, \Psi_A}_{\mathcal{D}\Psi_A} \, + \, {\rm i} \, g \,  V^a \,
  \Gamma_a \, \wedge \, \Psi_A\, + \, 4 \,{\rm i} \,  g \,  \Psi_B \,
  \wedge \, \hat{\mathcal{A}}^\Lambda \, \sigma^{\Lambda|B}_{\phantom{\Lambda|B} A} \,   \, \nonumber\\
   \hat{\mathfrak{F}}^\Lambda
    &\equiv& \underbrace{\mathrm{d}\hat{\mathcal{A}}^\Lambda
    \, - \, 4 g \, \varepsilon^{\Lambda\Gamma\Delta}
   \hat{\mathcal{A}}^\Gamma
   \, \wedge \, \hat{\mathcal{A}^\Delta}}_{{\hat{F}}^\Lambda} \, - \, {\rm i}\,
   \ft 12 \,  \sigma^{\Lambda|B}_{\phantom{\Lambda|B} A} \, \overline{\Psi}^A \, \wedge \, \Psi_B
  \label{osp262}
\end{eqnarray}
where $g$ is a dimensionful parameter that can be identified with the inverse of the anti de Sitter radius.
\par
The above curvatures  are obtained by introducing the following $(8+2)\times (8+2)$ graded matrix of
one-forms:
\begin{equation}\label{Qmatrozza}
    \mathcal{Q} \, = \, \left( \begin{array}{c|c}
                                 \Delta^\alpha_{\phantom{\alpha}\beta} &
                                 2 \, e^{{\rmi}\ft \pi 4}\,\sqrt{g} \, \Psi_B^\alpha \\
                                 \null & \null  \\
                                 \hline
                                 \null & \null  \\
                                 2 \, e^{{\rmi}\ft \pi 4}\,\sqrt{g}
                                 \, \overline{\Psi}^{A}_\beta &  \mathcal{O}^A_{\phantom{A}B}
                               \end{array}
    \right)
\end{equation}
where:
\begin{equation}\label{boseconnecti}
    \Delta \, \equiv \, - \, \ft 14 \, \omega^{ab} \, \Gamma_{ab}
    \, + \,{\rm i}\, g \, V^a \, \Gamma_a \quad ;
    \quad  \mathcal{O}^A_{\phantom{A}B}
    \, = \, 4 \, {\rm i} \, g\, \hat{\mathcal{A}}^\Lambda \, \sigma^{\Lambda|A}_{\phantom{\Lambda|A} B}
\end{equation}
and then by setting:
\begin{equation}\label{sobillus}
 \mathcal{R} \, \equiv \,  \mathrm{d}\mathcal{Q} \, + \, \mathcal{Q}\,
 \wedge \, \mathcal{Q} \, \equiv \, \left( \begin{array}{c|c}
                                - \, \ft 14 \, \hat{\mathfrak{R}}^{ab}
                                \, \Gamma_{ab} \, + \, {\rm i} \, g \,
                                \hat{\mathfrak{T}}^a \, \Gamma_a  & 2
                                \, e^{{\rmi}\ft \pi 4}\, \sqrt{g} \, \rho_B \\
                                 \null & \null  \\
                                 \hline
                                 \null & \null  \\
                                 2 \, e^{{\rmi}\ft \pi 4}\, \sqrt{g} \,
                                 \overline{\rho}^{A} & \, 4 \, {\rm i} \, g\,
                                 \sigma^{\Lambda|A}_{\phantom{\Lambda|A} B} \, \hat{\mathfrak{F}}^\Lambda
                               \end{array}\right)
\end{equation}
Note that the matrix $\Delta$ is $\so(2,6)$ Lie algebra valued, yet it is not in the vector representation of
$\so(2,6)$, rather it is in its spinor representation which is also $8$-dimensional. Indeed $\Delta $ is an
antisymmetric matrix and hence an element of the $\so(8,\mathbb{C})$ complex Lie algebra. The appropriate
location of the ${\rm i}$-factors makes $\Delta$ an element of the real algebra $\so(2,6)$ in the
$\mathbf{8}_s$ representation.
\par
The Poincar\'e superalgebra is obtained by setting the coupling constant $g$ to zero.
\par
Besides the Lorentz covariant differential it is convenient to introduce also the $\so(1,6) \times \usp(2)$
covariant differential acting on the fermions:
\begin{equation}\label{uspdiff}
  \nabla \, \Psi_A \, = \,  \mathrm{d}\Psi_A \, - \, \ft 14 \, \omega^{ab} \,
  \Gamma_{ab} \, \Psi_A \,  + \, 4 \,{\rm i} \,  g \,  \Psi_B \,
  \wedge \, \hat{\mathcal{A}}^\Lambda \, \sigma^{\Lambda|B}_{\phantom{\Lambda|B} A} \,
\end{equation}
Utilizing such a notation the gravitino curvature is rewritten as follows:
\begin{equation}\label{usprhoCurv}
\rho_A \, = \,   \nabla \, \Psi_A \, - \, {\rm i} \, g \,   \Gamma_a \, \Psi_A\wedge V^a \,
\end{equation}
\subsection{The FDA in the Poincar\'e case and its $\mathrm{AdS_7}$ -fate}
Let us name $\overline{\osp\mathrm(2,6|2)}$ the contracted superalgebra  obtained by letting $g\to 0$ in the
Maurer Cartan equations corresponding to the vanishing of the  (\ref{osp262}) curvatures.
\par
As we anticipated few lines above the algebra $\overline{\osp\mathrm(2,6|2)}$ has two Chevalley cocycles
respectively of degree $3$ and $4$ that we show below:
\begin{eqnarray}\label{Pcohomology}
   \mathbb{K}^{[3]} & = & - \, {\rm i} \, \ft 12 \,
   \sigma^{\Lambda|A}_{\phantom{\Lambda|A}B} \,
   \overline{\Psi}^B \, \wedge \,   \Psi_A \,
   \wedge \, \mathcal{A}^\Lambda \, - \, \ft 12 \,\overline{\Psi}^A \, \wedge \, \Gamma_a \,\Psi_A \, \wedge \, V^a
   \label{bergcocycle}\\
    \mathbb{K}^{[4]} & = &  \,{\rm i} \, \ft 12 \,\overline{\Psi}^A \,
    \wedge \, \Gamma_{ab} \,\Psi_A  \wedge  V^a \wedge V^b
   \label{riccarcocycle}
\end{eqnarray}
The first cocycle is closed ($\mathrm{d}\mathbb{K}^{[3]} \, = \, 0$)
as a consequence of the fundamental Fierz
identity (\ref{bergsFierz}). The second cocycle is closed  ($\mathrm{d}\mathbb{K}^{[4]} \, = \, 0$) as a
consequence of the fundamental Fierz identity (\ref{bergsFierz}).
\par
The most general FDA is obtained by adjoining to the set of 1--forms $V^a, \, \omega^{ab}, \,
\mathcal{A}^\Lambda, \,\Psi_A$ a 2--form $\mathbf{B}^{[2]}$ and a 3-form $\mathbf{B}^{[3]}$ and by enlarging
the set of the super  Poincar\'e curvatures in the following way:
\subsubsection{Definition of the curvature $p$-forms in the Poincar\'e case}
\begin{eqnarray}
  \mathfrak{T}^a &\equiv& \underbrace{\mathrm{d}V^a
  \, - \, \omega^{ab} \, \wedge \,V^b \,}_{\mathcal{D}V^a} -
  \, \ft 12 \, \overline{\Psi}^A \, \wedge \, \Gamma^a \,\Psi_A \label{torsiondefi}\\
  \mathfrak{R}^{ab} &\equiv& \mathrm{d}\omega^{ab}
  \, - \, \omega^{ac} \, \wedge \, \omega^{cb} \label{curvadefi}\\
  \rho_A &\equiv& \underbrace{\mathrm{d}\Psi_A
  \, - \, \ft 14 \, \omega^{ab} \, \Gamma_{ab} \, \Psi_A}_{\mathcal{D}\Psi_A} \label{rhodefi} \\
\mathfrak{F}^\Lambda   &\equiv& \mathrm{d}\mathcal{A}^\Lambda
\, - \,{\rm i} \ft 12 \, e^{-\ft 12 \, \phi}
\,\sigma^{\Lambda|B}_{\phantom{\Lambda|B}A} \, \overline{\Psi}^A \, \wedge \,\Psi_B\label{Fdefi}  \\
\mathfrak{G}^{[3]}  &\equiv& \mathrm{d}\mathbf{B}^{[2]}
\, + \, \mathfrak{F}^\Lambda\, \wedge\,\mathcal{A}^\Lambda
\, + \, q \, e^{-\,\delta\,\phi} \,\mathfrak{T}^a \, \wedge \, V_a \, \nonumber\\
 &\null & + \,{\rm i} \ft 12 \, e^{-\ft 12 \, \phi}
 \,\sigma^{\Lambda|B}_{\phantom{\Lambda|B}A} \, \overline{\Psi}^A \,
 \wedge \,\Psi_B\,\wedge \, \mathcal{A}^\Lambda \, + \, \ft 12
 \,e^{- \, \delta\,\phi}\, \overline{\Psi}^A \, \wedge \, \Gamma_a \,\Psi_A \, \wedge \, V^a \label{3Gdefi}\\
\mathfrak{G}^{[4]}  &\equiv& \mathrm{d}\mathbf{B}^{[3]}
\,- \,{\rm i} \, \ft 12 \,e^{- \,\theta \, \phi}\,
\overline{\Psi}^A \, \wedge \, \Gamma_{ab} \,\Psi_A \wedge V^a \wedge V^b \label{4Gdefi}\\
  \mathrm{d}\phi &\equiv& \mathrm{d}\phi \label{phicurvadefi}\\
  \mathcal{D}\chi_A &\equiv& \mathrm{d}\chi_A \, - \, \ft 14 \, \omega^{ab} \,
  \Gamma_{ab} \, \chi_A \label{chicurvadefi}
\end{eqnarray}
where $q,\delta , \theta$ are  numerical parameters.
\par
Some comments are in order in relation with the above definitions. The basis for the construction of any FDA
is provided by the two fundamental structural theorems by Sullivan for whose discussion we refer the reader to
\cite{pietrobook}. The zeroth order step is provided by the minimal algebra which, as stated by the second of
Sullivan's theorems, requires a Chevalley  cohomology class of the superalgebra defined by the Maurer Cartan
equations. In the present case the minimal FDA is simply given by:
\paragraph{The minimimal FDA}
\begin{eqnarray}
  0 &=& \mathrm{d}V^a \, - \, \omega^{ab} \, \wedge \,V^b \, - \, \ft 12 \,
  \overline{\Psi}^A \, \wedge \, \Gamma^a \,\Psi_A \label{torsiondefi0}\\
  0 &=& \mathrm{d}\omega^{ab} \, - \, \omega^{ac} \, \wedge \, \omega^{cb} \label{curvadefi0}\\
  0 &=& \mathrm{d}\Psi_A \, - \, \ft 14 \,
  \omega^{ab} \, \Gamma_{ab} \, \Psi_A  \label{rhodefi0} \\
0 &=& \mathrm{d}\hat{\mathcal{A}}^\Lambda
\, - \,{\rm i} \,\sigma^{\Lambda|B}_{\phantom{\Lambda|B}A} \,
\overline{\Psi}^A \, \wedge \,\Psi_B\label{Fdefi0}  \\
0 &=& \mathrm{d}\mathbf{B}^{[2]} \, - \,  \mathbb{K}^{[3]} \label{3Gdefi0}\\
0 &=& \mathrm{d}\mathbf{B}^{[3]} \, - \,  \mathbb{K}^{[4]} \label{4Gdefi0}
  \end{eqnarray}
where the cohomology classes $\mathbb{K}^{[3,4]}$ were singled out above  in
eq.s(\ref{bergcocycle}-\ref{riccarcocycle}). The transition from the minimal FDA to the complete one encoded
in eq.s (\ref{torsiondefi}-\ref{chicurvadefi}) is related to Sullivan's first theorem stating that the most
general FDA is the semidirect sum of a contractible FDA with a minimal one. As it was observed many years ago
by one of us in \cite{FDAgauge}, this mathematical theorem has a deep meaning relative to the gauging of
algebras:
\begin{enumerate}
  \item The \textit{contractible generators} $\Omega^{A(p+1)}$ of any given FDA are
  to be physically identified with the \textit{curvatures}.
  \item The Maurer Cartan equations that begin with $\mathrm{d}\Omega^{A(p+1)}$ are
  the \textit{Bianchi identities.}
  \item The algebra which is gauged is the minimal subalgebra.
  \item The Maurer Cartan equations of the minimal subalgebra are consistently
  obtained by those of the full algebra by setting all contractible generators to zero.
\end{enumerate}
When a minimal FDA contains only one-forms, namely when it describes an ordinary Lie (super)-algebra, its
corresponding \textit{decontracted gauged} version is uniquely determined. Indeed the contractible generators,
\textit{i.e.} the curvatures, are  introduced  deforming the Maurer Cartan equations by means of new $2$-forms
that replace the $0$ on the left hand side. Instead when the minimal FDA is proper, namely when it contains
$p$-forms with $p>1$, the gauging is not unique. The contractible generators, namely the curvatures, can be
introduced not only on the left-hand side of the generalized Maurer Cartan equations, but also in appropriate
combinations on the right hand side. This involves the appearance of new coefficients that have to be selected
by the use of other principles. This is what happens in the case under consideration. There are three
modifications involved in the gauging procedure that leads from eq.s (\ref{torsiondefi0}-\ref{4Gdefi0}) to
eq.s (\ref{torsiondefi}-\ref{chicurvadefi}).
\par
The first modification corresponds to the introduction of the dilaton field $\phi$ which we know should be
there since it is comprised in the graviton multiplet. This is trivially done by rescaling the field
$\hat{\mathcal{A}}^\Lambda \, \rightarrow \, \exp\left[ \ft 12 \, \phi \right] \, \mathcal{A}^\Lambda $. The
normalization of the dilaton is arbitrarily fixed at this level in the pure (super) Lie algebra subsector;
then a relative coefficient to be later fixed by Bianchi consistency of the rheonomic parameterizations has to
be introduced in the curvatures of the $\mathbf{B}^{[2,3]}$-forms. Such coefficient has been named $\delta$.
\par
The second modification is precisely related with the introduction of curvature terms in the definition of the
$\mathfrak{G}^{[3]}$-curvature. Taking into account Lorentz invariance and scale dimensions we write:
\begin{eqnarray}
\mathfrak{G}^{[3]}  &\equiv& \mathrm{d}\mathbf{B}^{[2]} \, + \, \alpha \,\mathfrak{F}^\Lambda\,
\wedge\,\mathcal{A}^\Lambda \, + \, q \, e^{-\,\phi} \,\mathfrak{T}^a \, \wedge \, V_a \, \nonumber\\
 &\null & + \,{\rm i} \ft 12 \, e^{-\ft 12 \, \phi} \,\sigma^{\Lambda|B}_{\phantom{\Lambda|B}A}
 \, \overline{\Psi}^A \, \wedge \,\Psi_B\,\wedge \, \mathcal{A}^\Lambda
 \, + \, \ft 12 \,e^{- \,\delta\, \phi}\, \overline{\Psi}^A \,
 \wedge \, \Gamma_a \,\Psi_A \, \wedge \, V^a \label{Gdefinizioncina}
\end{eqnarray}
which at $\phi \, = \, 0$ and at zero-curvatures reduces to eq.(\ref{3Gdefi0}). The coefficient $\alpha$ is
fixed to $\alpha\, = \, 1$ by the requirement that in the Bianchi identities do not appear any bare
$\mathcal{A}^\Lambda$ fields, on the other hand the coefficient $q$ should be fixed later by the requirement
that the Bianchi identities admit a consistent rheonomic solution. In this respect we should remind ourselves
that from the physical point of view, the graviton multiplet just contains the degrees of freedom of a
$2$--form, or in a dual formulation of a $3$-form. Hence, when writing the ansatz for the rheonomic
parameterization of the FDA curvatures in (\ref{3Gdefi}-\ref{4Gdefi}), we should write their inner components
in the following way:
\begin{eqnarray}
  \mathfrak{G}^{[3]} &=& \mathcal{G}_{a_1a_2a_3} \, V^{a_1} \wedge V^{a_2} \wedge V^{a_3}
  \, + \, \mbox{outer part}\nonumber\\
  \mathfrak{G}^{[4]} &=& \nu \, e^{(1-\theta)\phi} \, \varepsilon_{b_1\dots b_3 a_1\dots a_4 }
  \, \mathcal{G}^{b_1\dots b_3} \, V^{a_1} \wedge V^{a_2} \wedge V^{a_3}\wedge V^{a_4}+
  \, \mbox{outer part}\label{g4g3duality}
\end{eqnarray}
As we are going to see the parameter $\theta$ will remain a free parameter up to the very end in the solutions
of Bianchi identities and it will be fixed only at the level of the Lagrangian, requiring that this latter
includes the following topological term:
\begin{equation}\label{fantasticus}
    \mathcal{L} \, \supset \, \mathfrak{G}^{[4]}\wedge \mathfrak{G}^{[3]}
\end{equation}
with no factor in front which depends on the dilaton. It will be particularly rewarding that such a condition
will set the other coefficients to the values utilized in \cite{bershoffo1} and \cite{PvNT}, which constitutes
a very powerful check on the consistency of our solution of the Bianchi identities. It should also be noted
that at the purely bosonic level the above term reduces to the following:
\begin{eqnarray}\label{filiotta}
    \mathfrak{G}^{[4]}\wedge \mathfrak{G}^{[3]}&
    \stackrel{\mbox{bosonic limit}}{\Longrightarrow} &
    \mathrm{d}\mathbf{B}^{[3]} \wedge \mathrm{d}\mathbf{B}^{[2]} \,
    + \, \mathrm{d}\mathbf{B}^{[3]}\wedge \mathfrak{F}^{\Lambda} \wedge \ \mathcal{A}^{\Lambda}\nonumber \\
    & = & - \mathbf{B}^{[3]}\wedge \mathfrak{F}^{\Lambda} \wedge \ \mathfrak{F}^{\Lambda}
    \, + \, \mathrm{d}\left(\mbox{something}\right)
\end{eqnarray}
namely, up to a total divergence the term (\ref{fantasticus}) is the topological term whose presence was
advocated by the authors of \cite{PvNT}. Furthermore, as we have already stressed in section
\ref{flussibrane}, the term (\ref{filiotta}) is the crucial one for the existence of flux $2$-branes with
Arnold-Beltrami fluxes, whose coefficient is to be precisely that one fixed by supersymmetry in the
supergravity lagrangian. Hence we can say that Arnold-Beltrami flux branes are a direct consequence of the FDA
structure analysed in the present section.
\section{Construction of Minimal $D=7$ Poincar\'e supergravity}
\label{construzia} In this section we perform the construction \textit{ex novo} of minimal $D=7$ supergravity
using the rheonomic approach.
\par
As it is standard in such an approach we begin with the Free Differential Algebra and with its associated
Bianchi identities that we solve \textit{in toto} with a rheonomic parameterization of all the $p$-form
curvatures. Such rheonomic parameterization already implies the field equations that can be worked out from it
with some care. Alternatively one can construct the action whose consistency with the rheonomic
parameterizations already determined from the Bianchi identities imposes constraints on the relative coefficients of its
terms able to fix them completly. In this way the field equations of the theory can be worked out from the
action as well.
\subsection{The Free Differential Algebra in the Poincar\'e case}
We begin by writing the complete form of the Bianchi identities for the Poincar\'e FDA comprising both the
three-form and the two-form curvatures. Next we will solve the Bianchi identities rheonomically showing that a
consistent solution does indeed exist with uniquely fixed parameters.
\subsubsection{Bianchi Identities in the Poincar\'e case.} From the curvatures
defined in eq.s (\ref{torsiondefi}-\ref{chicurvadefi}), by exterior differentiations
we obtain the following Bianchi identities:
\begin{eqnarray}
  \mathcal{D}\,\mathfrak{T}^a &=&  - \, \mathfrak{R}^{ab} \, \wedge \,V^b
  \, + \, \overline{\Psi}^A \, \wedge \, \Gamma^a \,\rho_A \, = \, 0
  \quad \left(\mbox{If $\mathfrak{T}^a\, = \, 0$}\right) \label{torsobianchi}\\
  \mathcal{D}\,\mathfrak{R}^{ab} &=& 0 \label{Riebianchi}\\
  \mathcal{D}\rho_A &=& - \, \ft 14 \, \mathfrak{R}^{ab} \, \wedge \, \Gamma_{ab} \,
  \Psi_A \label{rhobianchi} \\
\mathrm{d}\,\mathfrak{F}^\Lambda   &=&  {\rm i}\, \ft 14 \, e^{-\ft 12 \, \phi} \, \mathrm{d}\phi \,\wedge
\,\overline{\Psi}^A \, \wedge \, \sigma^{\Lambda|B}_A \,\Psi_B \, +\, {\rm i}\, e^{-\ft 12 \, \phi}
\,\overline{\Psi}^A \,
\wedge \, \sigma^{\Lambda|B}_A \,\rho_B \label{fbianchi} \\
\mathrm{d}\,\mathfrak{G}^{[3]}  &=& \, q \, e^{- \,\delta\, \phi} \, \mathfrak{R}^{ab}
\wedge  V_a  \wedge  V_b \, + \, q\, e^{- \,\delta\, \phi}\,\mathfrak{T}^a
\wedge \mathfrak{T}_a \, + \, \mathfrak{F}^\Lambda \wedge\mathfrak{F}^\Lambda\,
\, - \, q\, e^{- \, \delta \, \phi} \, d\phi  \wedge \mathfrak{T}^a \wedge  V_a\nonumber\\
 &\null& + \, \ft{q+1}{2} \, e^{- \,\delta\, \phi}\, \mathfrak{T}^a\,\wedge
 \, \overline{\Psi}^A\, \wedge \, \Gamma_a \Psi_A \,
 \, - \,\ft \delta 2 \, e^{- \,\delta \, \phi}\, \mathrm{d}\phi  \wedge  \overline{\Psi}^A \wedge
 \Gamma_a \Psi_A  \wedge  V^a\,\nonumber\\
 &\null & +\, {\rm i}\, e^{-\ft 12 \, \phi}\,\overline{\Psi}^A \, \wedge
 \, \sigma^{\Lambda|B}_A \,\Psi_B \, \wedge \, \mathfrak{F}^\Lambda
 \, + \, (q-1) \, e^{- \, \delta \, \phi}\, \overline{\Psi}^A \wedge
 \Gamma_a \rho_A  \wedge  V^a\,  \label{g3bianchi}\\
 \mathrm{d}\,\mathfrak{G}^{[4]}  &=& \,{\rm i} \, \left(
-\, e^{- \,\theta\, \phi}\,  \overline{\Psi}^A\, \wedge \, \Gamma_{ab}
\Psi_A \, \wedge \,\mathfrak{T}^a\, \wedge \, V^b \, + \,\ft \theta 2
\, e^{- \,\theta \, \phi}\, \mathrm{d}\phi  \wedge  \overline{\Psi}^A \wedge
\Gamma_{ab} \Psi_A  \wedge  V^a\, \wedge V^b \right. \nonumber\\
 &\null &\left. +  \, e^{- \, \theta \, \phi}\, \overline{\Psi}^A
 \wedge  \Gamma_{ab} \rho_A  \wedge  V^a\, \wedge V^b \right) \label{g4bianchi}
 \end{eqnarray}
 and
 \begin{eqnarray}
 \mathrm{d}\, \mathrm{d}\phi &=& 0 \label{dilabianchi}\\
  \mathcal{D}\,\mathcal{D}\chi_A &=& - \, \ft 14 \, \mathfrak{R}^{ab} \, \wedge \, \Gamma_{ab}
  \, \chi_A \label{gravitelbianchi}
\end{eqnarray}
Let us now turn to study the rheonomic solution of the Bianchi identities.
\subsection{Ansatz for the rheonomic parameterization of the curvatures in the Poincar\'e case}
First of all let us write a complete rheonomic ansatz for the curvature parameterizations.
\par
We begin by writing a rheonomic parameterization of all the curvatures for the forms of degree $p\le 1$ that
correspond to a standard superalgebra enlarged with the dilaton and the dilatino zero-forms. In such a
rheonomic parameterization we introduce also a three-index antisymmetric tensor $\mathcal{G}_{abc}$ which
later can be identified with the space-time components of either  the three-form or the four-form curvature.
Explicitly we set:
\begin{eqnarray}
  \mathfrak{T}^a &=& 0 \label{torsO}\\
  \mathfrak{R}^{ab} &=&\mathcal{R}^{ab}_{\phantom{ab}cd} \, V^c
  \wedge  V^d \, + \, \overline{\Theta}^{ab|A}_{c} \, \Psi_A \wedge V^c
  + \, \lambda_1 \,  e^{\delta \,\phi}\,\mathcal{G}^{abc} \,  \overline{\Psi}^A  \wedge  \Gamma_c \Psi_A \nonumber\\
  &\null& + \, \lambda_2 \,  e^{\delta \,\phi}\, \mathcal{G}_{pqr}
  \, \overline{\Psi}^A  \wedge   \Gamma^{abpqr} \Psi_A \, + \,
 {\rm i}\, \mu_1 \, e^{\ft 12 \phi}\, \mathcal{F}^{\Lambda|ab}
 \, \sigma^{\Lambda|B}_{\phantom{\Lambda|B}A} \,\overline{\Psi}^A  \wedge  \Psi_B \,\nonumber\\
  &\null&{\rm i}\, \mu_2 \, e^{\ft 12 \phi}\, \mathcal{F}^{\Lambda}_{pq}
  \, \sigma^{\Lambda|B}_{\phantom{\Lambda|B}A} \,\overline{\Psi}^A  \wedge  \Gamma^{abpq}\,\Psi_B \label{Riepara}\\
  \rho_A &\equiv& \rho_{A|ab} \, V^a \, \wedge \, V^b \,
   + \, \left(\mathcal{M}_A^B \, \Gamma_a \, + \, \Gamma_a \,\mathcal{N}_A^B \right)\,
   \Psi_B \, \wedge \, V^a \nonumber\\
  &\null&+\, g_1 \, \Gamma_m \, \chi_A \, \overline{\Psi}^C \wedge \Gamma^m \Psi_C \,
  + \, g_2 \, \Gamma_{mn}  \chi_A \, \overline{\Psi}^C  \wedge  \Gamma^{mn} \,\Psi_C \nonumber\\
  &\null& -\, g_3 \,  \chi_B \, \sigma^{\Lambda|B}_{\phantom{\Lambda|B}A}
  \, \sigma^{\Lambda|D}_{\phantom{\Lambda|D}C}\, \overline{\Psi}^C  \wedge \Psi_D
  \, - \, g_4 \,\Gamma_{pqr} \chi_B \, \sigma^{\Lambda|B}_{\phantom{\Lambda|B}A}
  \, \sigma^{\Lambda|D}_{\phantom{\Lambda|D}C}\,  \overline{\Psi}^C  \wedge \Gamma^{pqr} \Psi_D  \label{rhopara}\\
\mathfrak{F}^\Lambda   &\equiv&  \mathcal{F}^\Lambda_{ab} \, V^a \, \wedge\, V^b \,
+ \,{\rm i} \, a_1 \, e^{-\ft 12 \, \phi} \,\sigma^{\Lambda|B}_{\phantom{\Lambda|B}A}
\,\overline{\Psi}^A\Gamma_{a} \, \chi_B \, \wedge \, V^a\label{Fpara}  \\
d\phi & = & \Phi_a \, V^a \, + \, \overline{\Psi}^A \, \chi_A \label{dilatpara}\\
  \mathcal{D}\chi_A &\equiv& \mathcal{D}_a \chi_A \, V^a \, + \, \mathcal{P}_A^B \, \Psi_B\label{gravitelpara}
\end{eqnarray}
where $\overline{\Theta}^{ab|A}_{c}$ is a spinor-tensor linear in the gravitino field strength $\rho_{A|ab}$
and where the matrices appearing in the fermionic curvatures are the following ones:
\begin{eqnarray}
  \mathcal{M}_A^B &=& \delta^A_B \, \left(b_1 \, e^{\delta \phi}\, \Gslat \, + \, \kappa_1 \,
  \Phislat\right) \, - \, {\rm i} \, d_1 \, e^{\ft 12 \phi}\,\Fslat^B_A \label{gravitinM}\\
  \mathcal{N}_A^B &=& \delta^A_B \, \left(b_2 \,e^{\delta \phi}\,  \Gslat \, + \, \kappa_2
  \, \Phislat\right) \, - \,{\rm i} \, d_2  \, e^{\ft 12 \phi}\,\Fslat^B_A \label{gravitinN}\\
\mathcal{P}_A^B &=& \delta^A_B \, \left(c_1 \, e^{\delta \phi}\,  \Gslat \, + \, c_3 \, \Phislat\right) \, -
\,{\rm i} \, c_2 \, e^{\ft 12 \phi}\, \Fslat^B_A \label{gravitelP}
\end{eqnarray}
having defined
\begin{eqnarray}
  \Gslat \, \equiv \,  \mathcal{G}_{abc} \, \Gamma^{abc} \quad ; \quad
  \Fslat_A^B \, \equiv \, \mathcal{F}^\Lambda_{ab} \, \Gamma^{ab}
  \, \sigma^{\Lambda|B}_{\phantom{\Lambda|B}A} \quad ; \quad
  \Phislat \,  \equiv \, \Phi_a \, \Gamma^a \label{slatti}
\end{eqnarray}
The above paramerization involves the following set of  19 numerical coefficients \footnote{Actually the last
coefficient $\delta$ is already contained in the FDA comprising either  the three-form or the four-form
curvature. However when we consider only the curvatures of the curvatures of degree $p\le 2$, then $p$ is some
parameter appearing only in the rheonomic parameterizations.}:
\begin{equation}\label{parammi}
 \mbox{coeff}_{\mathrm{Lie}} \, = \,   \left\{a_1,b_1,b_2,d_1,d_2,c_1,c_2,c_3,\kappa_1,
 \kappa_2,g_1,g_2,g_3,g_4,\lambda_1,\lambda_2,\mu_1,\mu_2 , \delta\right\}
\end{equation}
In addition to the above rheonomic parameterizations we introduce those of the higher-form curvatures, namely:
\begin{eqnarray}
  \mathfrak{G}^{[3]}  &\equiv& \mathcal{G}_{abc} \, V^a \, \wedge\, V^b \,
  \wedge\, V^c  + \, a_2 \, e^{-  \,\delta\, \phi} \,\overline{\Psi}^A \,
  \Gamma_{ab} \, \chi_A \, \wedge \, V^a\,\wedge \, V^b\label{G3para}
\end{eqnarray}
\begin{eqnarray}
  \mathfrak{G}^{[4]}  &\equiv&\nu \, e^{(1-\theta)\phi} \,
  \varepsilon_{a_1\dots a_3 b_1 \dots b_4} \, \mathcal{G}^{a_1a_2 a_3}
  \, V^{b_1} \, \wedge\, \dots \wedge V^{b_4} \,
  + \,{\rm i} \, w \, e^{-  \,\theta \, \phi} \,\overline{\Psi}^A \,
  \Gamma_{abc} \, \chi_A \, \wedge \, V^a\,\wedge \, V^b \wedge \, V^c \nonumber\\
  \label{G4para}
\end{eqnarray}
If we consider the FDA that comprises only the three-form curvature the total set of numerical coefficients to
be determined is given by:
\begin{equation}\label{settusG3}
  \mbox{coeff}_{\mathrm{FDA}_3} \, = \,  \mbox{coeff}_{\mathrm{Lie}}\bigcup \,
  \underbrace{\{a_2 , q\}}_{\mbox{coeff}_{\mathfrak{G}_3}}
\end{equation}
If instead we consider the FDA that comprises only the four-form curvature, the total set of numerical
coefficients to be determined is given by:
\begin{equation}\label{settusG4}
  \mbox{coeff}_{\mathrm{FDA}_4} \, = \,  \mbox{coeff}_{\mathrm{Lie}}\bigcup \,
  \underbrace{\{w , \nu, \theta\}}_{\mbox{coeff}_{\mathfrak{G}_4}}
\end{equation}
In the first case the total number of coefficients to be fixed is 21, while in the second is 22.
\par
In order for the three-form and four-form curvatures to coexist we should be able to determine consistently a
set of 24 parameters:
\begin{equation}\label{settusG34}
  \mbox{coeff}_{\mathrm{FDA}_{3\oplus4}} \, = \,  \mbox{coeff}_{\mathrm{Lie}}\bigcup
  \, \underbrace{\{a_2 , q,\nu, w, \theta \}}_{\mbox{coeff}_{\mathfrak{G}_3\oplus\mathfrak{G}_3}}
\end{equation}
In appendix \ref{dettaglione} we show that both solutions are available for the sets of 21 and 22 parameters,
respectively with a residual freedom of one parameter. The solution for the set of 24 parameters is also
available and fixes all parameters in function of a residual one that we choose to be $\theta$. The result
obtained in appendix \ref{bianchip34} is displayed in eq. (\ref{comExternB})and it is repeated here for the
reader's convenience:
\begin{equation}
 \begin{array}{llllllllllll}
 a_1 & = &
   -\frac{1}{2} & ;
   & a_2 & = &
   -\frac{1}{2} & ;
   & b_1 & = &
   -\frac{1}{8} & \nonumber
   \\
 b_2 & = & \frac{2
   \theta +1}{24-16 \theta } &
   ; & c_1 &
   = & \frac{1}{3-2
   \theta } & ; &
   c_2 & = &
   \frac{1}{3-2 \theta } &
   \nonumber \\
 c_3 & = &
   \frac{1}{2} & ;
   & d_1 & = &
   \frac{1}{4} & ;
   & d_2 & = &
   \frac{1-2 \theta }{8 \theta
   -12} & \nonumber \\
 g_1 & = &
   \frac{1}{64} (14 \theta +3)
   & ; & g_2 &
   = &
   \frac{1}{128} (2 \theta -3)
   & ; & g_3 &
   = & \frac{1}{64}
   (1-14 \theta ) &
   \nonumber \\
 g_4 & = &
   \frac{1}{384} (-2 \theta
   -1) & ; & \kappa
   _1 & = & 0 &
   ; & \kappa _2 &
   = & 0 &
   \nonumber \\
 \lambda _1 & = &
   \frac{3}{2}+\frac{3}{2
   \theta -3} & ; &
   \lambda _2 & = &
   \frac{1}{6-4 \theta } &
   ; & \mu _1 &
   = & \frac{1}{3-2
   \theta }-1 & \nonumber \\
 \mu _2 & = &
   \frac{1}{4 \theta -6} &
   ; & \delta  &
   = & 1 &
   ; & w &
   = &
   -\frac{\theta }{3} &
   \nonumber \\
 q & = & 1 &
   ; & \nu  &
   = &
   \frac{1}{12} &
   ; & \theta  &
   = & \theta  &
   \nonumber
\end{array}
\end{equation}
As usual the solution  is multiply checked since the constraints are many more than the parameters that can be
fixed.
\par
As we announced before the last parameter can be fixed requiring that the term (\ref{fantasticus}) can appear
in the Lagrangian without dilaton factor in front. For this to be possible it is necessary that after
substituting the rheonomic parameterization, the pure space time part of the term (\ref{fantasticus}) should
be proportional to the kinetic term of the $\mathbf{B}^{[2]}$-form, namely:
\begin{equation}\label{carondimonio}
    e^{2\phi} \, \mathcal{G}_{abc} \, \mathcal{G}^{abc} \, V^{a_1} \wedge \dots \wedge V^{a_7} \,
    \epsilon_{a_1 \dots a_7}
\end{equation}
This immediately fixes the value
\begin{equation}\label{cognatus}
    \theta \, = \, - \, 1
\end{equation}
Inserting such a value into eq. (\ref{comExternB}) we obtain the following final values of the coefficients:
\begin{equation}\label{calesse}
\begin{array}{llllllllllll}
 a_1 & = &
   -\frac{1}{2} & ;
   & a_2 & = &
   -\frac{1}{2} & ;
   & b_1 & = &
   -\frac{1}{8} & \nonumber
   \\
 b_2 & = &
   -\frac{1}{40} &
   ; & c_1 &
   = & \frac{1}{5}
   & ; & c_2 &
   = & \frac{1}{5}
   & \nonumber \\
 c_3 & = &
   \frac{1}{2} & ;
   & d_1 & = &
   \frac{1}{4} & ;
   & d_2 & = &
   -\frac{3}{20} & \nonumber
   \\
 g_1 & = &
   -\frac{11}{64} &
   ; & g_2 &
   = &
   -\frac{5}{128} &
   ; & g_3 &
   = &
   \frac{15}{64} & \nonumber
   \\
 g_4 & = &
   \frac{1}{384} &
   ; & \kappa _1 &
   = & 0 &
   ; & \kappa _2 &
   = & 0 &
   \nonumber \\
 \lambda _1 & = &
   \frac{9}{10} & ;
   & \lambda _2 & =
   & \frac{1}{10} &
   ; & \mu _1 &
   = & -\frac{4}{5}
   & \nonumber \\
 \mu _2 & = &
   -\frac{1}{10} &
   ; & \delta  &
   = & 1 &
   ; & w &
   = & \frac{1}{3}
   & \nonumber \\
 q & = & 1 &
   ; & \nu  &
   = &
   \frac{1}{12} &
   ; & \theta  &
   = & -1 &
   \nonumber
\end{array}
\end{equation}
It is extremely nice and reassuring that the condition (\ref{cognatus}) yields the same result as the
condition (\ref{g1Bergo}) which guarantees compatibility with the coefficients determined in
 \cite{bershoffo1} by means of the Noether coupling construction. This completely independent
 determination of the supersymmetry transformation rules confirms therefore from a
 pure algebraic viewpoint the Noether coupling calculations of both paper \cite{bershoffo1}  and paper \cite{PvNT}.
\par
It is now a question of constructing the geometrical action consistent with this rheonomic parameterization. This will be accomplished, up to four fermionic terms and for a generic number of vector multiplets, elsewhere. For the purpose of the present work, it suffices to define the precise dictionary between the fields and parameters on our rheonomic formulation and those in \cite{PvNT}.
\subsection{Construction of the bosonic action of ungauged minimal $D=7$ supergravity}
Following the standard procedures of the rheonomic approach we consider an ansatz for the action in terms of
differential forms living in superspace:
\begin{eqnarray}
  \mathcal{A}^{ungauged}_{\mbox{$D=7$ SUGRA}} &=& \int \, \mathcal{L}^{ungauged}_{tot} \\
  \mathcal{L}^{ungauged}_{tot} &=& \mathcal{L}^{ungauged}_{Bkin} \,
   + \,\mathcal{L}^{ungauged}_{Fkin}\, \, + \,\mathcal{L}^{ungauged}_{Pauli}\, + \,\mathcal{L}^{ungauged}_{4fermi}
\end{eqnarray}
where $\mathcal{L}^{ungauged}_{Bkin}$ is the bosonic Lagrangian containing the kinetic terms of the bosonic fields and the Chern-Simons term, $\mathcal{L}^{ungauged}_{Fkin}$ is the kinetic Lagrangian for the fermionic fields while the last two terms describe the Pauli interactions  and the quartic terms in the fermion fields. For the scope of the present work, we shall be only interested in $\mathcal{L}^{ungauged}_{Bkin}$ which has the general form:
\begin{eqnarray}
  \mathcal{L}^{ungauged}_{Bkin} &=& f_1 \,\mathfrak{R}^{a_1a_2} \wedge V^{a_3} \wedge
  \dots \wedge V^{a_7} \, \epsilon_{a_1\dots a_7}\nonumber\\
   &&\, + \, f_2 \,\Phi^{a_1} \, \left(\mathrm{d}\phi \, - \, \overline{\Psi}^A \,
   \chi_A \right)\wedge  V^{a_2} \wedge \dots \wedge V^{a_7} \, \epsilon_{a_1\dots a_7}  \nonumber\\
&& + f_3 \, e^\phi \, \mathcal{F}^{\Lambda|a_1a_2} \, \left(\mathfrak{F}^{\Lambda}
\, - \,{\rm i} a_1 e^{-\ft 12 \phi} \, \sigma^{\Lambda|B}_{\phantom{\Lambda B}A}
\, \overline{\Psi}^A \, \Gamma_a \, \chi_B \wedge V^a \right)\wedge V^{a_3}
\wedge \dots \wedge V^{a_7} \, \epsilon_{a_1\dots a_7} \nonumber\\
&& + \, f_4 \, \mathcal{G}_{abc} \, \left( \mathfrak{G}^{[4]} \,
- \, {\rm i} w \, e^\phi\, \overline{\Psi}^{A} \Gamma_{pqr} \,
\chi_A \wedge V^p \wedge V^q \wedge V^r \right) \wedge V^a \wedge V^b \wedge V^c\nonumber \\
&& + \, f_5 \, \left( \mathfrak{G}^{[3]} \, -\, a_2 \, e^{-\phi} \,
\overline{\Psi}^A \, \Gamma_{ab} \chi_A \wedge V^a\wedge V^b\right)
\wedge \left(\mathfrak{G}^{[4]} \,  + \,\frac{{\rm i}}{2} \, e^{\phi}
\overline{\Psi}^A \wedge \Gamma_{ab} \Psi_A \wedge V^a\wedge V^b\right)\nonumber\\
&&+ \, \left (- \,360 f_2 \, \Phi^a \, \Phi_a \, - \, 120 \, f_3
\, e^{\phi} \,\mathcal{F}^{\Lambda|ab} \mathcal{F}^{\Lambda}_{ab} - 6\, f_4
\, e^{2\phi} \, \mathcal{G}_{abc}\, \mathcal{G}^{abc} \right ) \, \mathrm{Vol}_7 \nonumber\\
\mathrm{Vol}_7 & \equiv & \frac{1}{7!} \, \epsilon_{a_1\dots a_7} \, V^{a_1} \wedge \dots \wedge V^{a_7}
\label{LBkin}
\end{eqnarray}
The coefficients $a_1,a_2,w$ appearing in the above action are those displayed in the rheonomic
parameterization of the curvatures and have already been determined through the solution of the Bianchi
identities. All the coefficients parametrizing $\mathcal{L}^{ungauged}_{tot}$, including $f_1,\dots ,f_5$ in the bosonic Lagrangian,
have to be fixed by considering the field equations from $\mathcal{A}^{ungauged}_{\mbox{$D=7$ SUGRA}}$ as differential form equations in
superspace that should be satisfied upon replacement of the previously determined Bianchi identities.
\par
Some observations can be immediately made. First of all let us note that in a similar way to the case of the
rheonomic formulation of $D=11$ supergravity \cite{D'Auria:1982nx} in the lagrangian we have both the
curvature $\mathfrak{G}^{[4]}$ and the curvature $\mathfrak{G}^{[3]}$, yet the second appears only in the
topological term $\mathfrak{G}^{[4]}\wedge\mathfrak{G}^{[3]} \oplus \mbox{more}$ having coefficient $f_5$. The
coefficient $f_5$ must be equal to $-\,f_4$: in this way when we vary the Lagrangian in $\delta
\mathbf{B}^{[3]}$ we obtain:
\begin{equation}\label{favoloso1}
    f_4 \, \left( \mathrm{d} \left [\mathcal{G}_{abc} V^a\wedge V^b\wedge V^c \, +\, a_2 \, e^{-\phi}
    \, \overline{\Psi}^A \, \Gamma_{ab} \chi_A \wedge V^a\wedge V^b\right] - \,
    \mathrm{d}\mathfrak{G}^{[3]} \right) \, = \,0
\end{equation}
which is nothing else but the statement that the rheonomic parameterization (\ref{G3para}) satisfies the
Bianchi identity (\ref{g3bianchi}) with the already determined coefficients (\ref{calesse}). At the same time
the variation of the Lagrangian in $\delta \mathbf{B}^{[2]}$ yields:
\begin{equation}\label{favoloso2}
    f_5 \,  \mathrm{d}\left[\mathfrak{G}^{[4]} \,  + \,\frac{{\rm i}}{2} \, e^{\phi} \overline{\Psi}^A
    \wedge \Gamma_{ab} \Psi_A \wedge V^a\wedge V^b   \right] \, = \,0
\end{equation}
which upon the substitution of the rheonomic parameterizations is identically satisfied. Indeed
\begin{equation}\label{identita}
    \mathfrak{G}^{[4]} \,  + \,\frac{{\rm i}}{2}  \, e^{ \phi} \overline{\Psi}^A \wedge \Gamma_{ab}
    \Psi_A \wedge V^a\wedge V^b \, =\, \mathrm{d}\mathbf{B}^{[3]} \,\Rightarrow
    \,\mathrm{d}^2\mathbf{B}^{[3]} \, = \, 0
\end{equation}
This means that $\mathbf{B}^{[3]}$ enters the Lagrangian only through a total derivative term.
\section{The bosonic lagrangian  and the embedding of flux $2$-branes in supergravity}
\label{boselagravaialetto} Next we consider the form of the bosonic lagrangian of minimal $D=7$ supergravity,
as it emerges from the rheonomic construction  and we address the embedding of the flux $2$-branes described
in section \ref{flussibrane} into solutions of supergravity field equations.
\par
As mentioned earlier, in a separate paper we plan to present the explicit derivation of the $D=7$ lagrangian utilizing the rheonomic
approach and completing the task with the inclusion of all $4$-fermi terms. Yet, as we stressed several times,
the field equations of the theory are already implicitly determined by the complete solution of the Bianchi
identities. In the spirit of such an observation we can already determine (up to an overall scale) all the
coefficients $f_{1,\dots,5}$ appearing in the bosonic action,  by considering the embedding of the $2$-brane
solutions; at the same time our embedding procedure provides a cross check of the rheonomic construction with
the Noether construction of \cite{PvNT}. Indeed we organize the embedding procedure in the following steps:
\begin{description}
  \item[A)] First, considering the  bosonic supergravity lagrangian as derived in \cite{PvNT},
  we easily work out  the rescalings that bring it to the standard flux $2$-brane form of eq. (\ref{fluxbraneaction}).
  \item[B)] Secondly, comparing the supersymmetry transformation rules derived in
  \cite{PvNT} with those that follow from our rheonomic solutions of the Bianchi identities,
  we work out the rescalings that connect our normalizations of the supergravity fields with those
  of \cite{PvNT} and of the standard flux $2$-brane form of eq. (\ref{fluxbraneaction}).
  \item[C)] Finally, knowing all relative normalizations we derive the constraints
  on the coefficients of the rheonomic lagrangian necessary for its bosonic sector
  to be identical (up to rescalings) with the $2$-brane form of eq. (\ref{fluxbraneaction})
  and hence to the action obtained in \cite{PvNT}. The direct verification that the rheonomic construction
  of the action yields precisely these coefficients $f_{1,\dots,5}$, and the determination of the remaining ones, will be presented in a future paper.
\end{description}
\subsection{Comparison of  minimal $D=7$ supergravity according to the \textit{TPvN} construction with the flux brane action.}
\label{compaTPvN} In this subsection we make a comparison between the action (\ref{fluxbraneaction}) and the
bosonic action of Minimal $D=7$ Supergravity as it was derived in \cite{PvNT}, which, for brevity we name
\textit{TPvN}.
\par
Since the authors of \cite{PvNT} use the Dutch conventions for tensor calculus with imaginary time, the
comparison of the lagrangians at the level of signs is difficult, yet at the level of absolute values of the
coefficients it is possible, by means of several rescalings. First we observe that the normalization of the
Einstein term in eq. (2) of \textit{TPvN} is the same, if we take into account the already stressed $\ft 12$
difference in the definition of the Ricci tensor and scalar curvature. Secondly we note that the normalization
of the dilaton kinetic term in eq.(2) of \textit{TPvN}, namely $\ft 12$ becomes that of the action
(\ref{fluxbraneaction}), namely $\ft 14$ if we define:
\begin{equation}\label{subillus1}
    \phi_{TPvN} \, = \, \ft{1}{\sqrt{2}} \, \varphi
\end{equation}
A check that this is the correct identification arises from inspection of the dilaton factor in front of the
three-form kinetic term. Using eq.(3) of \textit{TPvN}, we see that according to this construction such a
factor is:
\begin{equation}\label{dilfactus}
    \exp \left[- \, \ft{4}{\sqrt{5}} \, \phi_{TPvN} \right] \, = \, \exp \left[- 2 \, \ft{2}{\sqrt{5}}
    \, \varphi \right]
\end{equation}
This confirms the value $a \, = -\, 2 \, \ft{2}{\sqrt{5}}$ leading to the miraculous value $\Delta \, = \, 4$
of the dimensional reduction invariant. Thirdly we consider the necessary rescalings for the
$\mathbf{A}^{[3]}$ and $\mathbf{A}^{\Lambda}$ gauge fields. Taking into account the different strengths of the
exterior derivatives (see unnumbered eq.s of \cite{PvNT} in between eq.(1) and (2)) we see that in order to
match the normalizations of (\ref{fluxbraneaction}) we have to define:
\begin{eqnarray}
  {A}^{TPvN}_{\lambda\mu\nu} &=& \ft {1}{4\sqrt{2}} \, \mathbf{A}^{[3]}_{\lambda\mu\nu}
  \quad \Rightarrow \quad F^{TPvN}_{\lambda\mu\nu\rho} \, = \, \ft {1}{\sqrt{2}}
  \mathbf{F}_{\lambda\mu\nu\rho} \nonumber\\
  {A}^{\Lambda|TPvN}_\mu &=& \sqrt{\ft {\omega}{8}} \, \mathbf{A}^{\Lambda}_\mu
   \quad \Rightarrow \quad F^{\Lambda|TPvN}_{\lambda\mu} \, = \, \sqrt{\ft {\omega}{2} }
   \,\mathbf{F}^\Lambda_{\lambda\mu}
  \label{minotaurotto}
\end{eqnarray}
with these redefinitions  we can calculate the value of $\kappa$ according to \textit{TPvN}. We find:
\begin{equation}\label{gelsomino}
    \ft{1}{48 \, \sqrt{2}} \, F^{TPvN}_{\mu\nu\rho\sigma}F^{\Lambda|TPvN}_{\kappa\lambda} A^{\Lambda|TPvN}_\tau \, \epsilon^{\mu\nu\rho\sigma\lambda\kappa\tau} \, = \, \ft{\omega}{384} \, \mathbf{F}_{\mu\nu\rho\sigma}\mathbf{F}^{\Lambda}_{\kappa\lambda} \mathbf{A}^{\Lambda}_\tau \, \epsilon^{\mu\nu\rho\sigma\lambda\kappa\tau}
\end{equation}
which implies:
\begin{equation}\label{conservadipomodoro}
    \kappa \, = \, \ft{\omega}{384}
\end{equation}
In this way the bosonic action of supergravity, according to \textit{TPvN} is mapped into the flux brane
action (\ref{fluxbraneaction}) by means of the rescalings (\ref{gelsomino}) and (\ref{subillus1}). This shows
that Arnold Beltrami flux branes are solutions of Minimal $D=7$ supergravity and of no other theory of the
same type which is not supersymmetric.
\subsection{Comparison of \textit{TPvN} susy rules with the rheonomic solution of Bianchi identities}
The next step in our agenda is the comparison of the supersymmetry transformation rules derived in \cite{PvNT}
with those derived from our rheonomic solution of the Bianchi identities in order to find the appropriate
rescalings that map our normalizations of the supergravity fields into those of \cite{PvNT}. Combining the
results of the previous section \ref{compaTPvN} with the comparison explored in the present section we arrive
at the relation between the bosonic supergravity fields of our algebraic rheonomic construction and the fields
utilized in the flux-brane action (\ref{fluxbraneaction}), namely we achieve the desired embedding of flux
$2$-brane solutions into supergravity.
\par
Let us proceed systematically. We set:
\begin{eqnarray}
  \phi &=& \lambda \, \varphi \, = \, \sqrt{2} \, \lambda \, \phi^{TPvN} \nonumber\\
  \mathbf{B}^{[3]} &=& \tau \, \mathbf{A}^{[3]}  \nonumber\\
  \null & \Downarrow & \null \nonumber\\
  \mathcal{G}_{\lambda\mu\nu\rho} &= & \tau \, \mathbf{F}_{\lambda\mu\nu\rho} \quad \Rightarrow \quad
  \mathcal{G}_{\lambda\mu\nu\rho} \, = \, {\sqrt{2}\, \tau} \, {F}^{TPvN}_{\lambda\mu\nu\rho} \label{balbiez}
\end{eqnarray}
Our goal is to determine the rescaling factors $\lambda$ and $\tau$. The first is immediately determined by
comparison of the dilaton depending scaling factors in the transformation rules and it was already fixed by
the requirement $a \, = \, 2\,\sqrt{\frac{2}{5}}$. We have:
\begin{equation}\label{fescennino}
  \lambda \, = \, \sqrt{\frac{2}{5}}
\end{equation}
To fix the second we consider the supersymmetry transformation rules of the dilatinos displayed in eq.(4) of
\cite{PvNT}. We find:
\begin{equation}\label{coccodibenzina}
  \delta_{SUSY} \, \chi^{TPvN}_A \, = \, \left( \ft 12 \, \Dslat \,\phi^{TPvN} \,
  + \, \frac{1}{24\,\sqrt{10}} \, \exp\left[2\sqrt{\frac{2}{5}} \,\phi^{TPvN}\right]
  \,\Gamma^{\lambda\mu\nu\rho} \, F^{TPvN}_{\lambda\mu\nu\rho} \right) \, \epsilon_{A} \,
  + \, \mbox{$F^\Lambda_{\mu\nu}$ terms}
\end{equation}
\par
In the rheonomic approach the supersymmetry transformation of the dilatinos is obtained from the rheonomic
parametererization of their covariant differential encoded in eq.s (\ref{gravitelpara}) and (\ref{gravitelP}).
We obtain:
\begin{equation}\label{botticino}
  \delta_{SUSY}\, \chi_A \, = \, \mathcal{P}_A^B \, \epsilon_B
\end{equation}
which has to be compared with eq.(\ref{coccodibenzina}). An absolute comparison requires  the
relative normalizations of the dilatinos $\chi_A$ and $\chi^{TPvN}_A$, to be given below, although for the time being we may just  focus on the ratio of the coefficients of the $\Dslat \,\phi^{TPvN}$ and $\Fslat^{TPvN}$ terms.
Indeed this ratio is independent from the normalization of the dilatino field.
\par
First, recalling the duality relation (\ref{g4g3duality}) with $\nu \, = \, \frac{1}{12}$ we find:
\begin{equation}\label{g4slaccio}
  \Gamma^{a_1\dots a_4} \, \mathcal{G}_{a_1\dots a_4} \, = \, 2 \, \Gamma^{a_1\dots a_3} \, \mathcal{G}_{a_1\dots a_3}
  \, = \, 2 \, \Gslat
\end{equation}
Secondly utilizing the rescalings (\ref{balbiez}) and eq.(\ref{g4slaccio}) we convert
eq.(\ref{coccodibenzina}) to
\begin{equation}\label{coccodiesel}
  \delta_{SUSY} \, \chi^{TPvN}_A \, = \, \left( \frac{\sqrt{5}}{4} \, \Dslat \,\phi \,
  + \, \frac{ \exp{\phi}}{12\,\sqrt{20}\, \tau} \, \,\Gslat \right) \, \epsilon_{A} \,
  + \, \mbox{$F^\Lambda_{\mu\nu}$ terms}
\end{equation}
Consistency with our own result from Bianchi identities requires:
\begin{equation}\label{taudetermino}
  \frac{\frac{1}{12\,\sqrt{20}\, \tau}}{\frac{\sqrt{5}}{4}} \, = \, \frac{c_1}{c_3} \, = \,
  \frac{2}{5} \quad \Rightarrow\quad \tau \, = \, \frac{1}{12}
\end{equation}
In this way the embedding of the flux $2$-brane system in our rheonomic formulation of $D=7$ supergravity is
completly fixed. A summary of the conversion table is displayed below:
\begin{equation}\label{sigresca}
  \phi \, = \, \sqrt{\frac 2 5} \,\varphi \quad ; \quad \mathbf{B}^{[3]} \, = \,
  \frac {1}{12} \, \mathbf{A}^{[3]} \quad ; \quad
    \mathcal{A}^\Lambda \, = \, \sigma \, \mathbf{A}^\Lambda
\end{equation}
The reascaling of the supergravity vector fields encoded in the symbol $\sigma$ is not fixed so far since the
normalization of the vector fields is also adjustable in the flux-brane lagrangian by means of the free
parameter $\omega$.
\par
In appendix (\ref{costrettioneffi}) we show that the above comparisons imply the following prediction on the
coefficients of the supergravity bosonic action:
\begin{equation}\label{cofficini}
    f_2 \, = \, \frac{5}{12} \, f_1 \quad ; \quad f_3 \, = \, 2\, f_1 \quad ; \quad f_4 \, = \, - \, f_5 \, =
    \, - \, 60 \, f_1
\end{equation}
When these relations are fulfilled the bosonic action of supergravity (\ref{LBkin}) is mapped into the
flux-brane action (\ref{fluxbraneaction}) by means of the rescalings (\ref{sigresca}), the constraint $\kappa
\, = \, \frac{\omega}{384}$ is respected and the supersymmetry transformation rules in the background of any
brane solution can be worked out from the rheonomic parametrization of the FDA curvatures satisfying Bianchi
identities.
\par For the sake of completeness we also give the dictionary for the fermionic fields and the supersymmetry parameter:
\begin{eqnarray}
\chi^{TPvN}_A=\sqrt{\frac{5}{2}}\,\chi_A\,\,;\,\,\,\,\psi^{TPvN}_A=\sqrt{2}\,\Psi_A\,\,;\,\,\,\,\epsilon^{TPvN}_A=\sqrt{2}\,
\epsilon_A\,,
\end{eqnarray}
where we have renamed $\chi^{TPvN}_A$ the spin one-half fields denoted by $\lambda_i$ in
\cite{PvNT}.
\section{The Killing spinor equation}
\label{chilluspina} Let us now come to the central issue of the present paper that is the discussion of
preserved supersymmetries in the background of Arnold-Beltrami flux brane solutions. We start by writing the
Killing spinor equations in general terms.
\par
According to a well-established procedure, given a classical bosonic solution of supergravity, where the
fermion fields are set to zero, one considers the supersymmetry variation of the fermions in such a background
and imposes their vanishing. This yields a set of algebraic and first-order differential constraints on the
supersymmetry parameters $\epsilon_A$. By definition, the number of independent solutions to such equations is
the number of preserved supersymmetries and each solution is named \textit{Killing spinor}.
\par
 The supersymmetry variations of the gravitinos and of the spin one-half fermions (dilatinos) are
determined from the rheonomic parameterizations of the fermionic curvatures (\ref{rhopara}),
(\ref{gravitelpara}) using the definitions (\ref{gravitinM},\ref{gravitinN}) and (\ref{gravitelP}) and the
final values of the coefficients displayed in eq. (\ref{calesse}). In this way, for any supergravity bosonic
background, we obtain the following Killing spinor equations :
\begin{eqnarray}
0\, = \,\delta \psi_A&\equiv&\mathcal{D }\epsilon_A\, - \,\underbrace{e^{\phi}\, \left(  \frac{1}{40} \Gamma_a
\, \Gslat +\frac{1}{8}\, \Gslat \, \Gamma_a \right)\, V^a\,}_{\mathbf{S}}\,\delta_A^B \, \epsilon_B -
\,\underbrace{\mathrm{i}\,e^{\frac{\phi}{2}}\,\left( \frac{1}{4} \Fslat^{x} \,
\Gamma_a  - \frac{3}{20} \Gamma_a \,\Fslat^{x}  \right)\, V^a}_{\mathbf{\Omega^x}}\,\sigma_{A}^{x|B} \epsilon_B\nonumber\\
0 \, = \, \delta \chi_A&\equiv&\underbrace{\left(\frac{e^\phi}{5}\,\Gslat\,+ \, \frac{1}{2} \Phislat
\right)}_{\mathfrak{S}}\delta_A^B \,\epsilon_B-\underbrace{\frac{i\,e^{\frac{\phi}{2}}}{5}\,
\Fslat^{x}}_{\mathfrak{O}^x}\,\sigma_{A}^{x|B} \epsilon_B\,. \label{KSeque}
\end{eqnarray}
where:
\begin{equation}\label{Lorenzocovario}
    \mathcal{D }\epsilon_A\, \, \equiv \, \mathrm{d}\epsilon_A\, \, - \, \ft 14 \omega^{ab} \,\Gamma_{ab} \,
    \epsilon_A
\end{equation}
is the Lorentz covariant derivative ($\omega^{ab}$ being the spin connection) and where the operators
$\Gslat$, $\Phislat$ and $\Fslat^x$ have been defined in eq. (\ref{slatti}).
\par
In order to discuss the Killing equation in a general form it is convenient to adopt a Kronecker product
notation and put the candidate Killing spinors (\ref{splittiepsi}) into a 16-component row vector as it
follows:
\begin{equation}
\label{filangetto}
    \varepsilon \, \equiv \,
    \left(  \begin{array}{c}
\epsilon_1 \\
\hline \epsilon_2
\end{array}\right)
\end{equation}
and rewrite the two equations (\ref{KSeque}) in the following way:
\begin{eqnarray}
  0 &=& \nabla \, \varepsilon  \, \equiv \, d\varepsilon \, + \, \Theta \, \varepsilon \label{consequo1}\\
  0 &=& \mathfrak{P} \varepsilon \label{consequo2}
\end{eqnarray}
where the generalized connection $\Theta$ is a one-form valued $16\times 16$ matrix with the following
structure:
\begin{eqnarray}\label{thettona}
   \Theta & = & \left( \begin{array}{c|c}
                         \mathbf{\Sigma} +\mathbf{\Omega}^3 & \mathbf{\Omega}^1 \, + \, {\rm i} \,
                         \mathbf{\Omega}^2 \\
                         \hline
                         \mathbf{\Omega}^1 \, - \, {\rm i} \, \mathbf{\Omega}^2 & \mathbf{\Sigma}
                         \, - \, \mathbf{\Omega}^3
                       \end{array}
   \right)\nonumber\\
   \mathbf{\Sigma} & \equiv & - \,\ft 14 \omega^{ab} \,\Gamma_{ab} \, - \,\mathbf{S}
\end{eqnarray}
in terms of the previously introduced operators, while the $16 \times 16$ matrix $\mathfrak{P}$ is defined as
follows:
\begin{equation}\label{confienza}
    \mathfrak{P} \, \equiv \,\left( \begin{array}{c|c}
                         \mathfrak{S}\, +\,\mathfrak{O}^3 & \mathfrak{O}^1 \, + \, {\rm i} \, \mathfrak{O}^2 \\
                         \hline
                         \mathfrak{O}^1 \, - \, {\rm i} \, \mathfrak{O}^2 & \mathfrak{S}\, -\, \mathfrak{O}^3
                       \end{array}
   \right)
\end{equation}
Having rewritten the Killing spinor equations in the more abstract although much more transparent form
(\ref{consequo1}-\ref{consequo2}), the discussion of their solubility becomes much simpler. The first order
differential equation (\ref{consequo1}) has an integrability condition that reads as follows:
\begin{equation}\label{nulloide}
    \mathfrak{R}\, \varepsilon \, = \,0
\end{equation}
where $\mathfrak{R}\left[\Theta\right]$ denotes the $2$-form curvature of the generalized
connection (\ref{thettona}), namely:
\begin{equation}\label{curboide}
    \mathfrak{R} \, = \, d \Theta \, + \, \Theta \wedge \Theta
\end{equation}
Hence the necessary condition for the existence of Killing spinors is that both matrices
$\mathfrak{R}\left[\Theta\right]$ and $\mathfrak{P}$ should have rank smaller than $16$ in order to admit a
non-trivial Null-Space. Indeed the maximal possible number of Killing spinors is given by:
\begin{equation}\label{nullispizi}
    \# \,\mbox{of Killing spinors} \, \le \, \mbox{dim} \, \left[ \mbox{Null-Space}\left(\mathfrak{R}\right) \bigcap
    \mbox{Null-Space}\left(\mathfrak{P}\right)\right]
\end{equation}
In eq.(\ref{nullispizi}) the sign $\le$ is due to the fact that eq. (\ref{nulloide}) is a necessary but in
general not a sufficient condition. Once the candidate Killing spinor has been restricted to the space
$\mbox{Null-Space}\left(\mathfrak{R}\right) \bigcap
    \mbox{Null-Space}\left(\mathfrak{P}\right)$, the differential equation (\ref{consequo1}) has to be explicitly
integrated and, previous experience with this type of problem, suggests that new obstructions might arise. On
the contrary if the rank of $\mathfrak{R}$ is $16$ we can safely conclude that all supersymmetries are broken
by the considered background.
\par
Having anticipated this general discussion we consider the case of brane-solutions utilizing the split basis
of gamma matrices introduced in section \ref{splittorio}.
\par
We adopt the index convention (\ref{splittaggio}) and we summarize the flux-brane solution as follows:
\begin{eqnarray}
  \phi &=& - \, \frac{2}{5} \, \log \, \left [ H \right] \quad ; \quad H \, = \, H(y)\label{dilatoneB}\\
  V^{\underline{a}} &=& H^{-\ft 15} \, d\xi^{\underline{a}} \label{vielbeinIn}\\
  V^{P} &=& H^{\ft {3}{10}} \, dy^{P} \quad ; \quad y^P \, \equiv \, \left\{U,X,Y,Z \right\}\label{vielbeinOut}\\
  \mathbf{B}^{[3]} &=& \frac{1}{12} \, H^{-1} \, \frac{1}{3!} \, \epsilon_{\underline{abc}}d\xi^{\underline{a}}
  \wedge d\xi^{\underline{b}}\wedge d\xi^{\underline{c}}\label{Bforma}\\
  \mathcal{A}^\Lambda  &=& \frac{1}{2\sqrt{2}}\, \omega \, \lambda\, \exp[2\pi \,  \mu \, U] \, \mathbf{W}^\Lambda
  \label{arnoldini}
\end{eqnarray}
where the inhomogeneous harmonic function $H(U,X,Y,Z)$ satisfies eq. (\ref{sicurmorio}). Another essential
ingredient that we need is the spin-connection. For this latter we easily find:
\begin{eqnarray}
  \omega^{\underline{ab}} &=& 0 \\
  \omega^{\underline{a}}_P &=& - \ft 1 5 \,\mathrm{d}\xi^{\underline{a}} \,H^{-\ft 32} \partial_P H\\
  \omega^{P~}_{~Q} &=& \ft 3{10}\,H^{-1}\, \left(dy^P \, \partial_Q H \, - \, dy_Q \, \partial^P H\right )
\end{eqnarray}
Next let us analyze the structure of the algebraic matrix operators entering the definition of the projector
$\mathfrak{P}$ and of the connection $\Theta$.
 Let us begin with the structure of the operator $\mathfrak{S}$. We find:
\begin{eqnarray}\label{Sigmus}
  \mathfrak{S} & = & \mathbf{1}_{2\times 2}  \otimes \hat{\mathfrak{S}} \nonumber\\
  \hat{\mathfrak{S}} & = &  - \,\ft 2 5 \,\left(
\begin{array}{cccc}
 0 & 0 & \frac{\partial _1H+\mathrm{i} \partial _4H}{H^{13/10}} & \mathrm{i}\,\frac{ \left(\partial
   _2H+\mathrm{i} \partial _3H\right)}{H^{13/10}} \\
 0 & 0 & \mathrm{i}\, \frac{ \left(\partial
   _2H-i \partial _3H\right)}{H^{13/10}} & \frac{\partial _1H-\mathrm{i} \partial
   _4H}{H^{13/10}} \\
 0 & 0 & 0 & 0 \\
 0 & 0 & 0 & 0 \\
\end{array}
\right)
\end{eqnarray}
On the other hand the operators $\mathbf{\Omega}^x$ have the following structure:
\begin{eqnarray}
\mathbf{\Omega}^x &=& \mathbf{1}_{2\times 2}  \otimes \hat{\mathbf{\Omega}}^x \\
  \hat{\mathbf{\Omega}}^x &=& \lambda \, \left(\begin{array}{cccc}
 \mathfrak{A}^{x} & \mathfrak{B}^{x} & 0 & 0 \\
 \mathfrak{C}^{x} & \mathfrak{D}^{x} & 0 & 0 \\
 0 & 0 & 0 & 0 \\
 0 & 0 & 0 & 0 \\
\end{array}
\right)
\end{eqnarray}
the parameter $\lambda$ corresponding to that in front of Beltrami vector fields (see eq.(\ref{arnoldini})), so that
$\lambda = 0$ means pure branes without fluxes, and  the specific form of the submatrices
\begin{equation}
\mathfrak{M}^x \, = \,\left(\begin{array}{cc}
 \mathfrak{A}^{x} & \mathfrak{B}^{x}  \\
 \mathfrak{C}^{x} & \mathfrak{D}^{x} \\
\end{array}\right)
\end{equation}
depends on the specific form of the chosen Beltrami field.
\par
These informations are sufficient to conclude that the rank of the $16\times 16$ matrix $\mathfrak{P}$ is
always $8$ both in presence and in absence of fluxes, namely both with $\lambda \ne 0$ and with $\lambda \, =
\, 0$.
\subsection{The supersymmetry of pure $2$-branes}
If we do not introduce Arnold-Beltrami fluxes we have $2$-brane solutions of the form
(\ref{dilatoneB}-\ref{Bforma}), where $H$ is a harmonic function on $\mathbb{R}_+\otimes \mathrm{T^3}$ and
$\lambda \, = \,0$. In that case the Null-Space of $\mathfrak{P}$ is simply given by those $\epsilon_{1,2}$ in
eq.(\ref{splittiepsi}) where all the $\theta_i$ are set to zero. Next we can verify that
\begin{equation}\label{nulluspius}
    \mbox{Null-Space}(\mathfrak{P}) \, \subset \mbox{Null-Space}(\mathfrak{R})
\end{equation}
This suggests that there might be $8$ Killing spinors. Indeed making the following replacement in
eq.(\ref{splittiepsi}):
\begin{equation}\label{kulluspurus}
    \theta_i \, = \, 0 \quad ; \quad \xi_i \, = \,  H(y)^{\ft 1{10}} \, \chi_i \quad \quad (i=,1\dots,8)
\end{equation}
where $\chi_i$ are constant anticommuting spinors we can easily verify that the corresponding $\varepsilon$
defined in (\ref{filangetto}) satisfies both eq.s (\ref{consequo1}) and (\ref{consequo2}) for any choice of
the harmonic function $H$. Therefore we come to the conclusion that the pure $2$-branes described above
preserve $8$ supersymmetry charges, namely they are BPS states breaking $\ft 12$ of the supersymmetry charges
and preserving the other half.
\subsection{The supersymmetry of flux $2$-branes}
When we turn on  Arnold Beltrami Fluxes, things become much more complicated since the curvature matrix
$\mathfrak{R}$ has no longer a universal form and its structure critically depends on the choice of the vector
field triplet $\mathbf{W}$. A priori it is by no means clear whether flux-branes preserving any supersymmetry
can exist or any of them necessarily breaks all the supersymmetries. In order to decide this crucial point we
have considered many explicit solutions, in particular those already presented in \cite{arnoldtwobranes}. By
means of a specially developed code we have constructed the corresponding 2-form $\mathfrak{R}$
and then, since its form is in all cases too much involved for any analytical study we have resorted to
numerical calculations. An algorithm based on random number generation probes the rank of all the $16 \times
16$--matrices $\mathfrak{R}^{I}_{J|ab}$ obtained by expanding the curvature of the generalized spinor
connection $\mathfrak{R}^I_J$ along the vielbein:
\begin{equation}\label{curbalatti}
    \mathfrak{R}^I_J\, = \,\mathfrak{R}^{I}_{J|ab} \, V^a \wedge V^b
\end{equation}
Since we are in $7$-dimensions, for each randomly chosen point in $\mathbb{R}_+\times \mathrm{T^3}$ we obtain
a set of 21 matrices and the maximum rank displayed by this set is the rank of the curvature $2$-form. If this
rank is $16$ we conclude that there cannot be any Killing spinors and that supersymmetry is completely broken.
On the other hand, if the maximal rank is  less than $16$ for all the $21$ matrices mentioned in eq.
(\ref{curbalatti}) in a conveniently ample set of random points, this is a strong indication that the
curvature has a non vanishing Null-Space and one can attempt to calculate its form analytically. The result of
this numerical investigation was the following. All the models considered in \cite{arnoldtwobranes} and
several others that we have tested break supersymmetry entirely, leading to the conclusion that it is
generically  very hard and unlikely to hit a case where Killing vectors do exist. Actually we were strongly
tempted to assume that flux-brane break all supersymmetries always. Yet, by means of several trials and by
some educated guess, we were able to produce  counterexamples of an Arnold-Beltrami flux--brane which
respectively preserves $\ft 14$ and $\ft 18$ of the original supersymmetry. As we emphasize below the presence
of Killing spinors is entangled with the presence of  additional translational Killing vectors that are
instead absent in  generic flux-branes.
\par
Because of the relation between the Arnold-Beltrami flux-branes and the hydrodynamical models
\cite{arnoldus,ArnoldBook,Henon,contactgeometria,ABCFLOW} where the same three-dimensional vector fields are
used as flows (\textit{i.e.} velocity fields of a fluid) it is interesting to stress what follows.
\par
According to Arnold Theorem \cite{arnoldus,ArnoldBook} that of satisfying Beltrami equation is a necessary yet
not sufficient condition for a stationary flow to admit chaotic stream-lines. In particular if there are
additional continuous symmetries of the vector field, this introduces extra conserved charges that can lead to
integrability and bar the existence of any chaos. Furthermore if the integral curves of the vector field are
all planar, this also inhibits chaotic behavior on very general grounds. The so named $ABC$-flows
\cite{ABCFLOW} obtained from a particular truncation of the general solution of Beltrami equation with the
lowest eigenvalue $\mu \, = \, 1$ were extensively studied in the literature on mathematical hydrodynamics
since they have interesting and helpful discrete symmetries but no continuous ones.
\par
From our analysis of the Killing spinor equation it emerges that in order to have Killing spinors the flux
2-brane has to have some additional translational Killing vectors on the  torus $\mathrm{T^3}$. In particular
with two translational Killing vectors we obtain a flux $2$-brane that preserves $\ft 1 4$ of the
supersymmetry, with one additional Killing vector we obtain a flux $2$-brane that preserves $\ft 1 8$ of the
supersymmetry, while the request of three translational Killing vectors suppresses all the fluxes and
preserves  $\ft 12$ of the original supersymmetry (the maximal value for BPS states).
\par
Since the anticommutator of spinor charges produces translations, it is rather natural that the existence of
Killing spinors implies additional Killing vectors, besides those associated with the conformally flat
brane-world-sheet. From the point of view of the correspondence between supergravity flux $2$-branes and
hydro-models it is relevant that supersymmetry excludes chaotic stream-lines and vice-versa.
\par
Furthermore it is very much interesting to analyze the $2$-brane solutions from the point of view of
discrete/continuous symmetries. With just a discrete group of symmetries $\Gamma$ we break all
supersymmetries. When we preserve some supersymmetry, in addition to $\mathrm{U(1)}$ or $\mathrm{U(1)^2}$
(respectively corresponding to the $\ft 18$ and $\ft 14$ case), we have some residual discrete symmetry
$\Gamma$ that it is quite relevant to single out. Indeed $\Gamma$ is transmitted to the gauge theory on the
brane world-volume and the composite operators in the gauge/gravity correspondence have to be organized into
irreducible representations of such a $\Gamma$.
\par
In the next section we present a few examples of flux $2$-branes with and without supersymmetry where all such
symmetries are carefully analysed.
\section{Examples of flux $2$-branes  and their (super)-symmetries}
\label{unesempiopertutti} In this section we present just three explicit examples of Arnold-Beltrami flux
$2$-branes, one with no preserved supersymmetry, one with $\ft 14$, the last with $\ft 12$. We advocate the
relation of preserved supersymmetry with the presence of extra translational Killing vectors and we carefully
analyze the discrete symmetries of each of the considered branes.
\subsection{The Arnold-Beltrami flux $2$-brane with octahedral symmetry and no preserved supersymmetry}
In \cite{arnoldtwobranes} it was  presented the case of the $2$-brane solution where the triplet of
Arnold-Beltrami fields spans an irreducible tri-dimensional representation of a rather large discrete group,
namely the irreducible representation  $D_{12}$ of the group $\mathrm{GF_{192}}$ described both in \cite{arnolderie} and
\cite{arnoldtwobranes}. In the present section we reconsider that solution from a different standpoint and we
decode its symmetries in a more explicit way, moreover showing that it breaks all supersymmetries.
\par
The triplet of vector fields  that we want to consider is the following one:
\begin{equation}
\mathfrak{W}\left(\mathbf{X}\right)\, = \,\left\{\begin{array}{rcl}
  \mathbf{W}^1 &=& 2 \,\mathrm{d}X \,\cos (2 \pi  Z)-2\, \mathrm{d}Y \,\sin (2 \pi  Z)\\
  \mathbf{W}^2 &=& 2 \,\mathrm{d}X \,\cos (2 \pi  Y)+2\, \mathrm{d}Z \,\sin (2 \pi  Y) \\
  \mathbf{W}^3 &=& 2 \,\mathrm{d}Y \,\cos (2 \pi  X)-2 \,\mathrm{d}Z\, \sin (2 \pi  X)
  \end{array}\right.
  \label{D12WGF192}
\end{equation}
Any linear combination of these vector fields forms the celebrated $ABC$-flow of Hydrodynamics \cite{ABCFLOW}.
\par
Since the components of the vector field depend on all the three coordinates $X,Y,Z$ we have no continuous
translation symmetry on the three-torus and there are no further translational Killing vectors besides those
corresponding to the conformally flat directions of the $2$-brane world-volume:
\begin{equation}\label{conegrina}
    \xi^{\underline{a}} \, \rightarrow \, \xi^{\underline{a}} \, + \, c^{\underline{a}}
\end{equation}
There is however a residual global isometry forming a $\mathbb{Z}_2\times \mathbb{Z}_2 $ group. The reader can
easily verify that the following three substitutions leave each of the three  one-forms in eq.
(\ref{D12WGF192}) invariant:
\begin{equation}
\begin{array}{rcl}
  \mathcal{T}_1 &:& \left\{X\to -X-\frac{1}{2},\quad Y\to -Y-\frac{1}{2},\quad Z\to Z+\frac{1}{2}\right\} \\
  \mathcal{T}_2 &:& \left\{X\to -X,\quad Y\to Y+\frac{1}{2},\quad Z\to -Z-\frac{1}{2}\right\} \\
  \mathcal{T}_3 &:& \left\{X\to X+\frac{1}{2},\quad Y\to -Y,\quad Z\to -Z\right\}
  \end{array}\label{foschino}
\end{equation}
Each of the above translations squares to the identity, since it corresponds to some integral shift of the
coordinates $X,Y,Z$ which, on the $\mathrm{T^3}$ torus means no shift. In addition to these translational
symmetries, the supergravity solution generated by the vector field system (\ref{D12WGF192}) has a very
interesting symmetry:
\begin{equation}\label{fortebasso}
    \Gamma \, = \, \mathrm{O_{24}} \otimes \mathbb{Z}_2
\end{equation}
The octahedral group $\mathrm{O_{24}}$, which is isomorphic to the symmetric group $\mathrm{S_4}$, is one of
the exceptional finite subgroups of $\mathrm{SO(3)}$. Abstractly it can be described by two generators and
three relations:
\begin{equation}\label{corgu0}
    \mathrm{O_{24}} \quad = \quad \left( \mathrm{T},\mathrm{S} \,| \,\mathrm{T^3} \, = \, \mathbf{1} \, ,
    \,\mathrm{S^2} \, = \, \mathbf{1} \, , \,\left(\mathrm{S \, T}\right)^4 \, = \, \mathbf{1}
    \, \right)
\end{equation}
 An explicit representation by means of orthogonal integer valued $3\times 3$ matrices with unit
determinant is the following one:
\begin{equation}\label{ottageneri}
  \mathrm{D[T]}  \,=\, \left(
\begin{array}{ccc}
 0 & 0 & 1 \\
 -1 & 0 & 0 \\
 0 & -1 & 0 \\
\end{array}
\right) \quad ; \quad \mathrm{D[S]}  \,=\,\left(
\begin{array}{ccc}
 -1 & 0 & 0 \\
 0 & 0 & -1 \\
 0 & -1 & 0 \\
\end{array}
\right)
\end{equation}
The map $\mathrm{D}$ realizes an immersion of the octahedral group into the group $\mathrm{SO(3)}$:
\begin{equation}\label{isomorto}
   \mathrm{ D}\, : \, \mathrm{O_{24}} \, \hookrightarrow \,\mathrm{SO(3)}
\end{equation}
If we add the matrix:
\begin{equation}\label{Zgener}
    \mathrm{D[Z]} \, = \, \left(
\begin{array}{ccc}
 -1 & 0 & 0 \\
 0 & -1 & 0 \\
 0 & 0 & -1 \\
\end{array}
\right)
\end{equation}
which has determinant $-1$ and commutes with both $\mathrm{D[T]}$ and $\mathrm{D[S]}$:
\begin{equation}\label{fannilobio}
   \left[ \mathrm{D[T]}\, ,\, \mathrm{D[Z]}\right]\,=\,\left[ \mathrm{D[S]}\, ,\, \mathrm{D[Z]}\right]\, = \,0
\end{equation}
we realize a homomorphic embedding:
\begin{equation}\label{circasso}
    \mathrm{D} \,: \, \mathrm{O_{24}} \times \mathbb{Z}_{2} \, \hookrightarrow \, \mathrm{O(3)}
\end{equation}
The claimed symmetry of the supergravity $2$-brane solution under the group (\ref{fortebasso}) stems from the
following identities that the reader can easily verify:
\begin{eqnarray}
  \mathfrak{W}\left(\mathfrak{T} \mathbf{X}\right) &=& \mathrm{D[T]}\cdot\mathfrak{W}\left(\mathbf{X}\right) \nonumber \\
  \mathfrak{W}\left(\mathfrak{S} \mathbf{X}\right) &=& \mathrm{D[S]}\cdot\mathfrak{W}\left(\mathbf{X}\right) \nonumber\\
  \mathfrak{W}\left(\mathfrak{Z} \mathbf{X}\right) &=&
  \mathrm{D[Z]}\cdot\mathfrak{W}\left(\mathbf{X}\right)\label{lamucello}
\end{eqnarray}
where the action of the three generators on the torus coordinates is defined below:
\begin{eqnarray}
  \mathfrak{T} \mathbf{X} &=& \left\{\frac{3}{4}-Y,Z+\frac{1}{4},-X-\frac{1}{2}\right\} \nonumber\\
  \mathfrak{S} \mathbf{X} &=& \left\{X+\frac{1}{2},Y+\frac{1}{2},Z+\frac{1}{2}\right\} \nonumber\\
  \mathfrak{Z} \mathbf{X} &=& \left\{\frac{1}{2}-Y,X,Z-\frac{3}{4}\right\} \label{sigul3}
\end{eqnarray}
It is important to stress that the three transformations (\ref{sigul3}) are defined modulo any additional
transformation of the $\mathbb{Z}_2\times \mathbb{Z}_2 $ group generated by the translations (\ref{foschino})
which leave the vector fields (\ref{D12WGF192}) invariant. From a group theoretical point of view the group
$\mathrm{GF_{192}}$ mentioned in \cite{arnoldtwobranes} and \cite{arnolderie} is the semidirect product:
\begin{equation}\label{squirta}
    \mathrm{GF_{192}} \, \sim \, \Gamma \, \ltimes \, \left(\mathbb{Z}_2\times \mathbb{Z}_2\right)
\end{equation}
 both $\Gamma$ and $\left(\mathbb{Z}_2\times \mathbb{Z}_2\right)$ being invariant subgroups. We can look at
the map $\mathrm{D}$ as a homomorphical embedding:
\begin{eqnarray}\label{consapevole}
    D & : & \mathrm{GF_{192}} \,\hookrightarrow \,\mathrm{O(3)} \nonumber\\
\mbox{ker}[D] &\sim & \mathbb{Z}_2\times \mathbb{Z}_2
\end{eqnarray}
the kernel of the homomorphism being the normal subgroup  generated by the translations (\ref{foschino}). This
way of thinking shows that the supergravity flux $2$-brane solution generated by the triplet of Beltrami
fields (\ref{D12WGF192}) has the large discrete symmetry $\mathrm{GF_{192}}$. Indeed it suffices to utilize
the global $\mathrm{O(3)}$ symmetry of supergravity and we can set:
\begin{equation}\label{cascuso}
    \forall \gamma \, \in \,\mathrm{GF_{192}} \quad : \quad \mathfrak{W}\left(\mathbf{X}\right)^\prime
    \, \equiv \, \mathrm{D[\gamma]^{-1}} \, \mathfrak{W}\left(\gamma \mathbf{X}\right) \, = \,
    \mathfrak{W}\left(\mathbf{X}\right)
\end{equation}
all the other fields, dilaton, metric and $3$-form, being already invariant.
\par
Indeed the inhomogeneous harmonic function produced by the choice (\ref{D12WGF192}) is the
following one:
\begin{equation}\label{harmonideD12}
   H(y) \, = \,  1-\frac{1}{8} \lambda ^2 e^{4 \pi  U}
\end{equation}
and all the other bosonic fields follow from eq.s (\ref{dilatoneB}-\ref{arnoldini}).
\par
Localized on this solution the projector $\mathfrak{P}$ has still rank $8$. The difference with the pure brane
case is just the following. In the eight null-vectors of $\mathfrak{P}$, the parameters $\theta_i$, instead of
being put to zero,  are forced to be point-dependent linear combinations of the $\xi_i$. Hence the dilatino
supersymmetry transformation rule can be nullified by eight independent spinors also in this case. However,
the situation is dramatically different at the level of the gravitino transformation rule. As our
computer code demonstrates, in any randomly chosen point, the rank of the curvature $\mathfrak{R}$ is always
$16$ which bars the existence of any Killing spinors. This brane solution has a large discrete symmetry but
breaks all supersymmetries.
\par
\begin{figure}[!hbt]
\begin{center}
\iffigs
\includegraphics[height=55mm]{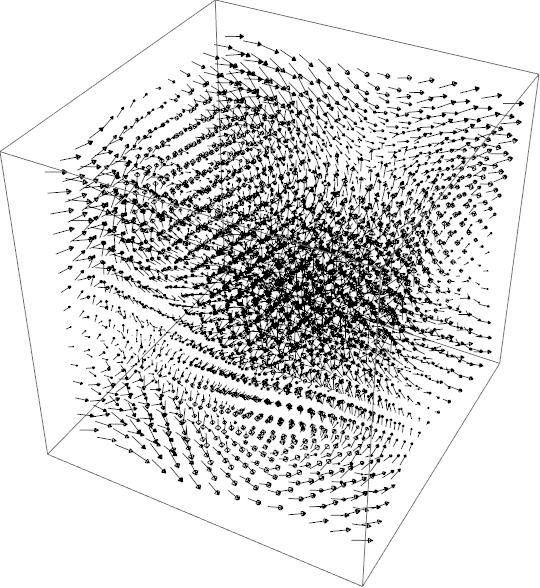}
\includegraphics[height=50mm]{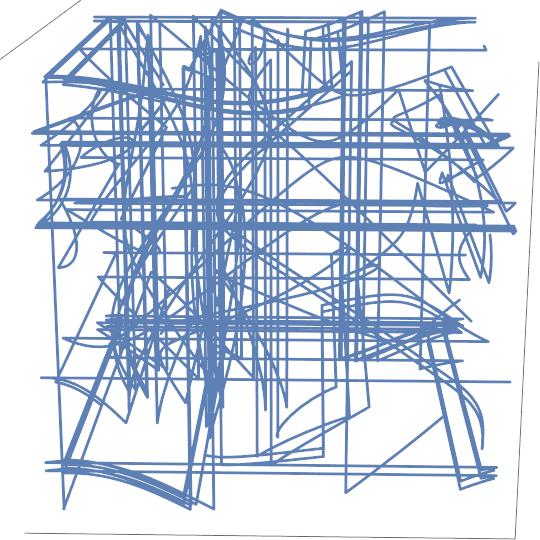}
\else
\end{center}
 \fi
\caption{\it On the left the plot of one arbitrarily chosen vector field in the Beltrami space defined by
eq.(\ref{D12WGF192}). On the right the plots of some of its integral curves in the 3-torus represented as a
cube with identified opposite faces. \label{fogliarus1}}
 \iffigs
 \hskip 1cm \unitlength=1.1mm
 \end{center}
  \fi
\end{figure}
In order to get a visual appreciation of the difference between Beltrami fields that lead to
non-supersymmetric and to supersymmetric $2$-branes we have produced some plots. In figure  \ref{fogliarus1}
you see the plot of an arbitrarily chosen vector field in the three-dimensional vector space spanned by
(\ref{D12WGF192}). On the right side a plot of some of its streamlines, namely of its integral curves, is shown.
\subsection{The  Arnold-Beltrami flux $2$-branes with bosonic symmetry
$\mathcal{D}_n \ltimes \left[\mathrm{U(1)}\times \mathrm{U(1)}\right]$ and 4 Killing spinors}
The next example we consider is a flux $2$-brane that preserves $1/4$ of the original supersymmetry, namely
possesses  $4$ Killing spinors. As discussed  above on general grounds we aspect in this case two
translational Killing vectors. This means that eq. (\ref{squirta}) defining the complete bosonic group of the
previously considered solution is replaced by:
\begin{equation}\label{squirta4}
    \mathrm{G}_{bosonic} \, \sim \, \Gamma \, \ltimes \, \left[\mathrm{U(1)}\times \mathrm{U(1)}\right]
\end{equation}
the two $\mathrm{U(1)}$'s being the continuous translation groups generated by the two additional Killing
vectors. The question remains: what is the discrete group $\Gamma$ in this case? We show that using a cubic
momentum lattice the answer is:
\begin{equation}\label{dihedrino4}
   \Gamma \, = \, \mathrm{\mathcal{D}_4}
\end{equation}
where $\mathcal{D}_4$ denotes a dihedral group. There is also a second solution based on the hexagonal lattice
which yields:
\begin{equation}\label{dihedrino6}
    \Gamma \, = \, \mathrm{\mathcal{D}_6}
\end{equation}
To see this let us consider the two cases together:
\begin{eqnarray}
\mathfrak{W}_{[4]}\left(\mathbf{X}\right)& = &\left\{\begin{array}{rcl}
  \mathbf{W}^1 &=& \mathrm{d}X \,\cos (2 \pi  Z)-\mathrm{d}Y \, \sin (2 \pi  Z)\\
  \mathbf{W}^2 &=& -\,\mathrm{d}Y \,\cos (2 \pi  Z)\,-\, \mathrm{d}X \,\sin (2 \pi  Z) \\
  \mathbf{W}^3 &=& 0
  \end{array}\right.
  \label{D4N4}\\
  \mathfrak{W}_{[6]}\left(\mathbf{X}\right)& = &\left\{\begin{array}{rcl}
  \mathbf{W}^1 &=& \mathrm{d}X \cos \left(\frac{4 \pi  Z}{\sqrt{3}}\right)-\mathrm{d}Y \sin \left(\frac{4
   \pi  Z}{\sqrt{3}}\right)\\
  \mathbf{W}^2 &=& -\,\mathrm{d}Y \,\cos \left(\frac{4 \pi  Z}{\sqrt{3}}\right)-\mathrm{d}X \sin
   \left(\frac{4 \pi  Z}{\sqrt{3}}\right) \\
  \mathbf{W}^3 &=& 0
  \end{array}\right.
  \label{D6N4}
\end{eqnarray}
Abstractly the dihedral group $\mathrm{\mathcal{D}_n}$ can be described by two generators and three relations:
\begin{equation}\label{corgu}
    \mathrm{\mathcal{D}_n} \quad = \quad \left( \mathrm{A},\mathrm{B} \,| \,
    \mathrm{A}^{n} \, = \, \mathbf{1} \, ,
    \,\mathrm{B^2} \, = \, \mathbf{1} \, , \,\left(\mathrm{B \, A}\right)^2 \, = \, \mathbf{1}
    \, \right)
\end{equation}
An explicit representation by means of orthogonal integer valued $3\times 3$ matrices with unit determinant is the following one :
\begin{equation}\label{ottageneri2}
\begin{array}{ccrclcrcl}
 \mathrm{\mathcal{D}_4} &:& \mathrm{D[A_4]}  &=& \left(
\begin{array}{ccc}
 0 & 1 & 0 \\
 -1 & 0 & 0 \\
 0 & 0 & 1 \\
\end{array}
\right) & ; & \mathrm{D[B]}  &=&\left(
\begin{array}{ccc}
 -1 & 0 & 0 \\
 0 & 1 & 0 \\
 0 & 0 & -1 \\
\end{array}
\right)\\
 \mathrm{\mathcal{D}_6} &:& \mathrm{D[A_6]}  &=& \left(
\begin{array}{ccc}
 \frac{1}{2} & \frac{\sqrt{3}}{2} & 0 \\
 -\frac{\sqrt{3}}{2} & \frac{1}{2} & 0 \\
 0 & 0 & 1 \\
\end{array}
\right) & ; & \mathrm{D[B]}  &=&\left(
\begin{array}{ccc}
 -1 & 0 & 0 \\
 0 & 1 & 0 \\
 0 & 0 & -1 \\
\end{array}
\right)
\end{array}
\end{equation}
The map $\mathrm{D}$ realizes an immersion of the two dihedral groups into the group $\mathrm{SO(3)}$:
\begin{equation}\label{isomorto2}
   \mathrm{ D}\, : \, \mathrm{\mathcal{D}_{4,6}} \, \hookrightarrow \,\mathrm{SO(3)}
\end{equation}
The claimed symmetry of the supergravity $2$-brane solution under the group (\ref{fortebasso}) stems from the
following identities :
\begin{eqnarray}
  \mathfrak{W}_{[4,6]}\left(\mathrm{D[A_{4,6}^{-1}]}\cdot \mathbf{X}\right) &=&
  \mathrm{D[A_{4,6}]}\cdot\mathfrak{W}_{[4,6]}\left(\mathbf{X}\right) \nonumber \\
  \mathfrak{W}_{[4,6]}\left(\mathrm{D[B^{-1}]}\cdot\mathbf{X}\right) &=&
  \mathrm{D[B]}\cdot\mathfrak{W}_{[4,6]}\left(\mathbf{X}\right)
   \label{lamuscote}
\end{eqnarray}
where the action of the two generators on the torus coordinate is given, this time, by standard matrix
multiplication.  Hence, just as in the previous case, the complete semidirect product group:
\begin{equation}\label{cocococ}
    G_{bosonic} \, = \, \mathrm{\mathcal{D}_{4,6}}\ltimes\left(\mathrm{U_X(1)}\times \mathrm{U_Y(1)}\right)
\end{equation}
is an isometry group for the supergravity solution since the matrices $\mathrm{D[A]}$ and $\mathrm{D[B]}$ are
orthogonal and $\mathrm{O(3)}$ is a global symmetry of the supergravity lagrangian.
\par
The inhomogeneous harmonic functions for these brane--solutions are the following ones:
\begin{equation}\label{harmoniden4}
\begin{array}{rcl}
    H_4(y) & = & 1-\frac{1}{48} \lambda ^2 e^{4 \pi  U}\\
    H_6(y) & = & 1-\frac{1}{48} \lambda ^2 e^{\frac{8\pi}{\sqrt{3}}  U}
    \end{array}
\end{equation}
and the rest of the solution is obtained from eq.s (\ref{dilatoneB}-\ref{arnoldini}).
\par
 Calculating the $\mathfrak{R}$ curvature associated with
this solution we find that in any point the rank of its $21$ vielbein components is  bounded from above by
$12$. Indeed, with little effort, we find a set of $4$ null vectors which surprisingly are null-vectors also
of the matrix $\mathfrak{P}$. In this four dimensional subspace the Killing spinor equation is easily
integrated by taking all the non vanishing components proportional to $H^{\ft 1{10}}$ where $H$ is the
inhomogeneous harmonic function. Finally we arrive at the following  explicit form of $4$ indipendent Killing
spinors:
\begin{equation}\label{killispinati}
    \epsilon_1 \, = \, H_{4,6}(y)^{\ft {1}{10}} \,\left(
\begin{array}{c}
 0 \\
 \chi _4 \\
 0 \\
 0 \\
 0 \\
 \chi _3 \\
 0 \\
 0 \\
\end{array}
\right) \quad ; \quad \epsilon_2 \, = \, H_{4,6}(y)^{\ft {1}{10}} \, \left(
\begin{array}{c}
 \chi _2 \\
 0 \\
 0 \\
 0 \\
 \chi _1 \\
 0 \\
 0 \\
 0 \\
\end{array}
\right)
\end{equation}
The considered  flux-brane solution preserves $\ft 14$ of the original supersymmetry.
\par
\begin{figure}[!hbt]
\begin{center}
\iffigs
\includegraphics[height=55mm]{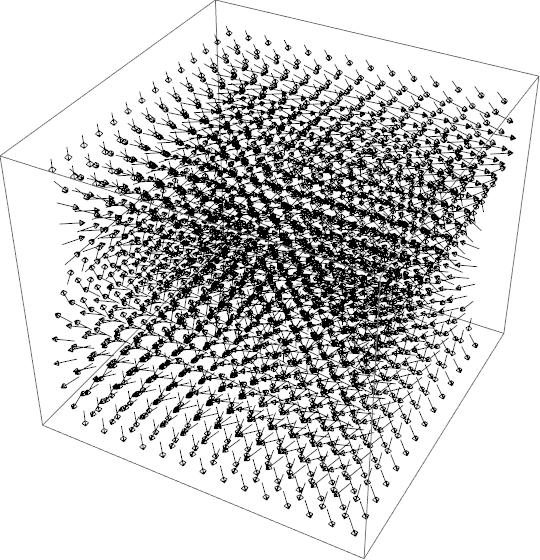}
\includegraphics[height=50mm]{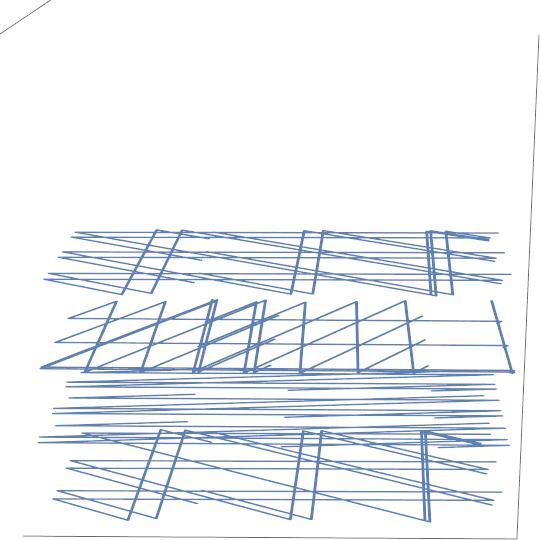}
\else
\end{center}
 \fi
\caption{\it On the left the plot of one arbitrarily chosen vector field in the Beltrami space defined by
eq.(\ref{D4N4}). On the right the plots of some of its integral curves in the 3-torus represented as a cube
with identified opposite faces. It is evident from the picture that all the integral curves are planar. This a
consequence of the two Killing vectors in the $X$ and $Y$ directions.\label{fogliarus4}}
 \iffigs
 \hskip 1cm \unitlength=1.1mm
 \end{center}
  \fi
\end{figure}
In the spirit of comparison with the previous case that breaks all  supersymmetries, in fig. \ref{fogliarus4}
we have displayed the plot of an arbitrary vector field in the two-dimensional vector space defined by
eq. (\ref{D4N4}). The two Killing vectors in the $X$ and $Y$ directions imply that the integral curves are
always planar  for any element of this vector space and this is quite evident from the figure.
\par
A last comment on this solution concerns a question that might arise in relation with the structure of
equations (\ref{D4N4}) and (\ref{D6N4}). One might ask why we should not consider other dihedral groups with
$n\ne 4,6$. Indeed it suffices to write the same formulae with a different  angle namely:
\begin{equation}
\cos\left[\frac {4\pi}{\sqrt{3}}Z\right] \,\rightarrow \, \cos\left[\frac {2\pi}{\sqrt{\mathbf{m}}}Z\right]
 \label{proposta}
\end{equation}
The answer why the replacement (\ref{proposta}) is generically forbidden comes from classical results of
crystallography. The coordinates $X,Y,Z$ are supposed to span a torus $\mathbb{R}^3/\Lambda$ and in the
present case it suffices to consider the planar projection of the lattice $\Lambda$ which produces a
tessellation of the plane. Hence the considered dihedral group must be in the list of the so named
\textit{Wall Paper Point Groups} which is finite. Besides $\mathcal{D}_4$ and $\mathcal{D}_6$ we might still
have $\mathcal{D}_3$ and $\mathcal{D}_2$. We have not explicitly constructed the corresponding supergravity
solutions but it is rather clear that they are bound to be completely analogous.
\subsection{The Arnold-Beltrami flux $2$-brane with
$\left[\mathcal{D}_4\otimes \mathbb{Z}_2\right]\ltimes \mathrm{U(1)}$ bosonic symmetry and 2 Killing spinors}
The next example we consider is a flux $2$-brane that preserves $1/8$ of the original supersymmetry, namely
possesses  $2$ Killing spinors. On general grounds in this case we expect  just one additional Killing vector.
This means that eq.s (\ref{squirta}) and (\ref{squirta4}) defining the complete bosonic groups of the
previously considered solutions should now be replaced by:
\begin{equation}\label{squirta2}
    \mathrm{G}_{bosonic} \, \sim \, \Gamma \, \ltimes \, \mathrm{U(1)}
\end{equation}
the $\mathrm{U(1)}$ factor being the continuous translation group generated by the unique additional Killing
vector. The question is the same as in the previous case: what is the discrete group $\Gamma$ here? We show
that using a cubic momentum lattice the answer is:
\begin{equation}\label{dihedrino2}
   \Gamma \, = \, \mathrm{\mathcal{D}_4} \times \mathbb{Z}_2
\end{equation}
where $\mathcal{D}_4$ denotes once again the dihedral group.  To see this let us consider the following
triplet of Beltrami vector fields:
\begin{eqnarray}
\widehat{\mathfrak{W}}\left(\mathbf{X}\right)& = &\left\{\begin{array}{rcl}
  \mathbf{W}^1 &=& \mathrm{d}X \cos (2 \pi  Z)-\mathrm{d}Y \sin (2 \pi  Z)\\
  \mathbf{W}^2 &=& \mathrm{d}X \cos (2 \pi  Y)+\mathrm{d}X \sin (2 \pi  Z)+\mathrm{d}Z \sin (2 \pi  Y)+\mathrm{d}Y \cos
   (2 \pi  Z) \\
  \mathbf{W}^3 &=& \mathrm{d}X \sin (2 \pi  Y)-\mathrm{d}Z \cos (2 \pi  Y)
  \end{array}\right.\nonumber\\
  \label{D4N2}
\end{eqnarray}
Abstractly the dihedral group $\mathrm{\mathcal{D}_n}$ is described in eq. (\ref{corgu}). In this case,
relevant to us  is the following representation by means of orthogonal integer valued $3\times 3$ matrices
with unit determinant:
\begin{equation}\label{ottageneri3}
\begin{array}{ccrclcrcl}
 \mathrm{\mathcal{D}_4} &:& \mathrm{D[A]}  &=& \left(
\begin{array}{ccc}
 0 & 0 & -1 \\
 0 & 1 & 0 \\
 1 & 0 & 0 \\
\end{array}
\right) & ; & \mathrm{D[B]}  &=&\left(
\begin{array}{ccc}
 -1 & 0 & 0 \\
 0 & -1 & 0 \\
 0 & 0 & 1 \\
\end{array}
\right)
\end{array}
\end{equation}
The map $\mathrm{D}$ realizes an immersion of the dihedral group $\mathcal{D}_4$ into the group
$\mathrm{SO(3)}$:
\begin{equation}\label{isomorto3}
   \mathrm{ D}\, : \, \mathrm{\mathcal{D}_{4}} \, \hookrightarrow \,\mathrm{SO(3)}
\end{equation}
If we add the matrix:
\begin{equation}\label{Zgener2}
    \mathrm{D[Z]} \, = \, \left(
\begin{array}{ccc}
 -1 & 0 & 0 \\
 0 & -1 & 0 \\
 0 & 0 & -1 \\
\end{array}
\right)
\end{equation}
which has determinant $-1$ and commutes with both $\mathrm{D[A]}$ and $\mathrm{D[B]}$:
\begin{equation}\label{fannilobio2}
   \left[ \mathrm{D[A]}\, ,\, \mathrm{D[Z]}\right]\,=\,\left[ \mathrm{D[B]}\, ,\, \mathrm{D[Z]}\right]\, = \,0
\end{equation}
we realize a homomorphic embedding:
\begin{equation}\label{circasso2}
    \mathrm{D} \,: \, \mathcal{D}_{4} \times \mathbb{Z}_{2} \, \hookrightarrow \, \mathrm{O(3)}
\end{equation}
The claimed symmetry of the supergravity $2$-brane solution under the group (\ref{dihedrino2}) stems from the
following identities that the reader can easily verify:
\begin{eqnarray}
  \widehat{\mathfrak{W}}\left(\mathfrak{A} \mathbf{X}\right) &=&
  \mathrm{D[A]}\cdot\widehat{\mathfrak{W}}\left(\mathbf{X}\right) \nonumber \\
  \widehat{\mathfrak{W}}\left(\mathfrak{B} \mathbf{X}\right) &=& \mathrm{D[B]}\cdot
  \widehat{\mathfrak{W}}\left(\mathbf{X}\right) \nonumber\\
  \widehat{\mathfrak{W}}\left(\mathfrak{Z} \mathbf{X}\right) &=&
  \mathrm{D[Z]}\cdot\widehat{\mathfrak{W}}\left(\mathbf{X}\right)\label{pastrognetto}
\end{eqnarray}
where the action of the three generators on the torus coordinate is defined below:
\begin{eqnarray}
  \mathfrak{A} \mathbf{X} &=& \left\{X,\frac{1}{4}-Z,Y-\frac{3}{4}\right\}\nonumber\\
  \mathfrak{B} \mathbf{X} &=& \{-X,-Y,Z\} \nonumber\\
  \mathfrak{Z} \mathbf{X} &=& \left\{X,Y+\frac{1}{2},Z+\frac{1}{2}\right\} \label{sigullus2}
\end{eqnarray}
Hence, just as in the previous case, the complete semidirect product group (\ref{squirta2}) is an isometry
group for the supergravity solution since the matrices $\mathrm{D[A]}$ and $\mathrm{D[B]}$, $\mathrm{D[Z]}$
are orthogonal and $\mathrm{O(3)}$ is a global symmetry of the supergravity lagrangian.
\par
The inhomogeneous harmonic function for this brane--solution is the following one:
\begin{eqnarray}
    \hat{H}(y) & = & 1-\frac{1}{96} \lambda ^2 e^{4 \pi  U} \left(4 \,- \,2 \sin \left[2 \pi
   (Y-Z)\right]+2 \sin \left[2 \pi  (Y+Z)\right]\right)\label{harmoniden2}
\end{eqnarray}
and the rest of the solution is obtained from eq.s (\ref{dilatoneB}-\ref{arnoldini}).
\par
 Calculating the $\mathfrak{R}$ curvature associated with
this solution we find that in any point the rank of its $21$ vielbein components is  bounded from above by
$14$. Indeed, with little effort, we find a set of $2$ null vectors which miraculously are null-vectors also
of the matrix $\mathfrak{P}$. In such two--dimensional subspace the Killing spinor equation is easily
integrated by taking all the non vanishing components proportional to $\hat{H}^{\ft 1{10}}(y)$ where
$\hat{H}(y)$ is the inhomogeneous harmonic function (\ref{harmoniden2}). Finally we arrive at the following
explicit form of the two linearly independent Killing spinors:
\begin{equation}\label{killispinati2}
    \epsilon_1 \, = \, \hat{H}^{\ft {1}{10}}(y) \,\left(
\begin{array}{c}
 i \chi _2 \\
 0 \\
 0 \\
 0 \\
 i \chi _1 \\
 0 \\
 0 \\
 0 \\
\end{array}
\right) \quad ; \quad \epsilon_2 \, = \, \hat{H}^{\ft {1}{10}}(y) \, \left(
\begin{array}{c}
 0 \\
 \chi _2 \\
 0 \\
 0 \\
 0 \\
 \chi _1 \\
 0 \\
 0 \\
\end{array}
\right)
\end{equation}
In conclusion the considered  flux-brane solution preserves $\ft 18$ of the original supersymmetry.
\par
\begin{figure}[!hbt]
\begin{center}
\iffigs
\includegraphics[height=55mm]{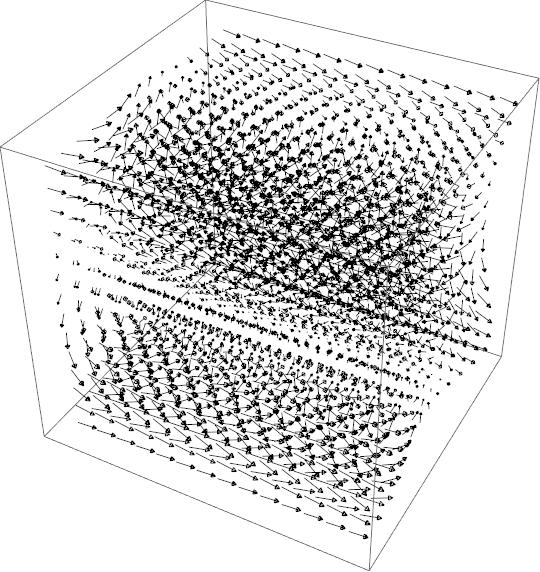}
\includegraphics[height=50mm]{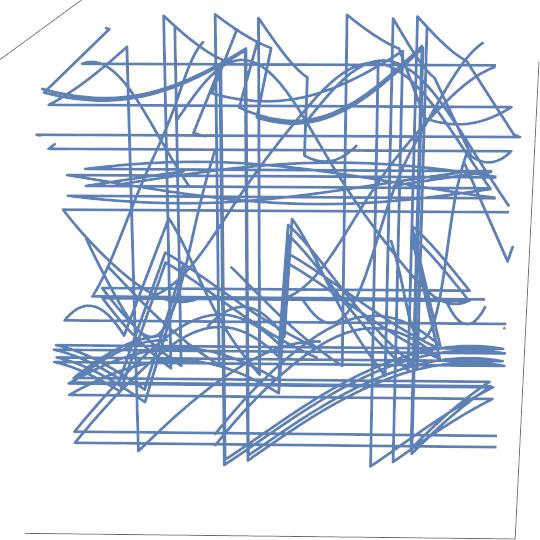}
\else
\end{center}
 \fi
\caption{\it On the left the plot of one arbitrarily chosen vector field in the Beltrami space defined by
eq.(\ref{D4N2}). On the right the plots of some of its integral curves in the 3-torus represented as a cube
with identified opposite faces. \label{fogliarus2}}
 \iffigs
 \hskip 1cm \unitlength=1.1mm
 \end{center}
  \fi
\end{figure}
In the spirit of comparison with the previous case that breaks all  supersymmetries, in fig.\ref{fogliarus2}
we have displayed the plot of an arbitrary vector field in the three-dimensional vector space defined by
eq. (\ref{D4N2}). The  Killing vector in the direction $X$ is visually appreciated by the shape of the vector
field plot.
\par
Let us finally comment on the structure of the inhomogeneous harmonic function (\ref{harmoniden2}). For the
first time among the considered examples this latter has a non trivial dependence on the $\mathrm{T^3}$ torus
coordinates. Obviously it has to be a function invariant under the action of the group (\ref{squirta2}).
Invariance under the continuous translation of the coordinate $X$ are guaranteed by the fact that $\hat{H}(y)$
does not depend on $X$. The invariance under the discrete part (\ref{dihedrino2}), whose action on the torus
is defined in eq. (\ref{sigullus2}) is a priori less obvious, yet it is indeed true, as it can be verified by
explicit calculation.
\begin{figure}[!hbt]
\begin{center}
\iffigs
\includegraphics[height=70mm]{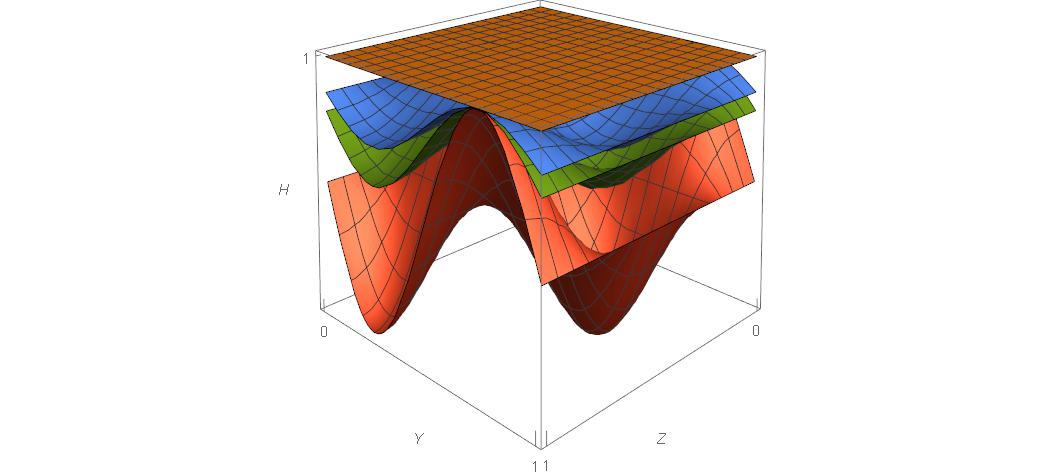}
\else
\end{center}
 \fi
\caption{\it Visualization of the inhomogeneous harmonic function $\hat{H}(y)$ defined by
eq.(\ref{harmoniden2}). The function $\hat{H}$ does not depend on $X$. It depends only on $U,Y,Z$. To
visualize it we have plotted the two argument  function $H(U_0,Y,Z)$ defined over the square $Y,Z$, for
various values of the  constant parameter $U_0$. When $U_0 \, = \, 0$ we have the most oscillating  surface.
As $U_0 \to - \infty $ the surface plot approaches that of a constant and this corresponds to the asymptotic
flatness of the supergravity solution. Note that we have also fixed one reference value of the parameter
$\lambda$, explicitly $\lambda = \sqrt{6}$\label{harmoniosa}}
 \iffigs
 \hskip 1cm \unitlength=1.1mm
 \end{center}
  \fi
\end{figure}
\par
In fig.\ref{harmoniosa} we present a visualization of this dihedral symmetric function.
\section{Uplift of the minimal $D=7$ model to $D=11$ supergravity}
\label{mitirosu}
In this section we illustrate how the minimal ungauged $D=7$ model, with no vector multiplets, is embedded, as a consistent truncation, in
eleven-dimensional supergravity. Consider the latter theory compactified on a 4-torus $\mathrm{T^4}$, which yields the maximal seven dimensional supergravity, and write the ${\rm SO}(4)$ symmetry of the internal manifold as: ${\rm SO}(4)={\rm SO}(3)_+\times {\rm SO}(3)_-$. The minimal $D=7$ supergravity with no vector multiplets describes \emph{the truncation of the maximal eleven dimensional theory to the ${\rm SO}(3)_-$-singlets.} This corresponds to an orbifold reduction from $D=11$ and it is a consistent truncation of the eleven dimensional supergravity.\par  To show this let us prove that the projection on the dimensionally reduced theory yields the right field content and amount of supersymmetry. Being a restriction to singlets with respect to a symmetry group of the maximal $D=7$ model, it is consistent. Let us denote by hatted indices the $D=11$ ones, so that
\begin{eqnarray}
\mbox{Rigid indices:}&&\,\,\,\hat{a}=0,\dots, 10\,\,;\,\,\,\hat{a}=(a,\,m)\,\,\,;\,\,\,\,a=0,\dots, 6\,\,;\,\,\,m=1,2,3,4\,,\nonumber\\
\mbox{Coordinate indices:}&&\,\,\,\hat{\mu}=0,\dots, 10\,\,;\,\,\,\hat{\mu}=(\mu,\,\alpha)\,\,\,;\,\,\,\,\mu=0,\dots, 6\,\,;\,\,\,\alpha=1,2,3,4\,,\nonumber
\end{eqnarray}
The ${\rm SO}(4)={\rm SO}(3)_+\times {\rm SO}(3)_-$ vector and spinor-representations, as usual, read:
\begin{equation}
V_\alpha\in {\bf\left(\frac{1}{2},\frac{1}{2}\right)}\,\,;\,\,\,\,\xi\in {\bf\left(\frac{1}{2},0\right)+\left(0,\frac{1}{2}\right)}\,.
\end{equation}
Restricting to the ${\rm SO}(3)_-$ - singlets, all tensors with an odd number of $m,\,n$ internal indices are projected out while spinors are halved. In particular the moduli of the internal metric on $\mathrm{T^4}$ are frozen to the origin of ${\rm GL}(4,\mathbb{R})/{\rm SO}(4)$, except the determinant of the internal vielbein, which is ${\rm SO}(4)$-invariant and corresponds to the dilaton. After the projection the internal vierbein therefore reads:
\begin{equation}
V_\alpha{}^m=e^{-\frac{5}{12}\phi}\,\delta_\alpha^m\,.
\end{equation}
By the same token the Kaluza-Klein vectors ${\bf B}^\alpha_\mu$ are truncated out.\par
The toroidal dimensional reduction of the 3-form yields:
\begin{eqnarray}
\hat{{\bf C}}^{[3]}_{\hat{\mu}\hat{\nu}\hat{\rho}}&\rightarrow & {\bf B}^{[3]}_{\mu\nu\rho}\,,\,\,{\bf C}^{[2]}_{\mu\nu\alpha}\,,\,\,\,{\bf C}^{[1]}_{\mu\alpha\beta}\,,\,\,\,{\bf C}^{[0]}_{\alpha\beta\gamma}\,.
\end{eqnarray}
Upon truncation to the ${\rm SO}(3)_-$-singlets, the only surviving fields are the 3-form ${\bf B}^{[3]}_{\mu\nu\rho}$ and the projection of the vector fields ${\bf C}^{[1]}_{\mu\alpha\beta}$ on the adjoint representation of ${\rm SO}(3)_+$. This projection is effected by restricting to the components of  ${\bf C}^{[1]}_{\alpha\beta}\equiv {\bf C}^{[1]}_{\mu\alpha\beta}\,dx^\mu$ along the basis $\omega^{(+)\,\Lambda}$ of self-dual 2-forms on the internal $\mathrm{T^4}$:
\begin{equation}
{\bf C}^{[1]}_{\alpha\beta}\vert_{{\rm proj}}\,\propto\,\mathcal{A}^\Lambda\,\omega^{(+)\,\Lambda}_{\alpha\beta}\,\,;\,\,\,\,\,\star_{{\tiny \mathrm{T^4}}}\omega^{(+)\,\Lambda}=\omega^{(+)\,\Lambda}\,\,;\,\,\,d\omega^{(+)\,\Lambda}=0\,.
\end{equation}
where summation over the repeated $\Lambda$ index is understood and
$$\omega^{(+)\,\Lambda}=J^{(+)\,\Lambda}_{\alpha\beta}\,dy^\alpha\wedge dy^\beta\,,$$
$J^\Lambda$ being the ${\rm SO}(3)_+$ generators:
\begin{equation}
J^{(+)\,1}=\left(
\begin{array}{llll}
 0 & \frac{1}{2} & 0 & 0 \\
 -\frac{1}{2} & 0 & 0 & 0 \\
 0 & 0 & 0 & \frac{1}{2} \\
 0 & 0 & -\frac{1}{2} & 0
\end{array}
\right)\,\,;\,\,\,J^{(+)\,2}=\left(
\begin{array}{llll}
 0 & 0 & -\frac{1}{2} & 0 \\
 0 & 0 & 0 & \frac{1}{2} \\
 \frac{1}{2} & 0 & 0 & 0 \\
 0 & -\frac{1}{2} & 0 & 0
\end{array}
\right)\,\,;\,\,\,J^{(+)\,3}=\left(
\begin{array}{llll}
 0 & 0 & 0 & \frac{1}{2} \\
 0 & 0 & \frac{1}{2} & 0 \\
 0 & -\frac{1}{2} & 0 & 0 \\
- \frac{1}{2} & 0 & 0 & 0
\end{array}
\right)\,.
\end{equation}
The three components $\mathcal{A}^\Lambda\equiv \mathcal{A}^\Lambda_\mu\,dx^\mu$ are the three vector fields of the seven-dimensional minimal model.
The dimensionally reduced eleven-dimensional six-form yields the following seven-dimensional fields:
\begin{equation}
\hat{{\bf C}}^{[6]}\,\,\longrightarrow\,\,{{\bf C}}^{[6]},\,{{\bf C}}^{[5]}_\alpha,\,{{\bf C}}^{[4]}_{\alpha\beta},\,{{\bf C}}^{[3]}_{\alpha\beta\gamma},\,{{\bf C}}^{[2]}_{\alpha_1\beta_1\alpha_2\beta_2}\,.
\end{equation}
Upon truncation, the only surviving fields are a two-form ${\bf B}^{[2]}$, dual to three-form ${\bf B}^{[3]}$, and the three 4-forms $\mathcal{A}^{[4]\Lambda}$ dual to the vector fields, defined as follows:
\begin{eqnarray}
{\bf B}^{[2]}&\propto &\frac{1}{4}\,{\bf C}^{[2]}_{\alpha_1\beta_1\alpha_2\beta_2}\,J^{(+)\,\Lambda\,|\alpha_1\beta_1}J^{(+)\,\Lambda\,|\alpha_2\beta_2}\,\,;\,\,\,\,
\mathcal{A}^{[4]\Lambda}\,\propto\frac{1}{2}\,{\bf C}^{[4]}_{\alpha\beta}\,J^{(+)\,\Lambda\,|\alpha\beta}\,.
\end{eqnarray}
From the above definitions and the form of the field strength of the eleven dimensional six-form, we find the correct expression of the field strength of ${\bf B}^{[2]}$:
\begin{equation}
\hat{{\bf F}}^{[7]}\equiv d\hat{{\bf C}}^{[6]}+\hat{{\bf F}}^{[4]}\wedge \hat{{\bf C}}^{[3]}+\dots\,\,\rightarrow\,\,\,\,
\mathfrak{G}^{[3]}=d{\bf B}^{[2]}+\mathfrak{F}^\Lambda\wedge \mathcal{A}^\Lambda+\dots
\end{equation}
Finally let us consider the fermionic sector. The $D=11$ gravitino yields:
\begin{eqnarray}
\hat{\Psi}\,&&\rightarrow\,\,\Psi_A,\,\,\Psi_{A'},\,\,\Psi_{A\alpha},\,\,\,\Psi_{A'\alpha}\,,\nonumber\\ \Psi_A&\in &{\bf\left(\frac{1}{2},0\right)},\,\,\,\Psi_{A'}\,\in\,\,{\bf\left(0,\frac{1}{2}\right)}\,,\nonumber\\
\Psi_{A\alpha}&\in &{\bf\left(\frac{1}{2},0\right)\otimes \left(\frac{1}{2},\frac{1}{2}\right)}={\bf\left(0+1,\,\frac{1}{2}\right)}\,\,;\,\,\,\Psi_{A'\alpha}\in {\bf\left(0,\frac{1}{2}\right)\otimes \left(\frac{1}{2},\frac{1}{2}\right)}={\bf\left(\frac{1}{2},\,0+1\right)}
\end{eqnarray}
 The projection singles out the $D=7$ gravitino field $\Psi^A$ and the spinors $\chi^A$ originating from the ${\bf\left(\frac{1}{2},\,0\right)}$-component of  $\Psi_{A'\alpha}$:
\begin{equation}
\chi_A\,\propto\,\,(\gamma^\alpha)_A{}^{A'}\, \Psi_{A'\alpha}\,.
\end{equation}
On the seven-torus $\mathrm{T^7}=\mathrm{T^3}\times \mathrm{T^4}$, product of the $\mathrm{T^3}$ in the seven-dimensional space-time and the internal $\mathrm{T^4}$, we can write the \emph{ Englert equation} for 3-forms ${\bf Y}^{[3]}$ defined on $\mathrm{T^7}$:
\begin{equation}
\star_{{\tiny \mbox{$\mathrm{T^7}$}}} d{\bf Y}^{[3]}=\mu\, {\bf Y}^{[3]}\,.\label{EnglertE}
\end{equation}
Upon restricting to 3-forms of the type ${\bf Y}^{[3]}={\bf Y}^{[1]\,\Lambda}({\bf X})\wedge \omega^{(+)\,\Lambda}$, the Englert equation reduces to the Arnold-Beltrami one considered in the present paper:
\begin{equation}
\star_{{\tiny \mbox{$\mathrm{T^7}$}}} d({\bf Y}^{[1]\,\Lambda}\wedge \omega^{(+)\,\Lambda})=\star_{{\tiny \mbox{$\mathrm{T^3}$}}}  d{\bf Y}^{[1]\,\Lambda}\wedge \star_{{\tiny \mbox{$\mathrm{T^4}$}}}\omega^{(+)\,\Lambda}=\mu\, {\bf Y}^{[1]\,\Lambda}\wedge \omega^{(+)\,\Lambda}\,\,\Leftrightarrow \,\,\,\,\star_{{\tiny \mbox{$\mathrm{T^3}$}}}  d{\bf Y}^{[1]\,\Lambda}=\mu\,{\bf Y}^{[1]\,\Lambda}\,.
\end{equation}
The dictionary defined in the present section allows to uplift any solution to the minimal $D=7$ supergravity, with no vector multiplets, to eleven dimensions, including the Arnold-Beltrami 2-branes extensively discussed in the previous sections, which describe $M2$-branes with fluxes. \par

\section{Conclusions}
\label{zakliuchenie} In the present paper we have presented the first half of the geometric reconstruction of
Minimal $D=7$ supergravity in terms of Free Differential Algebras and rheonomy. Indeed we have completely
solved Bianchi identities, fixing the precise form of the supersymmetry transformation rules to all orders in
the boson and including higher order terms in the fermion fields.
\par
This general result allowed us to embed Arnold-Beltrami flux $2$-branes into supergravity and study the
Killing spinor equation in their background. We have also presented four explicit examples of solutions
\begin{enumerate}
  \item One solution with no supersymmetry and a discrete symmetry
  \begin{equation}\label{bosuN0}
    G_{bosonic} \, = \, \underbrace{\left(\mathrm{O_{24}} \times \mathbb{Z}_2 \right)}_{\Gamma}\ltimes
    \underbrace{\left[\mathbb{Z}_2\times \mathbb{Z}_2\right]}_{\mbox{transl.}}
  \end{equation}
  where $\mathrm{O_{24}}$ denotes the octahedral group.
 \item One solution with $4$ Killing spinors and a discrete symmetry:
 \begin{equation}\label{bosuN4D4}
    G_{bosonic} \, = \, \underbrace{\mathrm{\mathcal{D}_{4}}  }_{\Gamma}\ltimes
    \underbrace{\left[\mathrm{U(1)}\times \mathrm{U(1)}\right]}_{\mbox{transl.}}
  \end{equation}
  where $\mathcal{D}_4$ denotes the dihedral group of index $4$.
  \item One solution with $4$ Killing spinors and a discrete symmetry:
 \begin{equation}\label{bosuN4D6}
    G_{bosonic} \, = \, \underbrace{\mathrm{\mathcal{D}_{6}}  }_{\Gamma}\ltimes
    \underbrace{\left[\mathrm{U(1)}\times \mathrm{U(1)}\right]}_{\mbox{transl.}}
  \end{equation}
  where $\mathcal{D}_6$ denotes the dihedral group of index $6$. (We have also advocated that similar solutions should
  exist for $\mathcal{D}_2$ and $\mathcal{D}_3$).
\item One solution with 2 Killing spinors and a discrete symmetry
  \begin{equation}\label{bosuN02}
    G_{bosonic} \, = \, \underbrace{\left(\mathrm{\mathcal{D}_{4}} \times \mathbb{Z}_2 \right)}_{\Gamma}\ltimes
    \underbrace{\mathrm{U(1)}}_{\mbox{transl.}}
  \end{equation}
  where $\mathrm{O_{24}}$ denotes the octahedral group.
\end{enumerate}
The perspectives of further investigations based on the results we have achieved so far are three-fold.
\begin{description}
  \item[A)] On the one hand we plan to complete our geometrical reconstruction of minimal $D=7$ supergravity, coupled to a generic number of vector fields and including higher order terms in the fermion fields, obtaining the action and after that studying the gaugings of the theory utilizing the method of the
   embedding tensor \cite{embedtensor1,embedtensor2,Dibitetto:2015bia}.
  \item[B)] On the other hand, from the point of view of flux $2$-branes we consider the present one as the
  first step in a logical path of development. We need now to construct the $\kappa$-supersymmetric $2$-brane
  actions in the background of the brane solutions and study the gauge/gravity correspondence between the $d=3$ gauge
  theory on the brane world-volume and the bulk supergravity. The discrete symmetries are expected to play a
  fundamental role in the classification and interactions of the composite operators.  Moreover the description of the solutions as $M2$-branes with fluxes in the eleven-dimensional theory suggests that the  dual CFT might be related to the ABJM model \cite{Aharony:2008ug}.
  \item[C)] A fully-fledged search of supersymmetric flux $2$-branes should be attempted considering all the
  crystallographic lattices and all their Point Groups. An ambitious aim would be to establish  more stringent
  a priori conditions for the existence of Killing spinors.
  \item[D)] Finally it would be interesting to study more general $M2$-branes in the eleven-dimensional supergravity characterized  by fluxes which are solutions to the Englert equation (\ref{EnglertE}).
\end{description}
\section*{Aknowledgements}
We are grateful to L. Andrianopoli  and R. D'Auria for their contributions to the early stages of the work and for enlightening discussions.
One of us (P.F.) is particularly grateful to his friend and collaborator A. Sorin
for many important discussions during the whole development of this research project.
\newpage
\appendix
\section{Detailed derivation of the rheonomic solution of Bianchi identities}
\label{dettaglione} 
In this appendix we present the detailed derivation of the unique rheonomic solution of Bianchi identities of
the relevant Free Differential Algebra. The determination of the 24 coefficients mentioned in the main text is
the absolute core of the supergravity theory. These numbers decide the explicit form of the supersymmetry
transformation rules and implicitly determine the field equations of supergravity, hence its classical
dynamics. We already stressed that the very existence of Arnold-Beltrami flux branes critically depends on the
precise numerical values of the lagrangian coefficients  which on their turn depend, in a one-to-one way, from
the coefficients found in the solution of Bianchi identities. Similarly the existence of Killing spinors for
given solutions of supergravity, in particular the flux branes studied in this paper, depends on the precise
values of 24 coefficients discussed here. Change one of them to a wrong value and the results change not
quantitatively but qualitatively. This is not surprising when you remind ourselves that we are talking about
the realization of an algebra of transformations. The fascination of supersymmetry and supergravity is that,
in this case, the algebra   is not kinematics, rather it is the very dynamics of the system.
\par
It follows from these considerations that the calculations presented in this appendix are not marginal rather
they are of the utmost relevance. Yet they are extremely tedious. The principle is simple and elegant. Its
implementation is desperately tedious, although essential. For this reason these important calculations are
relegated to an appendix.
\subsection{Rheonomic solution of the Bianchis for the curvatures of degree $p\le 2$}
According to the logic presented in the main test we start by solving completely the Bianchi identities of all
the curvatures of degree two or one associated with the standard superalgebra sector of the FDA. As we
demonstrate below the set of $19$ parameters $\mbox{coeff}_{\mathrm{Lie}}$  is reduced, after imposing the
constraints of these Bianchis to three free parameters, namely $c_1$, $g_1$ and $\delta$, all the others being
fixed in terms of these latter. Let us see how.
\subsubsection{Equations from the $3$$\Psi$ sector of the Torsion Bianchi}
At the level of $3$$\Psi$ the torsion Bianchi equation (\ref{torsobianchi}) is very simple. It reads:
\begin{equation}\label{godfather}
    \overline{\Psi}^A \wedge \Gamma^a \rho_A^{[\Psi\Psi]} \, = \, 0
\end{equation}
where we have named:
\begin{eqnarray}
  \rho_A^{[\Psi\Psi]} & = &  \, g_1 \, \Gamma_m \, \chi_A \, \overline{\Psi}^C \wedge \Gamma^m \Psi_C \,
  + \, g_2 \, \Gamma_{mn}  \chi_A \, \overline{\Psi}^C  \wedge  \Gamma^{mn} \,\Psi_C \nonumber\\
  &\null& -\, g_3 \,  \chi_B \, \sigma^{\Lambda|B}_{\phantom{\Lambda|B}A} \,
  \sigma^{\Lambda|D}_{\phantom{\Lambda|D}C}\, \overline{\Psi}^C  \wedge \Psi_D \,
   - \, g_4 \,\Gamma_{pqr} \chi_B \, \sigma^{\Lambda|B}_{\phantom{\Lambda|B}A} \,
    \sigma^{\Lambda|D}_{\phantom{\Lambda|D}C}\,  \overline{\Psi}^C  \wedge \Gamma^{pqr} \Psi_D \label{gulazzo}
\end{eqnarray}
Comparing eq.s (\ref{godfather}-\ref{gulazzo}) with eq.s (\ref{minestrina3psi}-\ref{elleb}) we realize that
eq. (\ref{godfather}) is nothing else but $\ell_b\, = \,0$ which is solved by eq.(\ref{baraccone}) expressing
$g_3$ and $g_4$ in terms of $g_{1,2}$. In this way we have reduced the 19  parameters we are dealing with to
seventeen. Let us also note in advance that once eq.(\ref{godfather}) is satisfied the contribution of
$\rho_A^{[\Psi\Psi]}$ to the Bianchi equation of $\mathfrak{G}^{[3]}$ (see eq.(\ref{g3bianchi})) vanishes
\textit{a fortiori.} This will we important in the sequel.

\subsubsection{Equations from the $2$$\Psi$-$1$$V$ sector of the Torsion-Bianchi}
Inserting the rheonomic parameterizations (\ref{torsO}-\ref{gravitelpara}) into the Bianchi identity
(\ref{torsobianchi}) and keeping only the terms proportional to $2$$\Psi$-$1$$V$, we obtain the following
equation:
\begin{equation}\label{cristolario}
   \begin{array}{rcl}
     0 & = & - \,R_{\Psi\Psi}^{ab} \wedge V^b \, +\, \mathcal{S}^{ab} \wedge V^b \\
   \end{array}
\end{equation}
where:
\begin{eqnarray}
  R_{\Psi\Psi}^{ab} & \equiv & \lambda_1 \,  e^{\phi}\,\mathcal{G}^{abc} \,  \overline{\Psi}^A  \wedge
  \Gamma_c \Psi_A \,+ \, \lambda_2 \,  e^{\phi}\, \mathcal{G}_{pqr} \, \overline{\Psi}^A  \wedge
   \Gamma^{abpqr} \Psi_A  \nonumber\\
  &\null&\, + \,
 {\rm i}\, \mu_1 \, e^{\ft 12 \phi}\, \mathcal{F}^{\Lambda|ab} \,
  \sigma^{\Lambda|B}_{\phantom{\Lambda|B}A} \,\overline{\Psi}^A  \wedge
   \Psi_B \, + \,{\rm i}\, \mu_2 \, e^{\ft 12 \phi}\, \mathcal{F}^{\Lambda}_{pq} \,
    \sigma^{\Lambda|B}_{\phantom{\Lambda|B}A} \,\overline{\Psi}^A  \wedge  \Gamma^{abpq}\,\Psi_B \label{duepsiR} \\
  \mathcal{S}^{ab} &=& \overline{\Psi}^A \, \wedge \,\left(\Gamma^a \, \mathcal{M}^{B}_{A} \Gamma^b
   + \Gamma^{ab} \,\mathcal{N}^{B}_{A} \right) \, \Psi_B \wedge V^b  \label{cossigus}
\end{eqnarray}
Equation (\ref{duepsiR}) is solved by setting first the antisymmetric part of $\mathcal{S}^{ab}$ to zero and
then by identifying the symmetric one with $R_{\Psi\Psi}^{ab}$. This yields the following equations on the
parameters:
\begin{equation}\label{Tparacqua}
    \begin{array}{rcl}
       0 & = & \kappa_1 \\
       0 & = & \kappa_2 \\
       0 & = & \lambda_1 +6\left( b_1 +b_2\right) \\
       0 & = & \lambda_2 + \left( b_1 -b_2\right)\\
       0 & = & \mu_1 +2\left( d_1 -d_2\right) \\
       0 & = & \mu_2 +\left( d_1 +d_2\right)
     \end{array}
\end{equation}
In this way the seventeen parameters  have been reduced to eleven.
\subsubsection{Equations from the Gravitino Bianchi at 3$\Psi$-level}
If we consider the gravitino Bianchi (\ref{rhobianchi}) and after insertion of the rheonomic parameterizations
(\ref{torsO}-\ref{gravitelpara}) we focus on the $3$$\Psi$-sector we obtain the following equation:
\begin{eqnarray}
  0 &=& -\, \ft 14 \, \Gamma_{ab} \Psi_A \wedge R^{ab}_{\Psi\Psi} \, +\, \ft 12 \,
  \left( \mathcal{M}^{B}_{A} \Gamma_b  + \Gamma_{b} \,\mathcal{N}^{B}_{A} \right) \,
  \Psi_B \wedge \overline{\Psi}^C\wedge\Gamma^b\Psi_C\nonumber \\
 \null &\null& -\, g_1 \, \Gamma_m \, \mathcal{P}^D_A \, \Psi_D \, \overline{\Psi}^C
 \wedge
 \Gamma^m \Psi_C \, - \, g_2 \, \Gamma_{mn}  \mathcal{P}^D_A \, \Psi_D \, \overline{\Psi}^C
 \wedge  \Gamma^{mn} \,\Psi_C \nonumber\\
  &\null& +\, g_3 \,  \mathcal{P}^E_B \,\Psi_E\, \sigma^{\Lambda|B}_{\phantom{\Lambda|B}A} \,
  \sigma^{\Lambda|D}_{\phantom{\Lambda|D}C}\, \overline{\Psi}^C  \wedge \Psi_D \, + \, g_4
  \,\Gamma_{pqr} \mathcal{P}^E_B \,\Psi_E \, \sigma^{\Lambda|B}_{\phantom{\Lambda|B}A} \,
  \sigma^{\Lambda|D}_{\phantom{\Lambda|D}C}\,  \overline{\Psi}^C  \wedge \Gamma^{pqr} \Psi_D \label{malandrina}
\end{eqnarray}
Separate cancellation of the terms proportional to $\mathcal{G}_{abc}$, $\mathcal{F}^\Lambda_{ab}$ and
$\Phi_a$ imposes on the parameters a set of conditions which together with those found in the previous two
subsections yields the following result:
\begin{eqnarray}\label{namus}
 b_1& = & 16 \, c_1 \, g_2 \nonumber \\
 b_2& = & \frac{4}{3}\,  c_1
   \left(g_1-2 \, g_2\right) \nonumber \\
 d_1& = & -32\, c_2\, g_2 \nonumber \\
 d_2& = & \frac{4}{3}\, c_2\,
   \left(g_1+10\, g_2\right) \nonumber \\
 g_3& = & \frac{1}{6} \left(-5\,
   g_1-14 \, g_2\right) \nonumber \\
 g_4& = & \frac{1}{36} \left(2\,
   g_2-g_1\right) \nonumber \\
 \kappa _1& = & 0 \nonumber \\
 \kappa _2& = & 0 \nonumber \\
 \lambda _1& = & -8 c_1\,
   \left(g_1+10\, g_2\right) \nonumber \\
 \lambda _2& = & \frac{4}{3} c_1
   \left(g_1-14 \, g_2\right) \nonumber \\
 \mu _1& = & \frac{8}{3}\,  c_2 \,
   \left(g_1+34 \, g_2\right) \nonumber \\
 \mu _2& = & -\frac{4}{3} \, c_2\,
   \left(g_1-14 \, g_2\right)
\end{eqnarray}
In this way the set of free coefficients among the 19 comprised in $\mbox{coeff}_{\mathrm{Lie}}$ is reduced to
seven, namely:
\begin{equation}\label{sopravvissuti}
  \{ a_1,\, c_1, \, c_2,\, c_3, \,  g_1, \, g_2, \, \delta\}
\end{equation}
\subsubsection{Equation for $c_3$ from the dilaton Bianchi}
The coefficient $c_3$ is easily and immediately determined from the dilaton Bianchi (\ref{dilabianchi}), upon
insertion of the rheonomic parameterization (\ref{dilatpara}). We immediately obtain:
\begin{equation}\label{pioggia}
    c_3 \, = \, \ft 12
\end{equation}
\subsubsection{Equations from the $2$$\Psi$-$1$$V$ sector of the $\mathfrak{F}^\Lambda$-Bianchi}
Inserting the rheonomic parameterizations (\ref{torsO}-\ref{gravitelpara}) into the Bianchi identity
(\ref{fbianchi}) and keeping only the terms proportional to $2$$\Psi$-$1$$V$, we obtain the following
equation:
\begin{eqnarray}
  0 &=& -\, \mathcal{F}^\Lambda_{ab} \overline{\Psi}^A \wedge \Gamma^a \Psi_A\wedge V^b \,
  + \, {\rm i} \, a_1 \, e^{-\ft 12 \phi} \,
  \sigma^{\Lambda|B}_{\phantom{\Lambda|B}A} \,\overline{\Psi}^A
  \wedge  \Gamma_a \mathcal{P}^C_B\Psi_C \wedge V^a\nonumber\\
  \null &\null& +{\rm i} \, e^{-\ft 12 \phi} \,
  \sigma^{\Lambda|B}_{\phantom{\Lambda|B}A} \,
  \overline{\Psi}^A \left( \mathcal{M}_B^C \, \Gamma_a +
  \Gamma_a \, \mathcal{N}_B^C \right) \Psi_C \wedge V^a\,
  +\, {\rm i} \ft 14 \, e^{-\ft 12 \phi} \, \sigma^{\Lambda|B}_{\phantom{\Lambda|B}A}
  \, \Phi_a\wedge \overline{\Psi}^A \, \wedge \, \Psi_B\wedge V^a\nonumber\\&&\label{salmonefritto}
\end{eqnarray}
Imposing the cancellation of all structures we obtain the following equations on the coefficients:
\begin{eqnarray}
\label{eccoF}
  a_1 &=&- \, \frac{1}{2} \nonumber \\
  c_2 &=& \frac{7}{24-64 \, g_1}\nonumber\\
 g_2 &=& \frac{1}{112}\left( 8\, g_1 \, - \, 3\right)
\end{eqnarray}
In this way the seven free parameters mentioned in eq. (\ref{sopravvissuti}) are reduced to the three mentioned
at the beginning of this subsection
\subsubsection{Equations from the 3$\Psi$-level of the $\mathfrak{F}^\Lambda$ curvature}
 At the  3$\Psi$-level the Bianchi identity of the $\mathfrak{F}^\Lambda$ curvature, namely  eq. (\ref{fbianchi}),
 reduces to the following statement:
\begin{eqnarray}
  0 &=& {\rm i}\, \ft 14 \, e^{-\ft 12 \, \phi} \,\overline{\Psi}^C \, \chi_C \,\wedge \,\overline{\Psi}^A \,
\wedge \, \sigma^{\Lambda|B}_A \,\Psi_B \, +\, {\rm i}\, e^{-\ft 12 \, \phi} \,\overline{\Psi}^A \,
\wedge \, \sigma^{\Lambda|B}_A \,\rho_B^{[\Psi\Psi]} \nonumber  \\
  \null &\null& +  \,{\rm i} \, \ft 12 \, a_1 \, e^{-\ft 12 \, \phi} \,
  \sigma^{\Lambda|B}_{\phantom{\Lambda|B}A} \,\overline{\Psi}^A\Gamma_{a} \, \chi_B \, \wedge \,
  \overline{ \Psi}^C  \wedge \Gamma^a \Psi_C \label{3psiFbianchi}
\end{eqnarray}
which, surprisingly imposes no new constraint and it is identically satisfied by the set of parameters
satisfying all the previous constraints, namely:
\begin{equation}\label{comintern}
  \begin{array}{rclcrclcrcl}
     a_1 & = & - \ft 12  & ; & b_1 & = & \frac{c_1}{7} \left(-3 + 8\, g_1\right ) & ; & b_2 &
     = &\frac{c_1}{7} \left(1 + 16\, g_1\right )  \\
     c_1 & = & c_1 & ; & c_2 &= & \frac{7}{24 - 64 \, g_1} & ; & c_3 & =  & \ft 12 \\
     d_1 & = & \ft 14  & ; & d_2 & = & \frac{5 - 32 \, g_1}{16(-3 + 8\, g_1)} & ; & g_1 & = & g_1 \\
     g_2 & = & \ft{1}{112}(-3+8\, g_1) & ; & g_3 & = & \ft {1}{16}- g_1 & ; & g_4 &
     = & - \, \ft{1}{672} \left(1+16 \, g_1\right) \\
     \kappa_1 & = & 0 & ; & \kappa_2 & = & 0 & ; & \lambda_1 &
     = &  - \, \frac{3}{7}\, c_1 \, \left( -5 + 32\, g_1\right)\\
      \lambda_2 & = & \frac{c_1}{2} & ; & \mu_1 & = & \frac{17-64 \, g_1}{8\left( -3 +8\, g_1\right)} &
       ; & \mu_2 & = & \frac{7}{16 \left( -3 +8\, g_1\right)}  \\
     \delta & = & \delta & ; & \null & \null &\null & ; & \null & \null & \null \\
   \end{array}
\end{equation}
\subsection{Solving the Bianchis for curvatures of degree $p=3,4$}
\label{bianchip34}
 Having completely solved the Bianchi identities for the curvatures of degree $p\le 2$ we
have been left with three parameters $\delta, c_1 $ and $g_1$ that parameterize all the others according to
eq.(\ref{comintern}). In the background of such parameterized curvatures we consider the Bianchi identities of
the higher degree curvatures.
\par
We begin with the Bianchi of the $\mathfrak{G}^{[3]}$ form corresponding to the formulation of
\cite{SalamSezgin} and \cite{bershoffo1}.
\subsubsection{Equations from the $2$$\Psi$-$2$$V$ sector of the $\mathfrak{G}^{[3]}$-Bianchi}
Inserting the rheonomic parameterizations (\ref{torsO}-\ref{gravitelpara}) into the Bianchi identity
(\ref{g3bianchi}) and keeping only the terms proportional to $2$$\Psi$-$2$$V$, we obtain the following
equation:
\begin{equation}\label{2psi2vG}
    \begin{array}{rcl}
       0 & = & \left(\lambda_1  q -\ft 32\right) \,G_{abc}\, \overline{\Psi}^A \wedge \Gamma^a \Psi_A
       \wedge V^b\wedge V^c \, + \, \lambda_2 q \, G_{pqr}\, \overline{\Psi}^A \wedge \Gamma^{pqrab}
       \Psi_A \wedge V_a\wedge V_b  \\
       \null & \null & + \, {\rm i}\left(\mu_1  q +1\right) \, e^{-\ft 12 \phi} \,
       \mathcal{F}^{\Lambda}_{ab}\,\sigma^{\Lambda|B}_{\phantom{\Lambda|B}A} \,
       \overline{\Psi}^A\wedge \Psi_B \wedge V^a\wedge V^b  \\
       \null & \null & + \, {\rm i}\mu_2  q   \, e^{-\ft 12 \phi} \,
       \mathcal{F}^{\Lambda}_{pq}\,\sigma^{\Lambda|B}_{\phantom{\Lambda|B}A} \,
       \overline{\Psi}^A\wedge \Gamma^{pqab}\Psi_B \wedge V_a\wedge V_b + \ft 12 \,e^{-\phi}\,\Phi_a \,
       \overline{\Psi}^A \wedge \Gamma_b\Psi_A\wedge V^a\wedge V^b \\
       \null & \null & -(q-1) \,e^{-\phi}\, \overline{\Psi}^A \wedge \left( \Gamma_a \,\mathcal{M}^B_A\,
       \Gamma_b + \Gamma_{ab} \mathcal{N}^B_A\right) \Psi_B \wedge V^a \wedge V^b\\
       &\null&+\, a_2 \,e^{-\phi}\,\overline{\Psi}^A \wedge \Gamma_{ab} \, \mathcal{P}^B_A \Psi_B \wedge V^a\wedge V^b \,
     \end{array}
\end{equation}
Imposing the identical cancellation of all type of terms and previously eliminating the parameters
$\lambda_{1,2},\mu_{1,2}$ via eq.s (\ref{Tparacqua}) we obtain the following equations on the remaining
parameters:
\begin{equation}\label{Gparacqua}
    \begin{array}{rcl}
       0 & = & -3 \left(4 b_1+4 b_2+4 a_2 c_1+1\right)\\
       0 & = & -b_1+b_2+a_2 c_1\\
       0 & = & 4 a_2 c_3+1 \\
       0 & = & a_2 c_2-(q-2) \left(d_1+d_2\right)\\
       0 & = & 2 a_2 c_2-2 d_1+2 d_2+1
     \end{array}
\end{equation}
Combining the above equations with those in eq.(\ref{comintern}) we obtain the final solution for the 21
parameters in eq. (\ref{settusG3}). Such a solution, which is displayed below, depends on a free parameter
that we have localized in $g_1$. All values of $g_1$ are permitted except $\ft 38$ for which the solution
becomes singular:
\begin{equation}\label{comExtern}
  \begin{array}{rclcrclcrcl}
     a_1 & = & - \ft 12  & ; & b_1 & = &- \ft 1 8& ; & b_2 & = & - \, \ft 1 8 + \frac{7}{48 - 128\, g_1}  \\
     c_1 & = & \frac{7}{24 - 64\, g_1} & ; & c_2 &= & \frac{7}{24 - 64 \, g_1} & ; & c_3 & =  & \ft 12 \\
     d_1 & = & \ft 14  & ; & d_2 & = & - \, \ft 1 4 + \frac{7}{48 - 128\, g_1} & ; & g_1 & = & g_1 \\
     g_2 & = & \ft{1}{112}(-3+8\, g_1) & ; & g_3 & = & \ft {1}{16}- g_1 & ; & g_4 & = & - \,
     \ft{1}{672} \left(1+16 \, g_1\right) \\
     \kappa_1 & = & 0 & ; & \kappa_2 & = & 0 & ; & \lambda_1 & = &   \frac{3}{8}\, \left( 4 +
     \frac{7}{-3+8\, g_1}\right)\\
      \lambda_2 & = & \frac{7}{48 - 128\, g_1} & ; & \mu_1 & = &-1 + \frac{7}{24 - 64\, g_1}  &
      ; & \mu_2 & = & \frac{7}{16 \left( -3 +8\, g_1\right)}  \\
     \delta & = &1 & ; & q & =&1 & ; &a_2 & = &- \ft 12 \\
   \end{array}
\end{equation}
It is now very interesting to compare the solution (\ref{comExtern}) with the supersymmetry transformation
rules derived by the authors of \cite{bershoffo1}. A comparison at the level of absolute values of the
coefficients is very laborius since it involves the normalization of the various fields, but there is a simple
and very significant test that is intrinsic and normalization independent. We refer to the ratio of the
coefficients $b_1/b_2$ and $d_1/d_2$ that appear in the gravitino curvature and that dictate the form of the
gravitino transformation rule. These ratios cannot be deformed by changing the normalization of any field and
hence are an intrinsic property of the susy algebra, \textit{i.e.} of the rheonomic parameterizations.
Comparing with eq.(2.9) of \cite{bershoffo1} we see that according to these authors the two ratios are
predicted to be:
\begin{equation}\label{berkohsez}
  \frac{b_1}{b_2} \, = \, 5 \quad ;  \quad \frac{d_1}{d_2} \, = \, - \, \frac{5}{3}
\end{equation}
It is non trivial and reassuring that the two above equations for the parameter $g_1$ are consistent and admit
the common solution:
\begin{equation}\label{g1Bergo}
  g_1 \, = \, -\frac{11}{64}
\end{equation}
In this way we have reconstructed the formulation by Bergshoeff et al of minimal $D=7$ supergravity, but we
have also learned that it admits a non trivial deformation encoded in the parameter $g_1$. Obviously the
parameter $g_1$ could not be seen by the authors of \cite{bershoffo1} since they did not consider quadratic
fermion terms in the transformation rules of the fermions and implicitly fixed a choice of $g_1$ adopting a
certain relative strength of the kinetic terms in the lagrangian.
\subsubsection{Equations from the $3$$\Psi$ sector of the $\mathfrak{G}^{[4]}$ Bianchi}
In the case we utilize the 3-form formulation we have to satisfy also the Bianchi identity of the
$\mathfrak{G}^{[4]}$-curvature. This latter has a $3$$\Psi$-sector that differently from the case of the
2-form is not identically satisfied by the solution of torsion Bianchi equation. This sector yields the
following equation:
\begin{eqnarray}\label{crisalide}
 0 & =& {\rm i} \, e^{-{\theta}\phi} \, \left(\overline{\Psi}^A  \wedge \Gamma_{ab} \rho^{[\Psi\Psi]}_A \,
 + \, \frac{3 }{2} \,  w \,
 \overline{\Psi}^A  \wedge \Gamma_{abc} \chi_A  \wedge \overline{\Psi}^B\wedge\Gamma^c\Psi_B\right.\nonumber\\
 &&\left.+ \frac{\theta}{2} \, \overline{\Psi}^A
 \wedge \chi_A  \wedge \overline{\Psi}^B\wedge\Gamma_{ab}\Psi_B\right)\wedge V^a \wedge V^b
\end{eqnarray}
which imposes the following two constraints on the coefficients:
\begin{eqnarray}
  \frac 67 \,  -\,  12 \, w \, - \, \frac{128}{7} \, g_1 &=& 0 \nonumber \\
  \frac 67 \,  + \,  12 \,\theta \, - \, \frac{128}{7} \, g_1 &=& 0 \label{g43psiEqua}
\end{eqnarray}
which are solved by the following conditions:
\begin{equation}\label{franziskus}
 w \, = \, - \,\frac{\theta}{3} \quad ; \quad g_1 \, = \, \frac{1}{64}\left( 3 \, + \, 14 \, \theta \right)
\end{equation}
\subsubsection{Equations from the $2$$\Psi$ sector of the $\mathfrak{G}^{[4]}$ Bianchi}
At this point  we have still  to consider the $2$$\Psi$ sector of the $\mathfrak{G}^{[4]}$-Bianchi which
yields the following equation:
\begin{equation}\label{2psi2vG4}
    \begin{array}{rcl}
       0 & = & - 2 \, \nu \,e^{(1-\theta)\phi}\, \varepsilon_{pqr abcd} \,G^{pqr}\, \overline{\Psi}^A \wedge
       \Gamma^d \Psi_A \wedge V^a \wedge V^b \wedge V^c \\
       \null & \null & +{\rm i}\left ( e^{-\theta \phi} \, \overline{\Psi}^A \wedge \left( \Gamma_{ab}
       \,\mathcal{M}^B_A\, \Gamma_c + \Gamma_{abc} \mathcal{N}^B_A\right) \Psi_B  \wedge V^a \wedge V^b
       \wedge V^c \right.\\
       &\null&\left. +\, w \, e^{-\theta \phi} \, \overline{\Psi}^A \wedge \Gamma_{abc} \, \mathcal{P}^B_A
       \Psi_B   \wedge V^a \wedge V^b \wedge V^c\right)-\frac{i}{2}\, e^{-\theta \phi}\,\Phi_c\,\overline{\Psi}^A\Gamma_{ab}\Psi_A\wedge V^a \wedge V^b \wedge V^c
     \end{array}
\end{equation}
It is very much reassuring that all the other structures cancel identically in eq.(\ref{2psi2vG4}) upon the
use of the coefficients that we have already determined and that those involving $G^{pqr}$ cancel also
identically upon fixing  the following value for the parameter $\nu$:
\begin{equation}\label{valoredinu}
  \nu \, = \, \frac{1}{12}
\end{equation}
In this way we have completely solved in a rheonomic way the Bianchi identities involving both the three-form
and the four-form curvatures whose space-time field strengths are dual to each other. Altogether we have found
the following set of coefficients parameterized by the single parameter theta:
\begin{equation}\label{comExternB}
 \begin{array}{llllllllllll}
 a_1 & = &
   -\frac{1}{2} & ;
   & a_2 & = &
   -\frac{1}{2} & ;
   & b_1 & = &
   -\frac{1}{8} & \nonumber
   \\
 b_2 & = & \frac{2
   \theta +1}{24-16 \theta } &
   ; & c_1 &
   = & \frac{1}{3-2
   \theta } & ; &
   c_2 & = &
   \frac{1}{3-2 \theta } &
   \nonumber \\
 c_3 & = &
   \frac{1}{2} & ;
   & d_1 & = &
   \frac{1}{4} & ;
   & d_2 & = &
   \frac{1-2 \theta }{8 \theta
   -12} & \nonumber \\
 g_1 & = &
   \frac{1}{64} (14 \theta +3)
   & ; & g_2 &
   = &
   \frac{1}{128} (2 \theta -3)
   & ; & g_3 &
   = & \frac{1}{64}
   (1-14 \theta ) &
   \nonumber \\
 g_4 & = &
   \frac{1}{384} (-2 \theta
   -1) & ; & \kappa
   _1 & = & 0 &
   ; & \kappa _2 &
   = & 0 &
   \nonumber \\
 \lambda _1 & = &
   \frac{3}{2}+\frac{3}{2
   \theta -3} & ; &
   \lambda _2 & = &
   \frac{1}{6-4 \theta } &
   ; & \mu _1 &
   = & \frac{1}{3-2
   \theta }-1 & \nonumber \\
 \mu _2 & = &
   \frac{1}{4 \theta -6} &
   ; & \delta  &
   = & 1 &
   ; & w &
   = &
   -\frac{\theta }{3} &
   \nonumber \\
 q & = & 1 &
   ; & \nu  &
   = &
   \frac{1}{12} &
   ; & \theta  &
   = & \theta  &
   \nonumber
\end{array}
\end{equation}
The solution as usual is multiply checked since the constraints are many more than the parameters that can be
fixed.
\section{Constraints on the rheonomic action coefficients from comparison with $TPvN$ and the flux brane
action} \label{costrettioneffi} We have shown that the second order bosonic lagrangian of \cite{PvNT} is
identical, after appropriate rescalings to the flux-brane lagrangian (\ref{fluxbraneaction}). On the other
hand the supersymmetry transformations of \cite{PvNT} agree, after appropriate rescalings, with those issuing
from the rheonomic parameterization of the Bianchi identities presented in the previous sections. Ergo the
bosonic sector of the action of $D=7$ supergravity streaming from the rheonomic approach must  map, after the
rescalings (\ref{sigresca}), into the flux-brane lagrangian (\ref{fluxbraneaction}). This happens if certain
relations on the coefficients $f_i$ of the bosonic action (\ref{LBkin}) are satisfied. In the present section
we derive these constraints postponing to a forthcoming publication their verification within the full
determination of all the coefficients of the full rheonomc action.
\par
Discarding the gravitino 1-forms and the dilatino $\chi$ the action $\mathcal{L}^{ungauged}_{Bkin}$ reduces
to:
\begin{eqnarray}
  \mathcal{L}^{ungauged}_{Bose} &=& f_1 \,\mathfrak{R}^{a_1a_2} \wedge V^{a_3} \wedge \dots
  \wedge V^{a_7} \, \epsilon_{a_1\dots a_7}\nonumber\\
   &&\, + \, f_2 \,\Phi^{a_1} \, \mathrm{d}\phi \, \wedge  V^{a_2} \wedge \dots \wedge V^{a_7} \,
   \epsilon_{a_1\dots a_7}  \nonumber\\
&& + f_3 \, e^\phi \, \mathcal{F}^{\Lambda|a_1a_2} \,  \mathfrak{F}^{\Lambda} \,  \wedge V^{a_3}
\wedge \dots \wedge V^{a_7} \, \epsilon_{a_1\dots a_7} \nonumber\\
&& + \, f_4 \, \mathcal{G}_{abc} \,  \mathfrak{G}^{[4]}  \wedge V^a \wedge V^b \wedge V^c\, + \, f_5 \,
\mathfrak{G}^{[3]} \wedge  \mathfrak{G}^{[4]}  \nonumber\\
&&+ \, \left (- \,360 f_2 \, \Phi^a \, \Phi_a \, - \, 120 \, f_3 \, e^{\phi} \,\mathcal{F}^{\Lambda|ab}
\mathcal{F}^{\Lambda}_{ab} - 6\, f_4 \, e^{2\phi} \, \mathcal{G}_{abc}\, \mathcal{G}^{abc} \right ) \,
\mathrm{Vol}_7 \nonumber\\
\mathrm{Vol}_7 & \equiv & \frac{1}{7!} \, \epsilon_{a_1\dots a_7} \, V^{a_1} \wedge \dots \wedge V^{a_7}
\label{LBkinBosu}
\end{eqnarray}
Eliminating the auxiliary fields that realize the first order formalism we can rewrite the second order form
of the  above lagrangian which reads as follows:
\begin{eqnarray}\label{osanna2}
   L_{Bose}^{ungauged} & = & \mbox{det}V \, \left( 240 \, f_1 \, R[g] \, + \, 360 \, f_2 \partial^\mu \phi \,
   \partial_\mu \phi\,
   + \, 6 \, f_4 e^{-2\phi} \, \mathcal{G}_{\lambda\mu\nu\rho}\,\mathcal{G}^{\lambda\mu\nu\rho} \,\right.\nonumber\\
   &&\left. +\,
   120 \, f_3 \, e^{\ft 12 \,\phi}\,\mathcal{F}^{\Lambda|\mu\nu} \,\mathcal{F}^{\Lambda}_{\mu\nu}  \right)
    \, d^7x  \, + \, f_5 \, \mathfrak{G}^{[4]} \wedge \mathfrak{G}^{[3]}
\end{eqnarray}
where $\mathcal{G}_{\lambda\mu\nu\rho}$ are the holonomic components of the field curvature
$\mathfrak{G}^{[4]}$:
\begin{equation}\label{Gstorto}
    \mathfrak{G}^{[4]} \, \equiv \, \mathrm{d}\mathbf{B}^{[3]} \, = \, \mathcal{G}_{\lambda\mu\nu\rho} \, dx^\lambda \wedge dx^\mu \wedge dx^\nu \wedge dx^\rho
\end{equation}
The anholonomic components of the same tensor with flat indices is related to $\mathcal{G}_{abc}$ by the
already established relation:
\begin{equation}\label{simplettus}
    \mathcal{G}_{a_1a_2a_3a_4}\, = \, \frac{1}{12} \, e^{2\phi} \, \epsilon_{a_1a_2a_3a_4pqr} \,\mathcal{G}^{pqr}
\end{equation}
An alternative way of writing the same lagrangian which is quite convenient while dealing with the equation of
motion is the following one:
\begin{eqnarray}\label{osanna}
   L_{Bose}^{ungauged} & = & \mbox{det}V \, \left( 240 \, f_1 \, \mathcal{R}[g] \, + \, 360
   \, f_2 \partial^\mu\phi \, \partial_\mu \phi\, \right) d^7x \nonumber\\
   && + \, \ft 14 \, f_4 e^{-2\phi} \, \mathfrak{G}^{[4]} \wedge \star\mathfrak{G}^{[4]}
   \, +\,  f_5 \, \mathfrak{G}^{[4]} \wedge \mathfrak{G}^{[3]} \, + \,
   60 \, f_3 \, e^{\phi} \, {\mathfrak{F}}^{\Lambda}\wedge \star\mathfrak{F}^{\Lambda}
\end{eqnarray}
Recalling that $\mathfrak{G}^{[3]}\, = \, \mathrm{d}\mathbf{B}^{[3]} \, + \,
{\mathfrak{F}}^{\Lambda}\wedge\mathcal{A}^\Lambda$ the field equations for the one-forms $\mathcal{A}^\Lambda$
and the three form $\mathbf{B}^{[3]}$ can be respectively written as follows\footnote{Here we use the a priori
information that $f_5 \, = \, -f_4$}:
\begin{eqnarray}
   \mathrm{d} \star\mathfrak{F}^{\Lambda} &=&  \frac{f_5}{60 \, f_3} \,{\mathfrak{F}}^{\Lambda}
   \wedge \mathfrak{G}^{[4]} \label{fequazia}\\
  \mathrm{d} \star\left[ e^{-2\phi} \, \star\mathfrak{G}^{[4]}\right] &=& 2
  \, \mathfrak{F}^{\Lambda} \wedge  \mathfrak{F}^{\Lambda}\label{gequazia}
\end{eqnarray}
while the equation for the dilaton takes the following form:
\begin{equation}\label{dilatonus}
    \Box \, \phi \, \mathrm{Vol}_7 = \, - \, \frac{f_4}{1440 \,f_2} e^{-2\phi} \, \mathfrak{G}^{[4]}
    \wedge \star\mathfrak{G}^{[4]} \, + \, \frac{ f_3}{12 \, f_2} \, e^{\phi}
    \, {\mathfrak{F}}^{\Lambda}\wedge \star\mathfrak{F}^{\Lambda}
\end{equation}
The Einstein equation for the metric can be finally written as follows:
\begin{eqnarray}
  \left(\mbox{Ric}_{\mu\nu} \, - \, \ft 12 \,g_{\mu\nu} \,  \mathcal{R}\right)
  &=& T^{\phi}_{\mu\nu} \, + \,T^{\mathcal{G}}_{\mu\nu}  \nonumber\\
  T^{\phi}_{\mu\nu} &=& - \, \ft 32 \, \frac{f_2}{f_1} \left(\partial_\mu\phi \,
  \partial_\nu\phi \, - \, \ft 12 \,g_{\mu\nu}\,\partial^\rho \phi \, \partial_\rho\phi\right) \label{dialstress}\\
 T^{\mathcal{G}}_{\mu\nu} &=& -\, \ft{1}{10} \, \frac{f_4}{f_1} \, \left(\mathcal{G}_{\mu...}
 \,\mathcal{G}_{\nu}^{\phantom{\nu}...}\, - \, \ft 18 \, g_{\mu\nu} \, \mathcal{G}_{....}\,\mathcal{G}^{....}\right)
\end{eqnarray}
where the dots denote saturated indices.
\subsection{Embedding the $2$-brane solution in supergravity}
\label{trovoDelta}
In order to embed the two brane solution discussed in section \ref{twobranastoria} into Minimal $D=7$
Supergravity one has to bring, by means of field redefinitions, the lagrangian (\ref{osanna}) to the standard
form of (\ref{braneaction}) or even (\ref{fluxbraneaction}) if we want to switch on Arnold Beltrami fluxes.
Let us divide the task in two parts. First we show that we can always embed the brane solution without fluxes,
next we consider the embedding of the flux brane solution and we work out the condition on the lagrangian
coefficients that has to be satisfied in order for such an embedding to be feasible.
\subsubsection{Matching with the pure brane action}
The first thing to do in order to compare (\ref{osanna}) with (\ref{braneaction}) is to truncate the gauge
fields $\mathcal{A}^\Lambda$ by setting them to zero, which is a consistent operation in the field equations
(\ref{fequazia}),(\ref{gequazia}) and (\ref{dilatonus}). Secondly we set the coefficient of the Einstein term
to the following value:
\begin{equation}\label{f1set}
    f_1 \, = \, - \, \frac{1}{5! 2!} \, = \, - \, \frac{1}{240}
\end{equation}
This is always possible since the overall constant in front of the lagrangian is a free parameter and
supersymmetry fixes all the other coefficients in terms of $f_1$. In the sequel the other coefficients
$\hat{f}_2,\hat{f}_3,\hat{f}_4$ are meant to attain the value predicted by supersymmetry when the Einstein
term is canonically normalized as in equation\footnote{For the reader not familiar with the rheonomy approach:
please remember that here the curvature 2-form is normalized to strength one so that the scalar curvature and
the Ricci tensor that we utilize are 1/2 of those utilized in traditional tensor calculus and standard
Relativity textbooks.}(\ref{f1set}):
\begin{equation}\label{hatti}
    \hat{f}_2 \, = \, \frac{f_2}{-240 \, f_1} \quad ; \quad \hat{f}_3 \, = \, \frac{f_3}{-240 \, f_1}
    \quad ; \quad \hat{f}_4 \, = \, \frac{f_4}{-240 \, f_1}
\end{equation}
The second and third steps consists of a rescaling of the dilaton and of the $\mathfrak{G}^{[4]}$-form. We
utilize the identifications provided by eq. (\ref{sigresca}), with the request that after rescaling the kinetic
terms become canonical namely:
\begin{equation}\label{colabrodo}
    6 \, \hat{f}_4 \, \tau^2 \, = \, \frac{1}{24}\,\hat{f}_4 \, = \, \frac{1}{96} \quad ;
    \quad 360 \, \hat{f}_2 \, \lambda^2 \, = \,360 \, \hat{f}_2 \,\frac 2 5 \, = \, - \, \frac{1}{4}
\end{equation}
The consistency of the above equations implies that when $f_1$ is negative, $f_2 <0$ should also be negative
and $f_4 >0$ should instead be positive. This requirement, although we have not yet fixed the coefficients by
supersymmetry, should be in any way respected, since it corresponds to positivity of the energy in the mostly
minus conventions for the metric signature. In this way we find:
\begin{eqnarray}
 \hat{ f}_4&=& \frac{1}{4} \label{taupara}\\
  \hat{ f}_2 &=& - \, \frac{1}{576} \label{bilancianera}
  \end{eqnarray}
\subsection{Matching with the flux brane action} In order for the flux brane action (\ref{fluxbraneaction})
to match the bosonic action of supergravity further conditions have to be satisfied by the action
coefficients. We presently derive them. First we consider the rescaling necessary to bring the kinetic term of
the gauge fields $\mathcal{A}^\mu$ to the normalization used in eq.(\ref{fluxbraneaction}). Referring to
eq.(\ref{sigresca}) we see that the necessary rescaling is given by:
\begin{equation}\label{sigval}
    \sigma^2 \, = \, - \, \frac{f_1}{4 \, f_3} \, \omega
\end{equation}
Then we can evaluate, in terms of $f_5$ the value of the parameter $\kappa$ appearing in the lagrangian
(\ref{fluxbraneaction}). We find the condition:
\begin{equation}\label{gospadi}
    \kappa \, = \, \frac{f_5}{240 \, f_1} \, \tau \, \sigma^2 \, = \,\frac{\omega}{384}
\end{equation}
Utilizing $\tau \, = \, \ft{1}{12}$, the identification (\ref{sigval}) and:
\begin{equation}\label{gospadi2}
   f_5 \, = \, - \, f_4 \, =\, \frac{1}{4} \times (240 \, f_1)
\end{equation}
we get:
\begin{equation}\label{gomitolodiseta}
 f_3 \, = \, 2 \, f_1 \quad \Rightarrow \quad\hat{ f}_3 \, = \,-\, \frac{1}{120}
\end{equation}

\section{Auxiliary items of the construction}
In this paper we utilize two different basis of gamma matrices in $D=7$. One basis, the antisymmetric ones is
the best suited to check identities in the general rheonomic construction of the theory. The second basis, the
split one, is instead well-adapted to brane solutions and it is best-suited for the analysis of Killing spinor
equations.
\subsection{$D=7$ gamma matrices in the antisymmetric basis}
\label{gammola} As mentioned in the main text the gamma matrices in $D=7$ Minkowski signature with mostly
minus metric:
\begin{equation}\label{clifford7}
    \left\{ \Gamma_a \, , \, \Gamma_b \right\} \, = 2\, \eta_{ab} \, \mathbf{1}_{8\times8} \quad ;
    \quad \eta_{ab} \, = \, \mbox{diag} \, \left\{+,-,-,-,-,-,-\right\}
\end{equation}
are all antisymmetric $\Gamma_a^T \, = \, - \, \Gamma_a$ and admit $C_{-} \, = \, \mathbf{1}_{8\times8}$ as
charge conjugation matrix. A convenient explicit representation is the following one: {\small
\begin{equation}\label{d7gamma}
    \begin{array}{ccccccc}
       \Gamma_0 & = & \left(
\begin{array}{llllllll}
 0 & 0 & 0 & 0 & 0 & 0 & i & 0
   \\
 0 & 0 & i & 0 & 0 & 0 & 0 & 0
   \\
 0 & -i & 0 & 0 & 0 & 0 & 0 &
   0 \\
 0 & 0 & 0 & 0 & 0 & 0 & 0 & i
   \\
 0 & 0 & 0 & 0 & 0 & -i & 0 &
   0 \\
 0 & 0 & 0 & 0 & i & 0 & 0 & 0
   \\
 -i & 0 & 0 & 0 & 0 & 0 & 0 &
   0 \\
 0 & 0 & 0 & -i & 0 & 0 & 0 &
   0
\end{array}
\right) & ; & \Gamma_1 & = & \left(
\begin{array}{llllllll}
 0 & 1 & 0 & 0 & 0 & 0 & 0 & 0 \\
 -1 & 0 & 0 & 0 & 0 & 0 & 0 & 0 \\
 0 & 0 & 0 & 0 & 0 & 0 & 1 & 0 \\
 0 & 0 & 0 & 0 & -1 & 0 & 0 & 0 \\
 0 & 0 & 0 & 1 & 0 & 0 & 0 & 0 \\
 0 & 0 & 0 & 0 & 0 & 0 & 0 & 1 \\
 0 & 0 & -1 & 0 & 0 & 0 & 0 & 0 \\
 0 & 0 & 0 & 0 & 0 & -1 & 0 & 0
\end{array}
\right) \\
       \Gamma_2 & = & \left(
\begin{array}{llllllll}
 0 & 0 & 0 & 0 & 0 & 0 & 0 & -1 \\
 0 & 0 & 0 & 0 & 0 & -1 & 0 & 0 \\
 0 & 0 & 0 & 0 & 1 & 0 & 0 & 0 \\
 0 & 0 & 0 & 0 & 0 & 0 & 1 & 0 \\
 0 & 0 & -1 & 0 & 0 & 0 & 0 & 0 \\
 0 & 1 & 0 & 0 & 0 & 0 & 0 & 0 \\
 0 & 0 & 0 & -1 & 0 & 0 & 0 & 0 \\
 1 & 0 & 0 & 0 & 0 & 0 & 0 & 0
\end{array}
\right) & ; & \Gamma_3 & = & \left(
\begin{array}{llllllll}
 0 & 0 & 0 & 0 & 1 & 0 & 0 & 0 \\
 0 & 0 & 0 & -1 & 0 & 0 & 0 & 0 \\
 0 & 0 & 0 & 0 & 0 & 0 & 0 & 1 \\
 0 & 1 & 0 & 0 & 0 & 0 & 0 & 0 \\
 -1 & 0 & 0 & 0 & 0 & 0 & 0 & 0 \\
 0 & 0 & 0 & 0 & 0 & 0 & -1 & 0 \\
 0 & 0 & 0 & 0 & 0 & 1 & 0 & 0 \\
 0 & 0 & -1 & 0 & 0 & 0 & 0 & 0
\end{array}
\right) \\
\end{array}
\end{equation}
 \begin{equation}\label{d7gammabis}
    \begin{array}{ccccccc}
       \Gamma_4 & = & \left(
\begin{array}{llllllll}
 0 & 0 & 0 & 1 & 0 & 0 & 0 & 0 \\
 0 & 0 & 0 & 0 & 1 & 0 & 0 & 0 \\
 0 & 0 & 0 & 0 & 0 & 1 & 0 & 0 \\
 -1 & 0 & 0 & 0 & 0 & 0 & 0 & 0 \\
 0 & -1 & 0 & 0 & 0 & 0 & 0 & 0 \\
 0 & 0 & -1 & 0 & 0 & 0 & 0 & 0 \\
 0 & 0 & 0 & 0 & 0 & 0 & 0 & -1 \\
 0 & 0 & 0 & 0 & 0 & 0 & 1 & 0
\end{array}
\right) & ; & \Gamma_5 & = & \left(
\begin{array}{llllllll}
 0 & 0 & 0 & 0 & 0 & -1 & 0 & 0 \\
 0 & 0 & 0 & 0 & 0 & 0 & 0 & 1 \\
 0 & 0 & 0 & 1 & 0 & 0 & 0 & 0 \\
 0 & 0 & -1 & 0 & 0 & 0 & 0 & 0 \\
 0 & 0 & 0 & 0 & 0 & 0 & -1 & 0 \\
 1 & 0 & 0 & 0 & 0 & 0 & 0 & 0 \\
 0 & 0 & 0 & 0 & 1 & 0 & 0 & 0 \\
 0 & -1 & 0 & 0 & 0 & 0 & 0 & 0
\end{array}
\right) \\
       \Gamma_6 & = & \left(
\begin{array}{llllllll}
 0 & 0 & -1 & 0 & 0 & 0 & 0 & 0 \\
 0 & 0 & 0 & 0 & 0 & 0 & 1 & 0 \\
 1 & 0 & 0 & 0 & 0 & 0 & 0 & 0 \\
 0 & 0 & 0 & 0 & 0 & 1 & 0 & 0 \\
 0 & 0 & 0 & 0 & 0 & 0 & 0 & 1 \\
 0 & 0 & 0 & -1 & 0 & 0 & 0 & 0 \\
 0 & -1 & 0 & 0 & 0 & 0 & 0 & 0 \\
 0 & 0 & 0 & 0 & -1 & 0 & 0 & 0
\end{array}
\right) & ; & \null & \null & \null
     \end{array}
\end{equation}}
\subsubsection{Pauli matrices}
\label{sigmotta} We also spell out the explicit form of the three Pauli matrices that we use in our
construction:
\begin{equation}\label{sigmallo}
    \sigma^{\Lambda=1,2,3} \quad: \quad \sigma^1 \, = \, \left(
\begin{array}{ll}
 0 & 1 \\
 1 & 0
\end{array}
\right) \quad ; \quad \sigma^2 \, = \, \left(
\begin{array}{ll}
 0 & -\, i \\
 i & 0
\end{array}
\right) \quad ; \quad \sigma_3 \, = \, \left(
\begin{array}{ll}
 1 & 0 \\
 0 & -1
\end{array}
\right)
\end{equation}
\subsection{$D=7$ gamma matrices in the split basis}
\label{splittorio}
 The gamma matrices in the split basis are devised to be well-adapted to the $2$-brane
solutions. To this effect we split the seven-dimensional flat indices according to the following notations:
\begin{equation}\label{splittaggio}
  {a,b,c,\dots}=\left\{\begin{array}{lcll}
                         \bar{a},\bar{b},\bar{c},\dots & = & \bar{1},\bar{2},\bar{3} & \mbox{brane world volume directions} \\
                         P,Q,R,\dots &= & 1,2,3,4 & \mbox{directions transverse to the brane}
                       \end{array}
   \right.
\end{equation}
Next we write the $7$-dimensional $8\times 8$ gamma  matrices as the following tensor products
\begin{eqnarray}
\Gamma _{\bar{a}} &=& \gamma _{\bar{a}}\otimes \tau _5 \nonumber\\
  \Gamma _P &=& \mathbf{1}_{2\times2}\otimes \tau _P \label{splittus1}
\end{eqnarray}
where
\begin{eqnarray}
  \left\{\gamma _{\bar{a}},\gamma _{\bar{b}}\right\} &=& 2 \eta _{\bar{a} \bar{b}} \quad ; \quad  \eta \, = \, \mbox{diag} \left(+,-,-\right) \nonumber \\
  \left\{\tau _P,\tau _Q\right\} &=& -2 \delta _{PQ}\nonumber\\
  \left\{\tau _5,\tau _Q\right\} &=& 0 \label{splittus2}
\end{eqnarray}
Explicitly, in terms of the Pauli matrices,  we can set:
\begin{equation}\label{gammettine}
  \gamma_{\bar{1}} \, = \, \sigma_2\quad ; \quad  \gamma_{\bar{2}} \, = \,  {\rm i} \, \sigma_1 \quad ; \quad  \gamma_{\bar{3}} \, = \, {\rm i} \, \sigma_3
\end{equation}
and
\begin{eqnarray}
  \tau_1 &=& \mbox{i$\sigma $}_1\otimes \mathbf{1}_{2\times2}\nonumber \\
 \tau_{1+i} &=& \mbox{i$\sigma $}_1\otimes \sigma_{i}\quad ; \quad (i=1,2,3)  \nonumber\\
 \tau_5 &=& \sigma_3\otimes \mathbf{1}_{2\times2} \label{taumette}
\end{eqnarray}
In this basis the charge conjugation matrix is not the identity matrix, rather it is the following symmetric
matrix:
\begin{eqnarray}
  C &=& \mbox{i$\sigma $}_2\otimes C_4 \\
  C_4 &=& \sigma_3 \otimes  \mbox{i$\sigma $}_2
\end{eqnarray}

\end{document}